\documentclass[12pt]{article}
\usepackage{graphicx}
\usepackage{url}

\usepackage{PRIMEarxiv}
\usepackage{hyperref,booktabs,amsfonts,nicefrac,microtype,lipsum,fancyhdr,booktabs,caption,amssymb,amsmath,subcaption,multirow,parskip,authblk,xpatch,cite,xcolor}
\usepackage[utf8]{inputenc} 
\usepackage[T1]{fontenc}    
\graphicspath{{media/}}
\usepackage[flushleft]{threeparttable}
\usepackage[labelfont=bf]{caption}
\usepackage[toc]{appendix}

\title{Towards Ultimate NMR Resolution with\\ Deep Learning}

\author[$\star$,1]{\textbf{Amir Jahangiri}}
\author[1,2]{\textbf{Tatiana Agback}}
\author[1]{\textbf{Ulrika Brath}}
\author[$\star$,1]{\textbf{Vladislav Orekhov\thanks{\textit{Corresponding authors: \href{mailto:amir.jahangiri@gu.se}{amir.jahangiri@gu.se} (Amir Jahangiri),
\href{mailto:vladislav.orekhov@nmr.gu.se}{vladislav.orekhov@nmr.gu.se}}(Vladislav Orekhov)}}}

\affil[1]{Department of Chemistry and Molecular Biology, Swedish NMR Centre, University of Gothenburg, Box 465, Gothenburg, 40530, Sweden}
\affil[2]{Department of Molecular Sciences, Swedish University of Agricultural Sciences, Box 7015, Uppsala, 75007, Sweden}

\begin{document}

\maketitle


\begin{abstract} \bfseries \boldmath
    In multidimensional NMR spectroscopy, practical resolution is defined as the ability to distinguish and accurately determine signal positions against a background of overlapping peaks, thermal noise, and spectral artifacts. In the pursuit of ultimate resolution, we introduce Peak Probability Presentations ($P^3$)—a statistical spectral representation that assigns a probability to each spectral point, indicating the likelihood of a peak maximum occurring at that location. The mapping between the spectrum and $P^3$ is achieved using MR-Ai, a physics-inspired deep learning neural network architecture, designed to handle multidimensional NMR spectra. Furthermore, we demonstrate that MR-Ai enables co-processing of multiple spectra, facilitating direct information exchange between datasets. This feature significantly enhances spectral quality, particularly in cases of highly sparse sampling. Performance of MR-Ai and high value of the $P^3$ are demonstrated on the synthetic data and spectra of Tau, MATL1, Calmodulin, and several other proteins. 
\end{abstract}


\section{Introduction}
\label{sec:Introduction}
    NMR spectroscopy has provided increasingly valuable insights into the behavior and properties of molecules. Over the past decades, it has established itself as an essential atomic-level tool in structural biology \cite{claridge2016high} and has played a crucial role in protein structural analysis \cite{hu2021nmr}. Despite its versatility, NMR spectroscopy of biological systems is often constrained by limited spectral resolution, where signal overlap complicates chemical shift assignment and interferes with the analysis of molecular dynamics and structure \cite{cavanagh1996protein}. Since the introduction of Fourier NMR spectroscopy in the middle of 1960s \cite{ernst1966application}, numerous signal processing methods have been developed and are now routinely used to enhance resolution. These include traditional techniques such as zero-padding \cite{lindon1980digitisation,bartholdi1973fourier}, apodization with weighting functions \cite{ebel2006effects}, linear prediction \cite{oschkinat1988three}, virtual decoupling by deconvolution \cite{delsuc1988application, shimba2004optimization,kazimierczuk2020resolution, qiu2023resolution}, and, more recently, spectral signal sharpening using artificial neural networks (NNs) \cite{shukla2024combined}.
    In response to the limitations of traditional approaches, Artificial Intelligence (AI), particularly Deep Learning (DL), has demonstrated significant potential across various areas of NMR research \cite{chen2020review,shukla2023biomolecular,luo2025deep}. AI-based tools not only surpass traditional NMR techniques in rapid and high-quality non-uniform sampling (NUS) reconstruction \cite{qu2020accelerated,hansen2019using,karunanithy2021fid,jahangiri2023nmr}, efficient homonuclear decoupling \cite{karunanithy2021virtual,jahangiri2023nmr}, pure shift spectra generation \cite{zheng2022fast,zhan2024fast,zhan2024accelerated}, spectra denoising \cite{lee2019intact,chen2023magnetic}, and automated peak picking \cite{klukowski2018nmrnet,li2022fundamental}, but also have the potential to push beyond the boundaries of traditional Magnetic Resonance processing, for example allowing reference-free assessment of the spectra quality and obtaining phase sensitive spectra without quadrature detection \cite{jahangiri2024beyond}.

    In this work, we address the challenge of achieving ultimate resolution and reducing data complexity in experimental NMR spectra affected by thermal noise, spectral artifacts, and signal overlap. Instead of relying on the relatively broad and loosely defined notion of spectral resolution, we redefine the problem in a more precise statistical framework—determining the probability of finding the center of a signal at any given point in the spectrum. This reformulation shifts the focus to statistical analysis using AI \cite{bengio2017deep}.

    The Bayesian approach has previously been used to define posterior probability distributions of spectral parameters in 1D metabolomic spectra, leveraging a templated library of expected underlying compounds \cite{astle2012bayesian,hao2014bayesian}. However, applying traditional Bayesian methods to unconstrained multidimensional protein spectra is computationally prohibitive due to the immense processing demands associated with Markov chain Monte Carlo (MCMC) algorithms \cite{andrieu1999sequential}. AI circumvents this computational bottleneck by exploiting the efficiency of deep learning in solving classification problems \cite{bengio2017deep}, enabling direct probability predictions. Reformulating the resolution problem as a classification task—peak or no peak—bridges the gap between statistical analysis and practical applications in multidimensional NMR.
    
    NMR spectra are commonly analyzed in the frequency domain using Traditional Intensity Presentation (TIP). While TIP is highly informative, it has notable drawbacks, including peak overlap exacerbated by the high dynamic range of signal intensities and difficulties in distinguishing genuine peaks from spectral artifacts. As a complementary alternative to TIP, we introduce Peak Probability Presentation ($P^3$), a statistical representation applicable to NMR spectra of any dimensionality. $P^3$ assigns, to each point in the spectrum, the probability of it being a peak maximum. This approach offers several advantages, including super-resolution, high sensitivity, a significantly reduced dynamic range, and effective suppression of spectral artifacts.

    Using simulated data and experimental spectra from several challenging systems including the globular MALT1 protein (45 kDa) and the intrinsically disordered Tau protein (45.8 kDa) we demonstrate that $P^3$ achieves near-ultimate spectral resolution based on the available information in 2D and 3D spectra.
    We introduce the newly developed Magnetic Resonance Processing with AI (MR-Ai) system, which is capable of handling both conventionally acquired spectra and non-uniformly sampled (NUS) data reconstructed using nonlinear compressed sensing algorithms \cite{jaravine2006removal,mobli2014nonuniform,qu2015accelerated,kazimierczuk2011accelerated,hyberts2012application,hassanieh2015fast,hyberts2010poisson,pustovalova2018xlsy,sibisi1984maximum,drori2007fast,holland2011fast,jiang2010gridding,ying2017sparse}.
    Furthermore, using a set of triple-resonance experiments for backbone assignment collected on the Calamondin protein, we demonstrate that $P^3$ can serve as a spectral quality metric, similar to the number of detected peaks in Targeted Acquisition (TA) data collection schemes \cite{jaravine2006targeted,isaksson2013highly}.
    
\section{Results and Discussion}
\label{sec:Results}

\subsection{Ultimate Resolution and dynamic range in $\mathbf{P}^\mathbf{3}$:}
        
\begin{figure}[htbp]
        \centering
        \includegraphics[width=\textwidth]{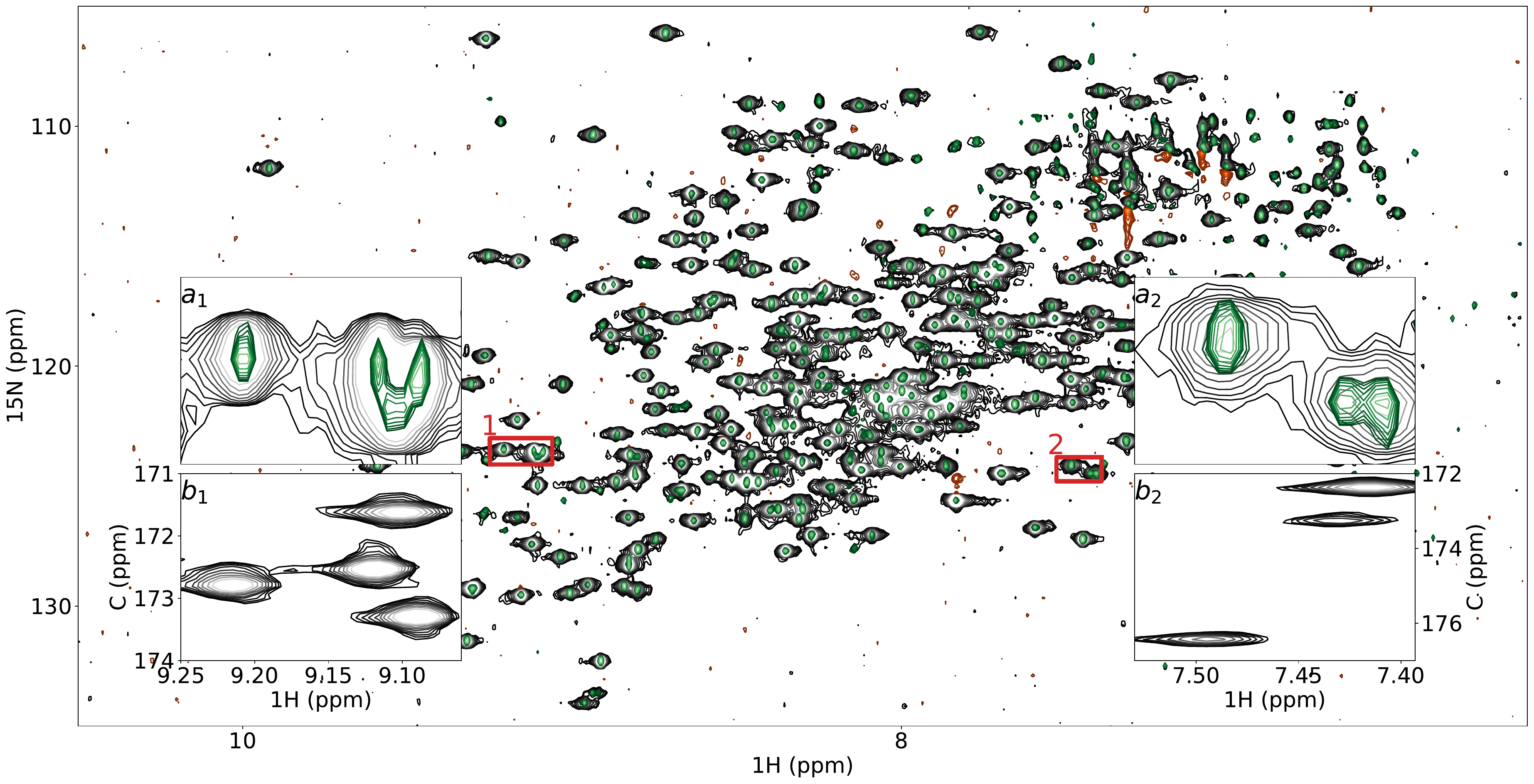}
        \caption{\textbf{The $P^3$ of experimental 2D NMR data.}
        2D $^{1}$H-$^{15}$N --- TROSY spectrum of MALT1 protein is shown with the black (positive) and orange (negative) contours, while the corresponding $P^3$ spectrum is depicted in green. Inset panels $a_1$ and $a_2$ highlight zoomed-in regions (marked with red rectangles) of the spectrum; $b_1$ and $b_2$ confirm the resolved peaks in $a_1$ and $a_2$ by showing the corresponding $^{1}$H-$^{13}$C slice from the 3D HNCO spectrum.}
        \label{fig:2D_Malt}
    \end{figure}

    Figure \ref{fig:2D_Malt} demonstrates the excellent performance of MR-Ai in generating $P^3$ for the 2D $^{1}$H-$^{15}$N correlation spectrum of MALT1 protein (45 kDa) \cite{unnerstaale2016backbone,han2022assignment}. Similar results for Ubiquitin (7 kDa) \cite{brzovic2006ubch5}, Azurin (14 kDa) \cite{korzhnev2003nmr}, and Tau (disordered, 45.8 kDa) \cite{lesovoy2021unambiguous} are provided in Supplementary Figures \ref{fig:HSQC_Ubiquitin}, \ref{fig:HSQC_Azurin}, and \ref{fig:HSQC_Tau}, respectively.
    
    The $P^3$ representation exhibits significant line narrowing across the 2D spectrum while accurately reproducing both strong and weak peaks. The two insets in Figure \ref{fig:2D_Malt} confirm the improved resolution by displaying the peaks resolved in $P^3$ alongside the corresponding $^{1}$H-$^{13}$C slice from the 3D HNCO spectrum.

    \begin{figure}[htbp] 
        \centering
        \includegraphics[width=0.85\textwidth]{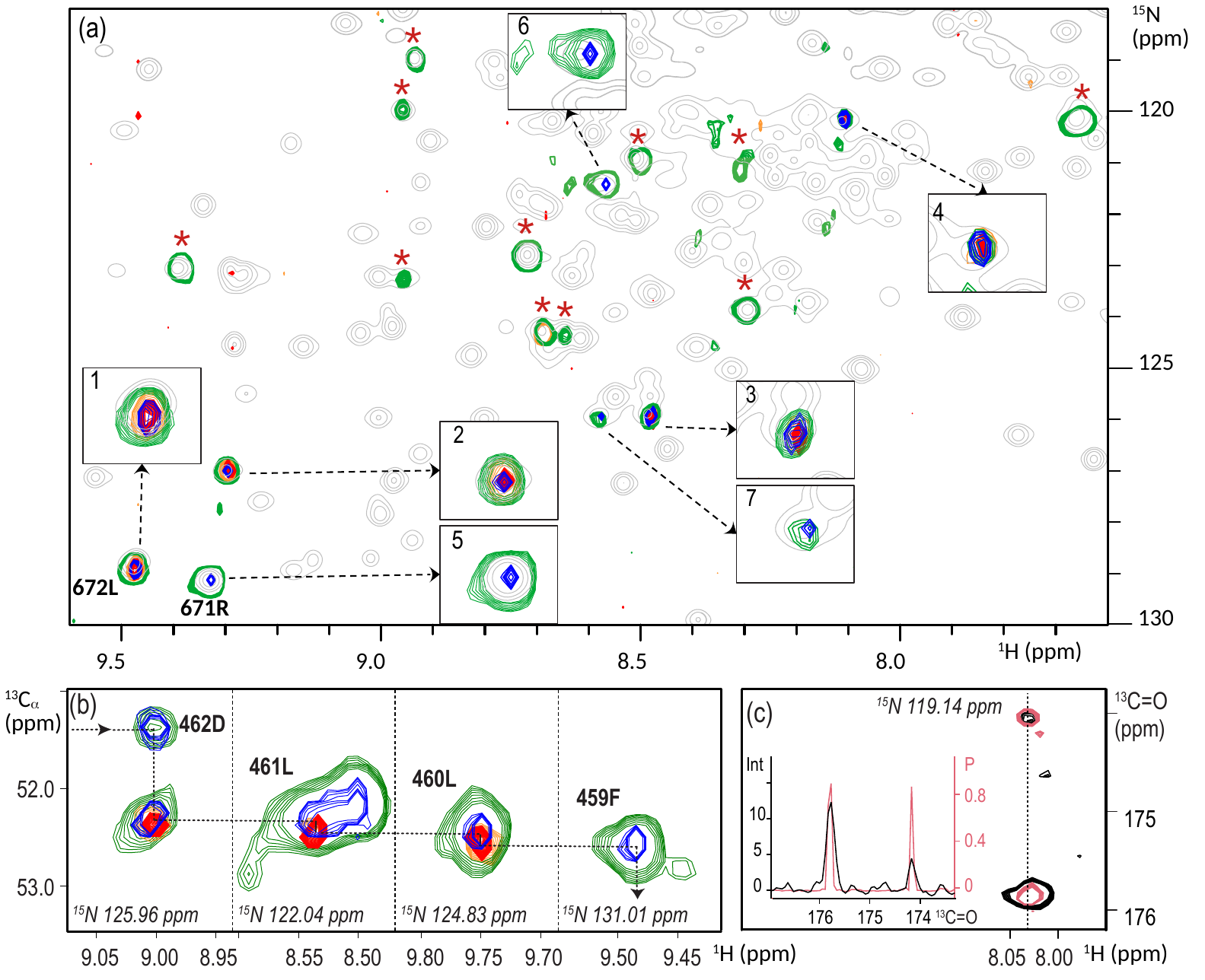}
        \caption{\textbf{$P^3$ and assignment procedure of 3D NUS reconstructed spectra with CS-IST for MALT1 Protein.}
        \textbf{(a)} Superposition of the $^{1}$H-$^{15}$N 2D planes, extracted at $^{13}$C$^\alpha$ frequency of $671R$: 55.01 ppm from 3D HNCA (green and blue for TIP and $P^3$) and HN(CO)CA (orange and red for TIP and $P^3$) spectra, and cross-peaks of the $^{1}$H-$^{15}$N 2D TROSY spectrum (grey): Sequential $(i)-(i-1)$ peak labeled $672L$ and enlarged in the inset box number one has the $^{1}$H and $^{15}$N chemical shifts of residue $672L$ and the $^{13}$C chemical shift of residue $671R$. The target $(i-1)$ cross-peak with the $^{1}$H, $^{15}$N and $^{13}$C chemical shifts of residue $671R$, is enlarged in box 5. Three additional sequential $(i)-(i-1)$ and two $(i-1)$ peaks appear in the plane because they exhibit $^{13}$C chemical shifts similar to that of $671R$. These peaks correspond to residues $679K-678L$ (peak 2), $457Q-456L$ (peak 3), $534E-533A$ (peak 4), $533A$ (peak 6), and one unidentified peak (7). Compared with $P^3$, due to limited $^{13}$C resolution, the 3D HNCA (green) and HN(CO)CA (orange) planes also include eleven additional cross-peaks, marked with red stars making the assignment procedure more complicated.
        \textbf{(b)} The superposition of four strips $^{1}$H-$^{13}$C 2D planes, extracted at $^{15}$N: 125.96, 122.04, 124.83 and 131.01 ppm, respectively, from 3D HNCA (green and blue for TIP and $P^3$) and HN(CO)CA (orange and red for TIP and $P^3$) spectra: The dashed line illustrates the flow of assignments through $(n)$ to $(n-1)$ cross peaks, starting with residue $462D$, and continuing through $461L$, $460L$ and $459F$.
        \textbf{(c)} The superposition of the $^{1}$H-$^{13}$C 2D planes, extracted at $^{15}$N: 119.14 ppm from 3D HN(CA)CO (black and pink for TIP and $P^3$) spectra: The inside panel displays 1D projections corresponding to the dash column at $^{1}$H: 8.038 ppm, representing the chemical shifts of residue $545K$ and $544G$ where the vertical axes (on the right and left) are scaled according to Intensity (Int) and Probability (P), respectively.}
        \label{fig:Malt_3D}
    \end{figure}
         
    Figure \ref{fig:Malt_3D} illustrates the application of $P^3$ to the backbone assignment of the 45 kDa MALT1 protein using 3D HNCA and HN(CO)CA spectra. The assignment progresses from residue $672L$ to its $(i-1)$ preceding residue $671R$ through visual inspection of the $^{1}$H-$^{15}$N 2D planes extracted from the two 3D spectra at $^{13}$C: 55.01 ppm, corresponding to the $^{13}$C$\alpha$ frequency of $(i-1)$ $671R$.
    In the crowded spectra of MALT1, this assignment is significantly complicated by the presence of multiple candidate cross-peaks for $(i-1)$ $671R$. Figure \ref{fig:Malt_3D}a shows numerous cross-peaks in HNCA (green contours) and HN(CO)CA (orange contours), all of which must be carefully evaluated to establish the correct sequential connection. The ambiguity, arising from insufficient resolution in the $^{13}$C dimension of traditional spectra, is largely alleviated in the $P^3$ representation (Figure \ref{fig:Malt_3D}a, blue for HNCA and red for HN(CO)CA). The enhanced resolution eliminates most of the irrelevant peaks (marked by red stars). Among the remaining seven peaks, only peak 5 (the correct $671R$ peak) and peaks 6 and 7 are retained in the $P^3$ of the HNCA spectrum.

    Another example of the superior resolution along the $^{1}$H and $^{13}$C dimensions in the $P^3$ representation of the MALT1 3D HNCA and HN(CO)CA spectra is shown in Figure \ref{fig:Malt_3D}b. This figure illustrates the assignment walk from $462D$ to $459F$, tracing $(i)$ to $(i-1)$ cross-peaks in four $^{1}$H-$^{13}$C 2D strips extracted at $^{15}$N: 125.96, 122.04, 124.83, and 131.01 ppm, respectively.
    The assignment walk is particularly challenging when relying solely on traditional 3D HNCA (green contours) and HN(CO)CA (orange contours) spectra, often necessitating additional experiments. In contrast, the walk based on better resolved  $P^3$, shown in blue (HNCA) and red (HN(CO)CA), is highly reliable, offering improved confidence in peak identification.

   A remarkable difference in signal dynamic range between the traditional intensity presentation and $P^3$ is illustrated in Figure \ref{fig:Malt_3D}c, which shows a 2D strip extracted from the 3D HN(CA)CO spectrum for residue $545K$ at $^{15}$N: 119.14 ppm. Two cross-peaks, observed at 174.07 ppm and 175.87 ppm, correspond to the carbonyl $^{13}$C frequencies of $544G$ and $545K$, respectively.
   A 1D slice taken through these two cross-peaks at $^{1}$H: 8.038 ppm reveals a significant difference in the relative peak amplitudes. In the $P^3$ representation, both peaks exhibit high and nearly equal probability (about 0.8), whereas in the conventional spectrum, their amplitude ratio is 3:1. Notably, in 3D HN(CA)CO spectrum, optimized for detecting $^{1}$H-$^{15}$N-$^{13}$C(i) cross-peaks corresponding to a residue’s own carbonyl carbon, the second peak, which corresponds to the carbonyl of the preceding $(i-1)$ residue, typically has lower intensity or may even disappear. The ability to simultaneously visualize all reliably detected signals, regardless of their intensity, significantly simplifies manual spectral analysis. 

\subsection{Validation of the $\mathbf{P}^\mathbf{3}$ on synthetic spectra}

\begin{figure}[htbp]
        \centering
        \includegraphics[width=0.75\textwidth]{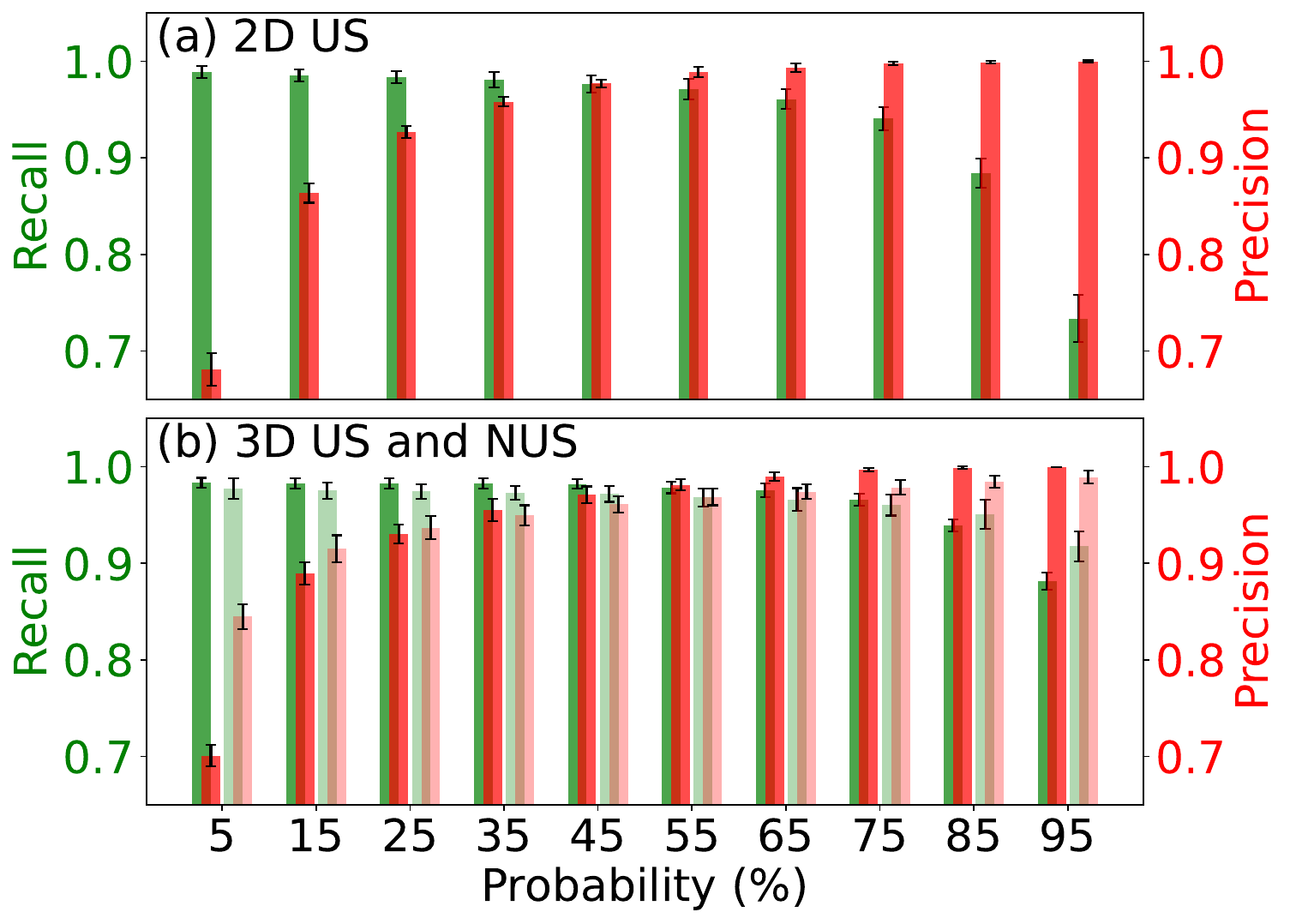}
        \caption{\textbf{Statistics on Peak Detection of $P^3$ in Synthetic 2D and 3D Spectra.} Green and red bars with error bars represent the mean and standard deviation of recall and precision quality metrics across 10 synthetic NMR spectra, each containing 256 peaks, for (a) 2D US, (b) 3D US (dark-colored bars), and 3D 15\% NUS reconstructed using CS-IST (light-colored bars). Recall is defined as the ratio of the correctly $detected$ pixels to all $detectable$ pixels, while precision is the ratio of correctly $detected$ pixels to all $detected$ pixels. A pixel is considered as $detected$ when its $P^3$ value is above the probability threshold indicated on the horizontal axis of the chart. A pixel is correctly $detected$ if it is found in the vicinity of a $detectable$ pixel. The $detectable$ pixels are defined as those near the maxima of the ground truth peaks with intensities higher than $2\sigma$-noise for the US spectra. To account for the shorter experiment time in the 15\% NUS spectra, a threshold of $5\sigma$-noise from the corresponding US spectra was used.}
        \label{fig:Metric}
    \end{figure}

    Despite the promising results obtained for the spectra of MALT1, Ubiquitin, Azurin, and Tau proteins, quantitatively assessing the performance of $P^3$ using experimental data remains challenging due to the absence of ground truth peak labeling. To address this limitation, we evaluated $P^3$ using synthetic spectra. Figure \ref{fig:Metric} presents the recall and precision metrics \cite{sammut2011encyclopedia} of $P^3$, calculated across ten 2D and ten 3D synthetic spectra. Each spectrum contains 256 peaks with varying degrees of overlap and amplitudes, spanning a dynamic range of 1:200, with the weakest peaks reaching one $\sigma$-noise. In all cases, including fully sampled 2D and 3D spectra as well as 3D NUS spectra, a favorable balance is observed between the number of spectral points correctly and incorrectly assigned to peak maxima. Thus, around a shallow optimum at approximately 50\% probability, most true peaks are accurately detected, while the false detection rate remains very low.

    Although 2D spectra exhibit a significantly higher degree of peak overlap compared to 3D spectra, $P^3$ demonstrates remarkably consistent performance in both cases. Furthermore, as shown in Figure \ref{fig:Metric}b, the quality of $P^3$ in 3D NUS-reconstructed spectra, which feature strongly non-Gaussian baseline noise, is comparable to that of uniformly sampled (US) data.
    
\begin{figure}[htbp]
        \centering             
        \includegraphics[width=\textwidth]{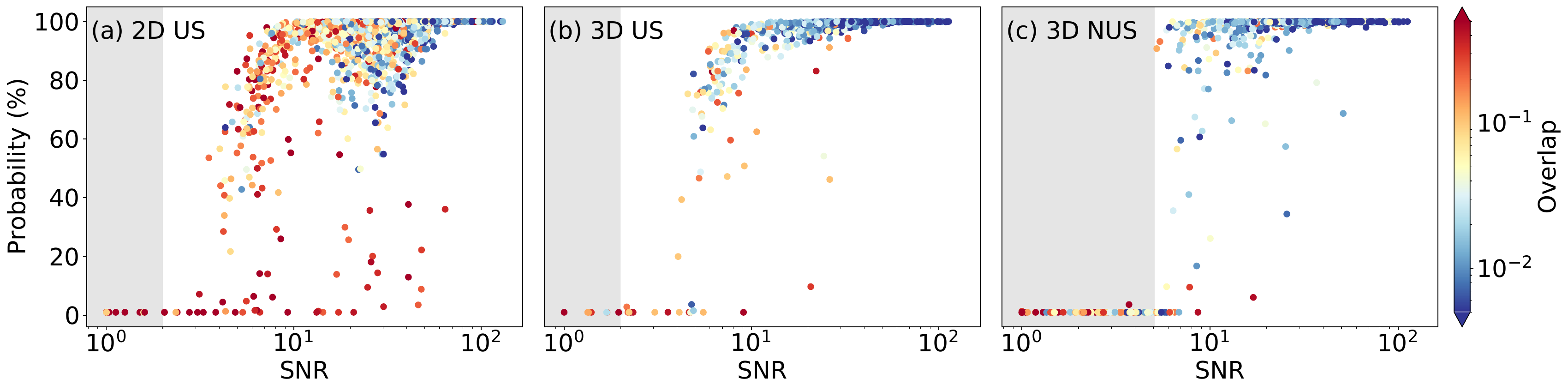}
        \caption{\textbf{Effect of peak intensity and overlap on the peak detection in the  $P^3$ for the synthetic 2D and 3D Spectra.} The probability of detecting peak maxima across 2,560 ground truth peaks in 10 synthetic spectra are plotted versus the signal-to-noise ratio (SNR) and overlap conditions (color scale) for  (a) 2D US, (b) 3D US, and (c) 3D 15\% NUS reconstructed using CS-IST spectra. For each ground truth peak with intensity $I_0$, SNR is calculated as $I_0$ divided by $\sigma$-noise in the US spectra (or the corresponding US spectra for NUS data). The overlap score is defined as $\Sigma_i \frac{I_i}{I_0 d^2}$, where $I_i$ is the intensity of a neighboring peak $i$ residing at a distance $d$ less than 16 pixels.
        The gray arias indicate a range of signal amplitudes where the ground truce peaks are theoretically undetectable with intensities below $2\sigma$ in the US spectra; in order to account for the shortened measurement time in the  15\% NUS spectra, the theoretical detention threshold was set at  $5\sigma$-noise in the corresponding US spectra.}
        \label{fig:SNR_P}
    \end{figure}

    Figure \ref{fig:SNR_P} illustrates the range of probability values for identifying peak maxima at spectral points corresponding to ground-truth peaks in synthetic 2D and 3D spectra. The $P^3$ values are clearly influenced by the peak signal-to-noise ratio (SNR) and the degree of peak overlap. The latter is determined by the number of neighboring peaks, their relative distances, and their amplitudes compared to the peak in question.
   
    In the 2D spectra (Figure \ref{fig:SNR_P}a), peaks with $SNR < 3–4$ have nearly zero probability values and are therefore not detected. This lower boundary for peak detection is close to the theoretical detection limit at a $5\%$ confidence level ($SNR = 2$) and can also be attributed to the substantially higher degree of interference and overlap with other peaks, particularly for low-intensity peaks, as indicated by the color scale in Figure \ref{fig:SNR_P}.
    Although most peaks with $SNR > 5$ are detected with high probability, a few medium-intensity peaks exhibit reduced probability values due to significant overlap. These peaks are located at the bottom of the chart. A similar pattern, although with less pronounced overlap, is observed in both 3D uniformly sampled (US) spectra (Figure \ref{fig:SNR_P}b) and 3D $15\%$-NUS spectra (Figure \ref{fig:SNR_P}c).
    A notable feature of $P^3$ in the $15\%$-NUS 3D spectra is that the apparent peak detection threshold is very close to the theoretical limit. This highlights the near-ideal performance of both the CS-IST NUS spectrum reconstruction algorithm and the $P^3$ produced by MR-Ai for synthetic spectra.

    The results obtained from experimental spectra of several representative proteins, supported by quantitative validation using synthetic data, establish $P^3$ as a powerful tool for significantly enhancing spectral resolution and simplifying analysis. Furthermore, $P^3$ highlights new opportunities in NMR signal processing and analysis enabled by AI-driven approaches.

\subsection{Reference-Free Quantitative Spectrum Quality Score with $\mathbf{P}^\mathbf{3}$, QSP:}

    In our previous work \cite{jahangiri2024beyond}, we demonstrated that MR-Ai can be trained to predict intensity uncertainties in spectra reconstructed by various methods. These predicted uncertainties can serve as a reference-free score for assessing spectrum quality.
    In this work, we introduce the integral of $P^3$ as a useful proxy for the number of peaks in a spectrum. This is reminiscent of traditional 1D NMR, where the integral of the spectrum is proportional to the number of spins in the sample. We further introduce a quantitative spectrum quality score (QSP) corresponding to the number of detectable peaks, which is defined as the number of clusters of spectral points with $P^3$ values exceeding a defined threshold ($20\%$).
    
\subsection{Targeted Acquisition with hyper-dimensional QSP:}

    Targeted Acquisition (TA) is a NUS-based incremental data acquisition strategy, in which the quality of processed spectra is concurrently assessed, typically in relation to a task-specific target, such as detecting a predefined number of peaks \cite{orekhov2011analysis}. TA provides essential feedback on experimental progress and enables timely termination of data collection, thereby significantly reducing measurement time for lengthy experiments. 
    The key ingredient of the TA procedure is a reliable and meaningful spectrum quality score that can be calculated at different levels of data completion. In our original implementation of the TA for the protein backbone assignment,  the peaks in the 3D NUS triple resonance experiments were detected at each TA step using a hyper-dimensional extension of the multi-dimensional decomposition \cite{isaksson2013highly,jaravine2008hyperdimensional}. 
    This approach was necessary to replace traditional peak-picking routines, which perform poorly on strongly under-sampled NUS spectra. However, generalizing this highly specialized and task-dependent method to other spectrum types and analysis tasks remains challenging.
    Leveraging ultimate resolution, high sensitivity, and accurate differentiation between genuine peaks and artifacts, $P^3$ and QSP provide a new, general, and reliable method for estimating the number of detectable peaks—without requiring explicit peak picking. This enables real-time assessment of spectrum quality during the TA process.
    As demonstrated below, $P^3$ also introduces a novel AI-based approach for hyper-dimensional co-processing \cite{kupče2006hyperdimensional,jaravine2008hyperdimensional} of multiple spectra. 

    \begin{figure}[htbp]
        \centering
        \includegraphics[width=0.75\textwidth]{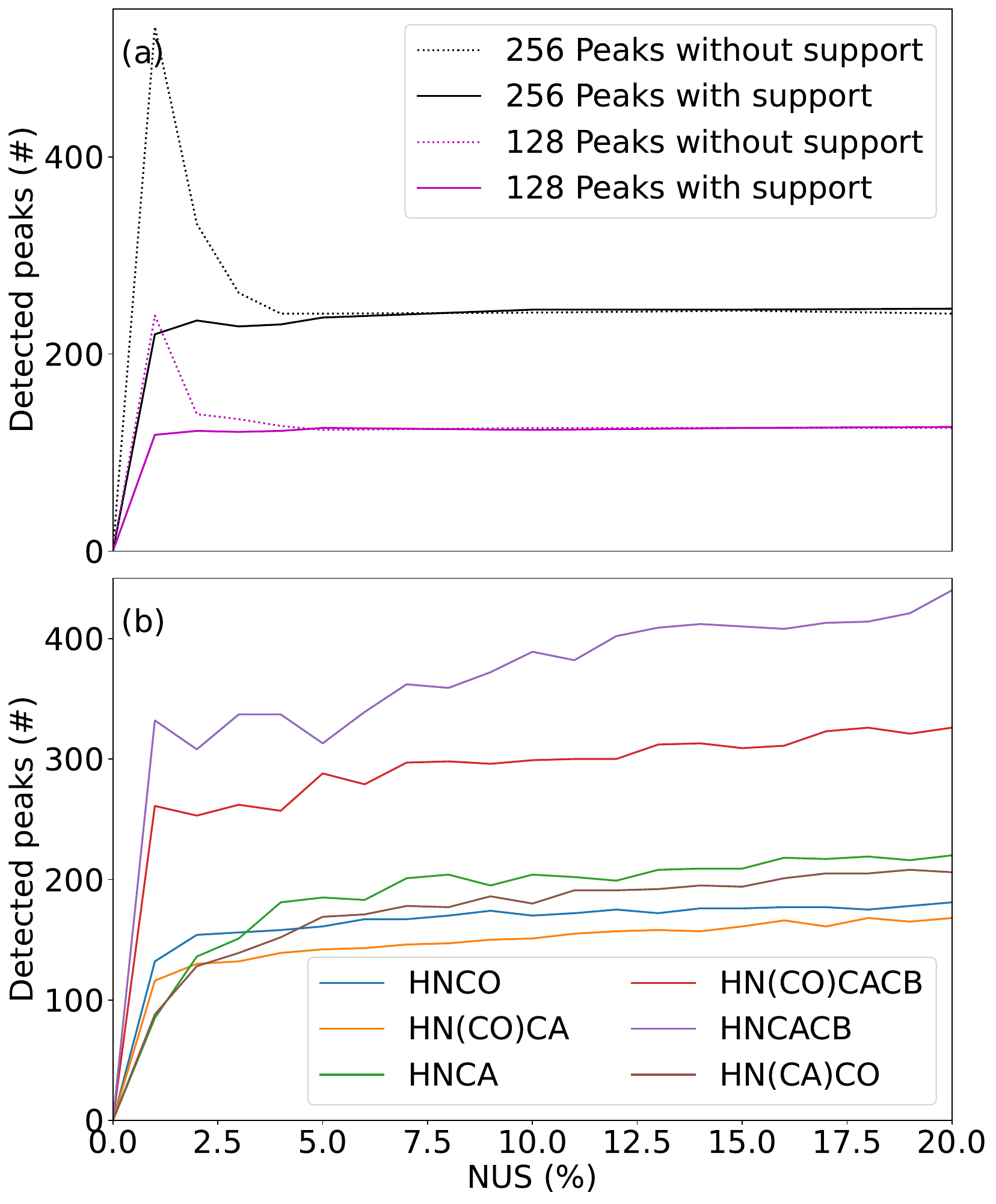}
        \caption{\textbf{Targeted Acquisition with the Reference-Free Quantitative Spectrum Quality Score with $P^3$ ($QSP$).} The number of detectable peaks estimated with the $QSP$ in different 3D spectra reconstructed using CS-IST is plotted versus NUS fraction build-up during in course of the TA acquisition. \textbf{(a}) Synthetic HNCO-type spectra reconstructed using CS-IST with 128 peaks (magenta) and 256 peaks (black). MR-Ai calculated $QSP$ values for standalone spectra (dashed lines) and using hyper-dimensional co-processing with a supporting 2D spectrum (solid lines).. \textbf{(b)} Experimental 3D spectra of Calmodulin protein (16.7 kDa). The $QSP$ was calculated using the 2D $^{1}$H-$^{15}$N projection of the most sensitive 3D 50\% NUS HNCO spectrum as support.}
        \label{fig:TA}
    \end{figure}

    Figure \ref{fig:TA}a demonstrates that QSP can be used to quantitatively predict the number of detectable peaks in 3D spectra reconstructed with CS-IST across a wide range of NUS rates. Increasing the number of peaks in the synthetic spectra from 128 to 256 results in a corresponding doubling of the QSP score, confirming its linear response to the number of true peaks. The figure also highlights the advantage of hyper-dimensional co-processing, which is particularly beneficial at low NUS fractions.  
    In Figure \ref{fig:TA}b, the hyper-dimensional QSP is applied to a set of experiments for backbone assignment in Calmodulin protein (16.7 kDa). The curves show a typical TA build-up of peaks, where the number of detected peaks increases as more NUS data is collected, with little further improvement beyond approximately $10\%$ NUS, as confirmed by visual inspection of the spectra. Notably, the plateau values of the curves correspond to the expected number of peaks for each experiment type in the protein. The TA buildup curves of peak numbers in individual 3D spectra of Calmodulin closely resemble those obtained using the original TA procedure \cite{isaksson2013highly} for the same spectra.
        
\section{Conclusion}
\label{sec:Conclusion}

    In this work, we leverage the power of AI to achieve the ultimate resolution attainable through signal processing of multidimensional NMR spectra. We introduce $P^3$, a new type of statistical spectral representation designed to enhance resolution while suppressing noise and spectral artifacts. We present a novel MR-Ai architecture based on a physics-aware cross-objective framework, generalized for any dimensionality. We demonstrate the high value of $P^3$ for the analysis of 3D spectra from several representative globular and disordered proteins. Furthermore, we illustrate its application in hyper-dimensional spectral analysis and Targeted Acquisition.

\section{Methods}
\label{sec:Materials and methods}

\subsection{MR-Ai Architecture for nD pattern:}
    
    We introduce a generalized version of our Magnetic Resonance processing with AI (MR-Ai) architecture, designed to handle NMR spectra of any dimensionality. The MR-Ai framework was originally developed \cite{jahangiri2024beyond} for capturing 2D spectral patterns, including phase-twisted peaks associated with P-type (or N-type) data in the frequency domain \cite{jahangiri2024beyond}. The original MR-Ai was, in turn, based on our earlier deep neural network architecture, WNN \cite{jahangiri2023nmr}, which was designed to analyze 1D NMR frequency-domain spectra. The WNN model effectively captures defined spectral features, such as specific patterns of NUS aliasing artifacts and peak multiplicities in homonuclear decoupling experiments. 
    
\begin{figure}[htbp]
        \centering
        \includegraphics[width=\textwidth]{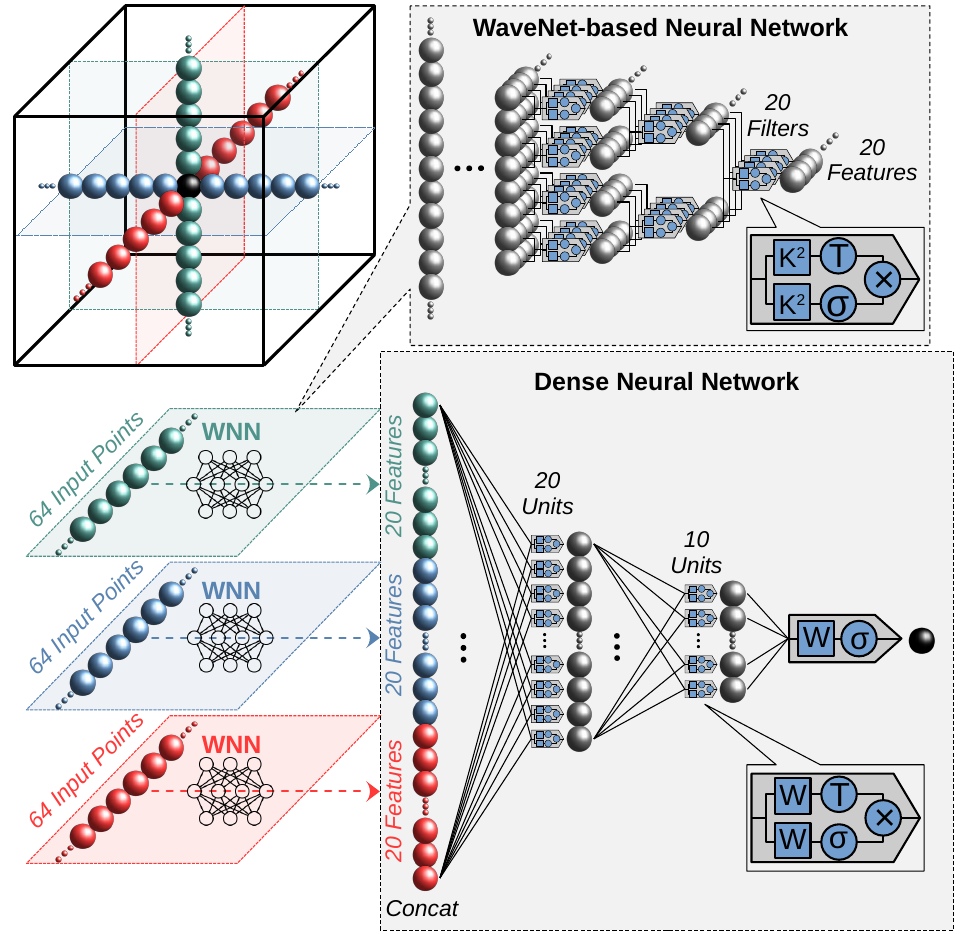}
        \caption{\textbf{Schematic of MR-Ai architecture for 3D}. Green, blue, and red points represent 64-point 1D cross-sections along the three spectral dimensions. Each of these vectors containing the examined point in the middle shown in black, is processed by an individual WaveNet-based Neural Network (WNN), resulting in an output feature vector with 20 elements. The outputs are subsequently concatenated and fed into a DNN for further processing. The WNN architecture within MR-Ai extracts 20 features based on the number of 20 convolution filters applied. Symbols $\times$, $T$, and $\sigma$ indicate element-wise multiplication operations, hyperbolic tangent activation, and sigmoid activation functions, respectively. The $K^2$ and $W$ denote learnable filter kernels of size 2 in 1D convolutional layers of the WNN and weights used in the DNN, respectively.}
        \label{fig:Cross}
    \end{figure}

    The new MR-Ai, referred to in this paper simply as MR-Ai, captures specific spectral patterns along the multidimensional cross in Cartesian coordinates, centered at a probed spectral point (Figure \ref{fig:Cross}). This cross-objective representation of the nD spectrum is inspired by the physical model of the NMR signal as a tensor product of 1D shapes \cite{jaravine2006removal}. 
    While notably compact, the cross-field of view retains most essential signal features, including line shape, phase distortions, $t_1$-noise, $J$-coupling multiplets, wiggles from spectral truncation, and NUS-related Point Spread Function (PSF) patterns. Using a larger field of view, for example, an nD box, would require more DNN parameters and introduce more noise than new information.
        
    Figure \ref{fig:Cross} depicts the architecture of MR-Ai, designed for generating $P^3$. Each vector within the cross-objective is first processed individually by the WNN module, which converts it into a 20-feature vector. 
    The WNN architecture consists of 1D convolutional layers with a stride of 2, effectively skipping one data point between each convolution operation. Each layer employs a kernel size of 2 ($K^2$) and contains 10 filters, using a combination of hyperbolic tangent activation functions ($T$) and sigmoid activation functions ($\sigma$), without padding between layers.
    The feature vectors generated by the final WNN layers for all spectral dimensions are concatenated and passed as input into a Dense Neural Network (DNN) with two hidden layers, containing 20 and 10 units, respectively. Each hidden layer applies the same combination of hyperbolic tangent ($T$) and sigmoid ($\sigma$) activation functions.

    A crucial aspect of deep neural network design is selecting an appropriate cost function, which must align closely with the chosen output unit activation function. Both the cost function and activation function depend on the specific task.
    For instance, in regression problems such as NUS reconstruction, Echo reconstruction, or virtual decoupling, a Rectified Linear Unit (ReLU) activation function \cite{agarap2018deep} combined with a Mean Square Error (MSE) loss function is commonly used \cite{jahangiri2023nmr}. Conversely, for uncertainty estimation, the Negative Log-Likelihood (NLL) loss function is required 
    \cite{jahangiri2024beyond}. 
    For $P^3$, determining the probability of a peak maximum is a binary classification problem with two classes: peak maxima points labeled as 1, and all other points labeled as 0 ($y \in {0, 1}$). In this context, the Binary Cross-Entropy (BCE) loss function, combined with a sigmoid activation function, provides an effective solution \cite{bengio2017deep}.
        
    The sigmoid function converts model outputs into probabilities by mapping values to the range $[0, 1]$:
    
    \begin{equation}
        \sigma(x) = \frac{1}{1 + e^{-x}}
    \end{equation}
    
    This property makes it ideal for binary classification tasks, where the output represents the likelihood of belonging to the peak maxima class. For a Bernoulli distribution, the probability of observing a label $y$ given a predicted probability $p$ is defined as:
        
    \begin{equation}
        P(y|p) = p^y(1-p)^{(1-y)}
    \end{equation}
        
    Taking the negative $\log$ of this likelihood yields the BCE loss function:
        
    \begin{equation}
        L = - \left[ y \log(p) + (1 - y) \log(1 - p) \right]
    \end{equation}
    
    By minimizing BCE, the model optimizes its outputs to assign probabilities close to 1 for peak maxima and close to 0 for all other regions. This approach corresponds to maximizing the likelihood under a Bernoulli distribution.
    The output layer of the DNN consists of a single unit with a sigmoid activation function and uses BCE as the loss function. This layer processes the concatenated feature vectors to generate the final output, representing the adjusted probability value for the central point within the cross-objective, informed by the surrounding spectral context.
    The network architecture and graphs were generated using the \textit{TensorFlow} Python library \cite{abadi2016tensorflow} with the \textit{Keras} front-end.
    The model was trained within TensorFlow using the stochastic ADAM optimizer \cite{kingma2014adam} with default parameters, a learning rate of 0.001, a mini-batch size of $2^{16}$, and up to $2^{10}$ epochs. Training was terminated early if the monitored metric failed to improve on the validation dataset.
    MR-Ai were trained on the \textit{NMRbox} server \cite{maciejewski2017nmrbox}, equipped with 128 cores, 2 TB of memory, and 4 \textit{NVIDIA A100 Tensor Core} GPUs.

\subsection{Training MR-Ai model for $\mathbf{P}^\mathbf{3}$:}

    The first challenge in training a deep neural network is acquiring a sufficiently large and diverse dataset. To effectively train the model, it is necessary to generate a substantial number of cross-objectives along with their corresponding labels (1 for peak maxima and 0 otherwise). This requires access to a large number of spectra with accurately annotated peak positions.
    Our previous studies demonstrated that synthetic data can serve as an effective proxy for realistic experimental NMR spectra \cite{jahangiri2024beyond,jahangiri2023nmr}. In this work, we train MR-Ai using synthetic spectra, generating approximately $2^{24}$ cross-objectives and their associated labels.
    While most previous efforts have primarily focused on signal modeling, we incorporate synthetic noise to better simulate spectra representative of real experimental conditions. We found that accurately modeling noise with an appropriate statistical distribution, as described below, is just as critical as modeling the signal itself, especially for NUS-reconstructed data, where low-intensity peaks are particularly affected.
        
\subsubsection{Synthetic nD spectra with associated labeling}
        
    For training, nD NMR time domain a hyper-complex signal $\mathbf{X}_\text{FID}\in\mathbb{H}_{nD}$, usually called free induction decay (FID), can be presented as a superposition of a small number of exponential functions:
        
    \begin{equation}
        \mathbf{X}_{FID}(t_1,...,t_n) = \sum_{j} A_j \prod_{n} e^{-t_n/\tau_{n_j}} e^{\pm\mathbf{i}(2 \pi \omega_{n_j} t_n+\phi_{n_j})}
    \end{equation}
            
    where $J$ and $N$ run over the number of exponentials and dimensions respectively where the $j$th exponential in $n$th dimensional has the amplitude $A_j$, phase $\phi_{n_j}$, relaxation time $\tau_{n_j}$, frequency $\omega_{n_j}$. The evolution time $t_n$ is given by the series 0, 1, ..., $T_n$-1, where $T_n$ is the number of complex points in $n$th dimension. The desired number of different FIDs for the training and testing set is simulated by randomly varying the above parameters in the ranges summarized in Table \ref{table:Table1} for 2D $^1$H-$^{15}$N and 3D $^1$H-$^{15}$N-$^{13}$C correlation spectra.
    
    \begin{table}[ht]
    \small
    \centering
    \caption{\textbf{Parameters for the synthetic nD FID}}
    \begin{tabular}{lcc}	
        \hline
        & Direct & Indirect\\
        \hline
        $T\in \mathbb{N}$ & 128 & 128\\
        $\omega_{n_j}\in \mathbb{R}$ & [-0.5,0.5] &	[-0.5,0.5]\\
        $\phi_{n_j}\in \mathbb{R}$ & [$-3^\circ$,$3^\circ$] & [$-3^\circ$,$3^\circ$]\\
        $\tau_{n_j}\in \mathbb{R}$ & [12.8,64] & [64,1280]\\    
        \hline
        $A_n\in \mathbb{R}$ & \multicolumn{2}{c}{[0.05,1]}\\
        $J\in \mathbb{N}$ & \multicolumn{2}{c}{256}\\
        \hline
    \end{tabular}
    \label{table:Table1}
    \end{table}

    We used Python libraries, including \textit{nmrglue} \cite{helmus2013nmrglue} and \textit{NMRPipe} \cite{delaglio1995nmrpipe}, for reading, writing, and processing NMR spectra, as well as \textit{mddnmr} \cite{jaravine2006removal}. The uniformly sampled spectra $\mathbf{S}_{nD}$ were obtained with the standard nD processing steps, which included apodization, zero-filling, Fourier Transform (FT), and phase correction.

    A binary label matrix, matching the dimensions of each processed synthetic spectrum, was generated based on the provided peak list. Matrix elements were assigned a value of one at indices corresponding to peak maxima and zero elsewhere. If a peak maximum fell between two data points, within a range of 0.25 to 0.75 units from each, both adjacent elements were assigned a value of one.

\subsubsection{Training model for 2D US spectra:}

    To train the model for 2D US spectra, we generated 640 synthetic noise-free 2D US spectra (using a 4:1 ratio for training and validation datasets) based on Table \ref{table:Table1}. The noise was simulated by a random Gaussian-distributed signal in the time domain and subsequently processed to the frequency spectrum using the same procedure as for synthetic spectra. After normalization of the noise in the frequency domain, so that its standard deviation matches to the smallest possible peak amplitude, the noise was added to the synthetic spectrum, $\mathbf{S}_{nD}$.

\subsubsection{Training model for 3D spectra:}
    
\begin{figure}[htbp]
        \centering
        \includegraphics[width=0.75\textwidth]{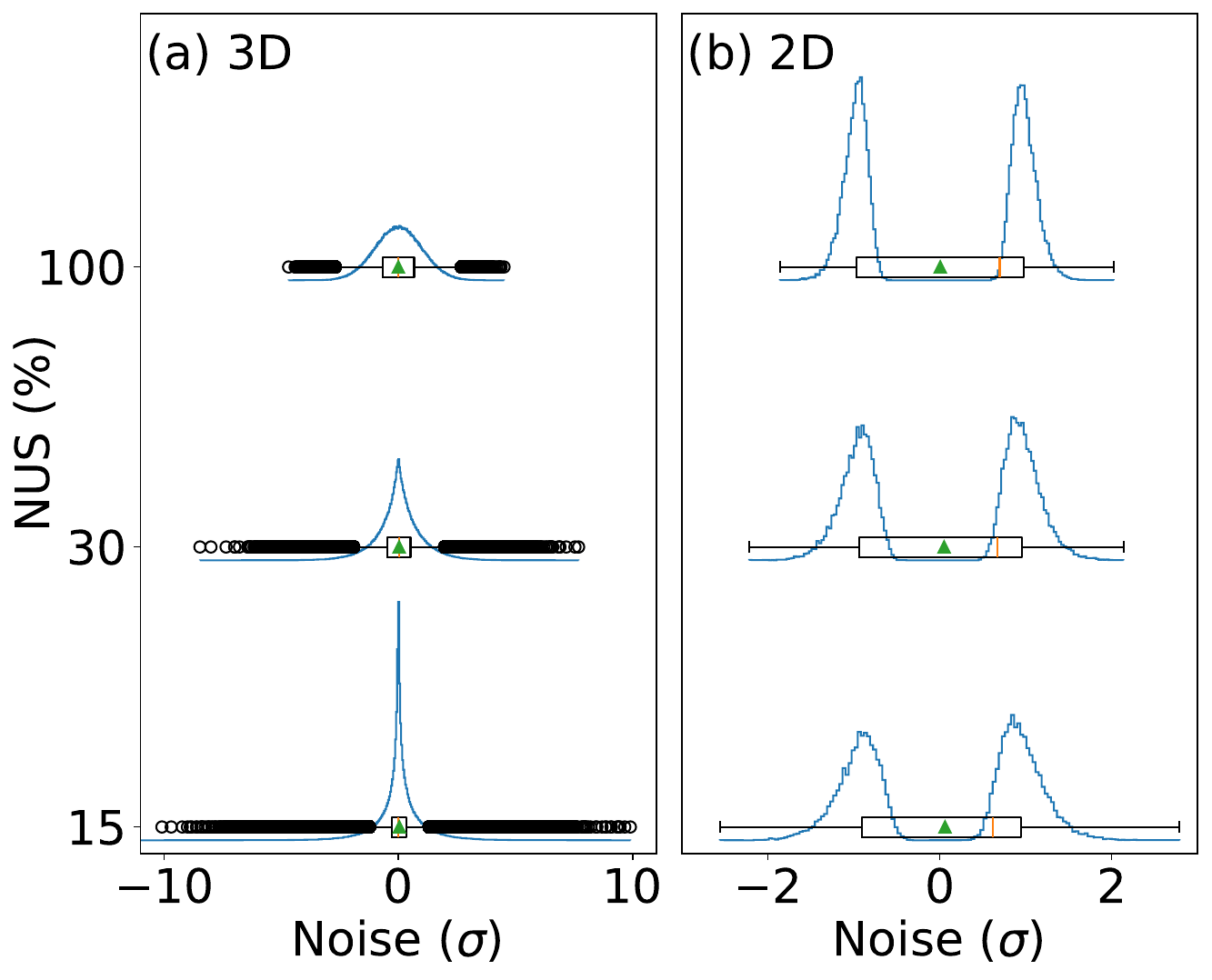}
        \caption{\textbf{Distribution of Noise and NUS Aliasing Artifact in synthetic 3D spectra and the 2D Sky projections.} The histograms (blue lines) and Box-plots for the baseline noise are shown for \textbf{(a)} synthetic 3D spectra sampled in full (US) and reconstructed using CS-IST from 15\% and 30\%, NUS. Note the broad Cauchy-like distribution with wide spread of outliers shown in the box-plot for NUS spectra \textbf{(b)} 2D sky projections form these 3D spectra.}
        \label{fig:Noise}
    \end{figure}

    To train the model for 3D spectra, we generated 1,280 synthetic noise-free 3D US spectra based on Table \ref{table:Table1}. Groups of four spectra were added with positive and negative signs. This increased the number of peaks and overlaps as well as resulted in both positive and negative peaks. 
    Similar to 2D US spectra, the 3D US spectra include Gaussian noise. However, in the case of 3D NUS spectra reconstructed using the CS-IST algorithm, the apparent baseline noise and artifacts do not follow a normal (Gaussian) distribution. To address this discrepancy, we empirically determined (Figure \ref{fig:Noise}a) that as the number of NUS points decreases, the noise distribution shifts toward a Cauchy-like distribution.
            
    To simulate this behavior, we modeled the noise as a combination of Cauchy and Gaussian distributions, where their relative contribution varies as a function of the NUS fraction. Using Cauchy-featured noise instead of a purely Gaussian distribution during training enhanced the precision of the trained network without significantly compromising its sensitivity, as measured by the recall score (see Supplementary Figure \ref{fig:3D_GC}).
    After normalizing the synthetic noise in the frequency domain, it was added to $\mathbf{S}_{nD}$ as previously described.
            
\subsection{Resampling and the Regions of interest:}
 
    In classification tasks, an imbalance in class representation, where one class significantly outnumbers another, can lead to biased models that perform poorly on the minority class \cite{krawczyk2016learning,sun2009classification}.
    In 2D spectra, where the class ratio (i.e., the number of points with and without spectral maxima) is approximately 1:100, this slight imbalance does not pose a significant issue. However, in 3D spectra, a severe imbalance arises due to the sparsity of 3D data, with class ratios exceeding $1:10^4$. Training the deep neural network on such an imbalanced dataset results in a model that is insensitive to low-intensity peaks, creating substantial challenges during training.
    To mitigate this issue, resampling techniques—which adjust the training data to balance class distributions—can significantly improve model performance. These techniques include oversampling the minority class or undersampling the majority class \cite{krawczyk2016learning,sun2009classification}. 
    For training the model in the 3D case, we specifically under-sampled data points associated with the background. This was achieved by selecting all points corresponding to peak maxima (labels with zeros) and a subset of non-peak points (labels with zeros), ensuring a class ratio of approximately 1:100. This approach significantly improved the model’s sensitivity to small peaks but also led to an increase in false positive hits. 
    This is unsurprising, as even in an ideal 3D spectrum of typical size filled with Gaussian noise, more than 100 peaks are expected to exhibit intensities exceeding four standard deviations of the baseline noise. The number of noise-induced peak-like features is even greater in NUS-reconstructed spectra.
    To mitigate this issue, we focus the analysis on \textit{regions of interest}, as described below.

\subsubsection{Training model for 2D Sky projections of 3D spectra:}
    
    Although the trained network for 3D spectra can successfully detect even very low SNR peaks, it is also highly sensitive to spurious peak-like noise features and spectral artifacts in NUS spectra. Distinguishing real peaks from intense noise features and artifacts is nearly impossible without additional prior knowledge.

    As it was noted above, the 2D spectra do not display a significant imbalance in the representation of the classes' peaks versus non-peak. This allows direct detection of the low-intensity peaks. Moreover, even if the baseline noise in a 3D spectrum has a strong Cauchy-like deviation from Gaussian distribution, the noise in the 2D projections has a distribution featuring favorable properties of the normal distribution due to the central limit theorem (CLT) \cite{ross2019first}.
    To leverage this property, we trained models to predict $P^3$ values for all three 2D sky projections of the 3D spectrum. The primary difference between 2D sky projections of 3D spectra and conventional 2D uniformly sampled (US) spectra lies in their noise distributions. While noise in 2D US spectra follows a Gaussian distribution, noise in 2D sky projections from 3D spectra resembles a symmetric double Gaussian distribution.
    Figure \ref{fig:Noise}b illustrates that across a broad range of NUS rates ($15\%$–$100\%$), artifacts and noise in 2D sky projections exhibit similar distributions, with slight variations in the middle gap and spread. To replicate this behavior during training, we modeled noise as a double Gaussian distribution. This approach significantly improves recall by capturing the characteristics of 2D sky projections from 3D spectra reconstructed under both US and different NUS rates (see also Supplementary Figure \ref{fig:2D_GC}).

\subsection{Production run of trained MR-Ai:}
        
\begin{figure}[htbp]
        \centering
        \includegraphics[width=\textwidth]{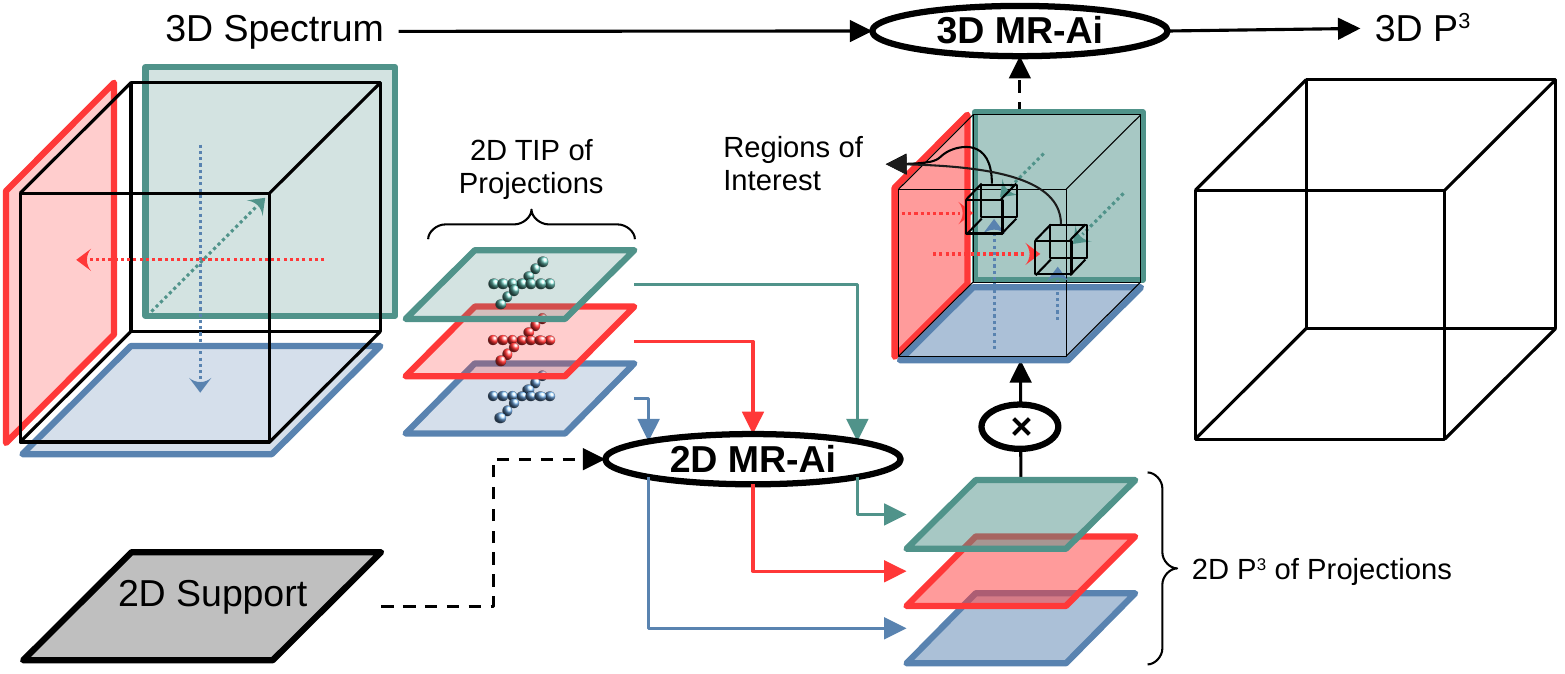}
        \caption{\textbf{Schematic of production of  3D $P^3$ with MR-Ai.} The input 3D spectrum is evaluated by the 3D version of MR-Ai at the points from the regions of interest, which are obtained from the $P^3$ of the 2D projections and the supporting spectrum (spectra).} 
        \label{fig:RoI}
    \end{figure}

    For 2D US spectra, the trained MR-Ai model, designed to handle Gaussian-distributed noise, can be applied directly to predict $P^3$. In the case of 3D spectra, $P^3$ reconstruction is preceded by an intermediate step to identify \textit{regions of interest}, as depicted in Figure \ref{fig:RoI}.

    First, $P^3$ values are calculated for the three orthogonal 2D sky projections using an MR-Ai model trained on 2D projections with noise modeled as a double Gaussian distribution. Then, each point in the original 3D spectrum is assigned a preliminary probability score, computed as the element-wise product of the corresponding probability values from the 2D $P^3$ projections.
    Points in the 3D spectrum with a score exceeding $10^{-5}$ are selected as regions of interest. This threshold corresponds to a minimum probability of approximately 2.5\% in the individual 2D projections. Finally, spectral points within the regions of interest are evaluated using the sensitive 3D MR-Ai model, which was trained on noise modeled as a Cauchy-Gaussian distribution. 
    It is worth noting that 2D projections can be directly measured in the time domain due to the Fourier Projection Theorem and have been successfully used in NMR for a long time, particularly in NUS spectrum reconstruction for multidimensional spectra \cite{coggins2010radial,freeman2004distant,pustovalova2018xlsy}.
         
    MR-Ai enables the hyper-dimensional analysis and co-processing of multiple spectra, leveraging the power of multi-spectral integration. \cite{kupče2006hyperdimensional,jaravine2008hyperdimensional,hiller2009coupled}. As shown in Figure \ref{fig:RoI}, MR-Ai incorporates supporting spectra during the definition of regions of interest.
    For instance, for enhanced signal detection, an alternative high-sensitivity 2D spectrum (e.g., 2D HSQC) can be used instead of a 2D sky projection from a processed 3D spectrum. As illustrated in Figure \ref{fig:TA}a, such spectral support is particularly beneficial in the sampling-limited regime, where the number of available NUS points is too low for reliably resolving all the spectra signals.

\subsection{Synthetic test data:}
    
    To test the trained MR-Ai model, we generated 2D and 3D US spectra based on Table \ref{table:Table1} and added Gaussian-distributed noise. For the 3D NUS case, the 3D US spectrum was first down-sampled to the desired number of NUS points. Subsequently, CS-IST was used for the spectrum reconstruction.

\subsection{Experimental test data:}
    To test trained MR-Ai performances, we used previously described 2D and 3D spectra for several proteins: Ubiquitin \cite{brzovic2006ubch5}, Azurin \cite{korzhnev2003nmr}, Tau (IDP) \cite{lesovoy2021unambiguous}, MALT1 \cite{unnerstaale2016backbone,han2022assignment}, and Calmodulin \cite{ikura1990novel}. The 2D US and 3D NUS experiments used in this study are described in Table \ref{table:Table2}. We used \textit{NMRPipe} \cite{delaglio1995nmrpipe}, Python package \textit{nmrglue} \cite{helmus2013nmrglue}, \textit{mddnmr} \cite{orekhov2011analysis}, and \textit{TopSpin} (Bruker Biospin) software for reading, writing, and traditional processing of the NMR spectra.
    We employed CS-IST \cite{kazimierczuk2011accelerated}, using the default \textit{mddnmr} parameters and the Virtual-Echo mode \cite{mayzel2014causality} for 3D NUS reconstruction.

    \begin{table*}[ht]
        \small
        \centering
        \caption{\textbf{Spectral parameters}}
        \begin{tabular}{l|ccccc}
            \hline
            Protein & Ubiquitin & Azurin & Tau & MALT1 & Calmodulin  \\
            \hline
            Size & 8.6 kDa & 14 kDa & 45.8 kDa & 44 kDa & 17 kDa\\
            Concentration & 0.6 mM & 1 mM & 0.5 mM & 0.5 mM & 1 mM \\
            \hline
            Spectrum &  \multicolumn{4}{c}{2D US ($^{1}$H, $^{15}$N)} & \\
            \cline{2-5}
            & HSQC & HSQC & TROSY &  TROSY &  \\
            & \multicolumn{3}{c}{} & \multicolumn{2}{c}{3D NUS ($^{1}$H, $^{13}$C,$^{15}$N)}\\
            \cline{5-6}
            &  &  & & HNCO & HNCO \\
            & &  &  & HNCA & HNCA \\
            & &  &  & HN(CA)CO & HN(CA)CO \\
            & &  &  & HN(CO)CA & HN(CO)CA \\
            & &  &  & HNCACB & HNCACB \\
            & &  &  & --- & HN(CO)CACB \\
            \hline
        \end{tabular}
        \label{table:Table2}
    \end{table*}
    
\newpage
\onecolumn

\clearpage 
\bibliographystyle{ieeetr} 
\bibliography{references}

\section*{Acknowledgement}
    The work was supported by the Swedish Research Council grants 2019-03561, 2023-03485, 2024-06251 to V.O. This study used NMRbox: National Center for Biomolecular NMR Data Processing and Analysis, a Biomedical Technology Research Resource (BTRR), which is supported by NIH grant P41GM111135 (NIGMS).


\newpage

\renewcommand{\thefigure}{S\arabic{figure}}
\renewcommand{\thetable}{S\arabic{table}}
\renewcommand{\theequation}{S\arabic{equation}}
\renewcommand{\thepage}{S\arabic{page}}
\setcounter{figure}{0}
\setcounter{table}{0}
\setcounter{equation}{0}
\setcounter{page}{1} 


\begin{center}
\section*{Supplementary Materials for \\ Towards Ultimate NMR Resolution with Deep Learning}

Amir~Jahangiri$^{1}$,
Tatiana~Agback$^{2\dagger}$,
Ulrika~Brath$^{1\dagger}$,
Vladislav~Orekhov$^{1\ast}$\\
\small$^\ast$Corresponding author. Email: vladislav.orekhov@nmr.gu.se\\
\small$^\dagger$These authors contributed equally to this work.
\end{center}

\subsection*{Supplementary Results}

\begin{figure}[htbp]
\centering
    \includegraphics[width=0.75\textwidth]{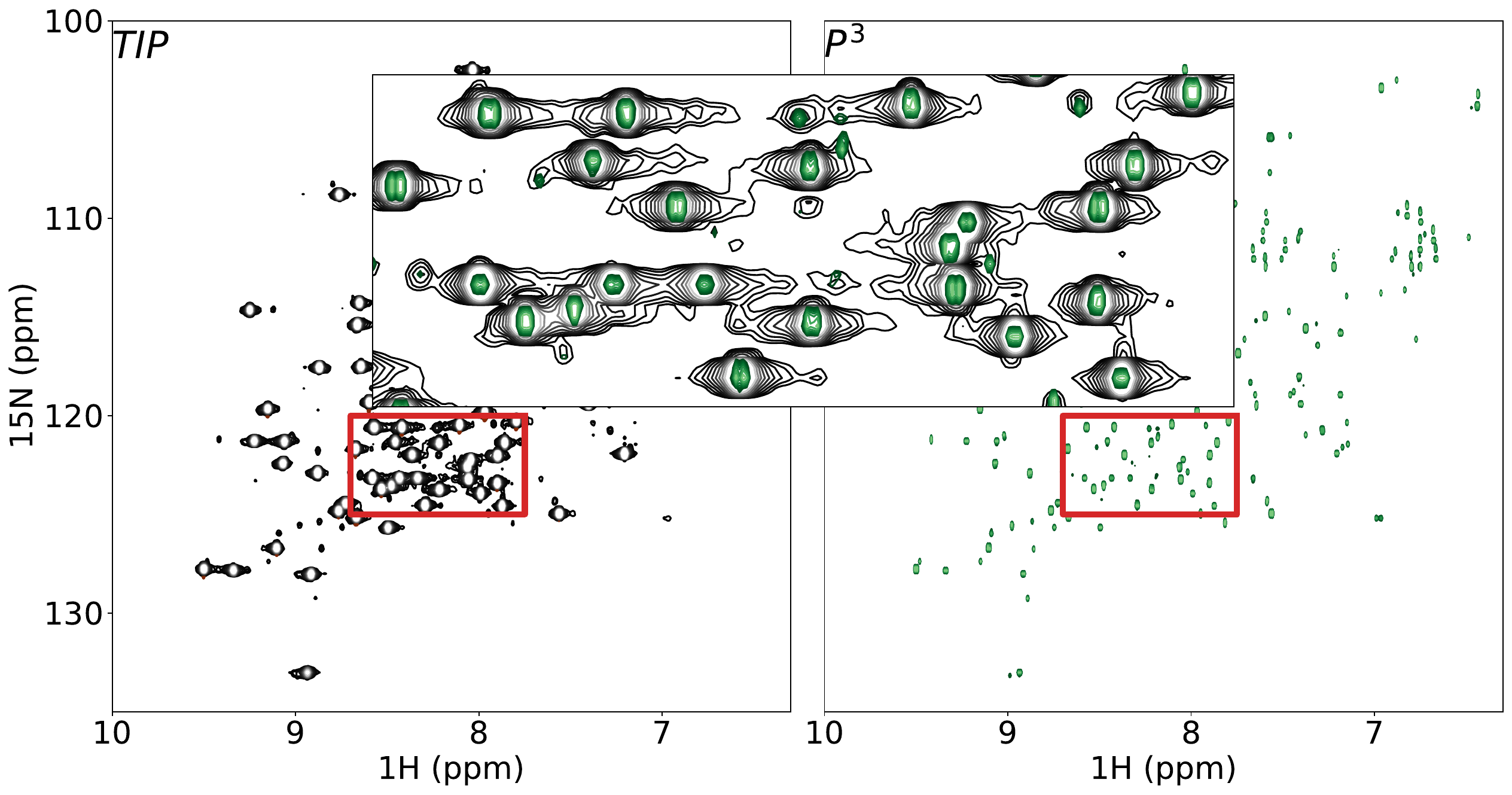}
    \caption{\textbf{2D $^{1}$H-$^{15}$N --- HSQC spectra of Ubiquitin.} Traditional intensity presentation, $TIP$, in black and peak probability presentations, $P^3$ by using MR-Ai, in green color. Note that many homonuclear couplings are visible in the $^1H$ dimension due to high resolution in the $P^3$ presentation.}
    \label{fig:HSQC_Ubiquitin}
\end{figure}

\begin{figure}[htbp]
\centering
    \includegraphics[width=0.75\textwidth]{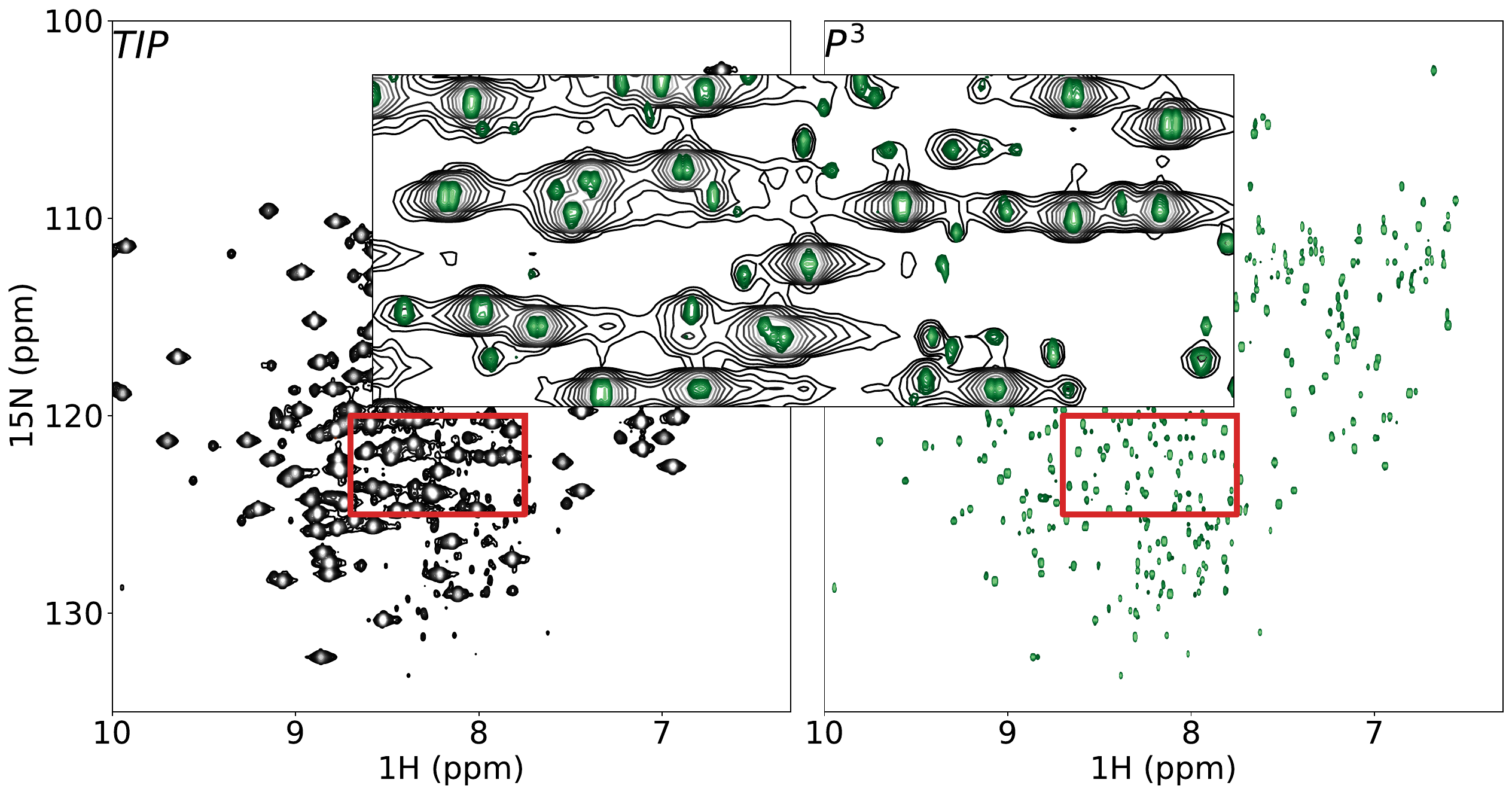}
    \caption{\textbf{2D $^{1}$H-$^{15}$N --- HSQC spectra of Azurin.} Traditional intensity presentation, $TIP$, in black and peak probability presentations, $P^3$ by using MR-Ai, in green color.}
    \label{fig:HSQC_Azurin}
\end{figure}

\begin{figure}[htbp]
\centering
    \includegraphics[width=0.75\textwidth]{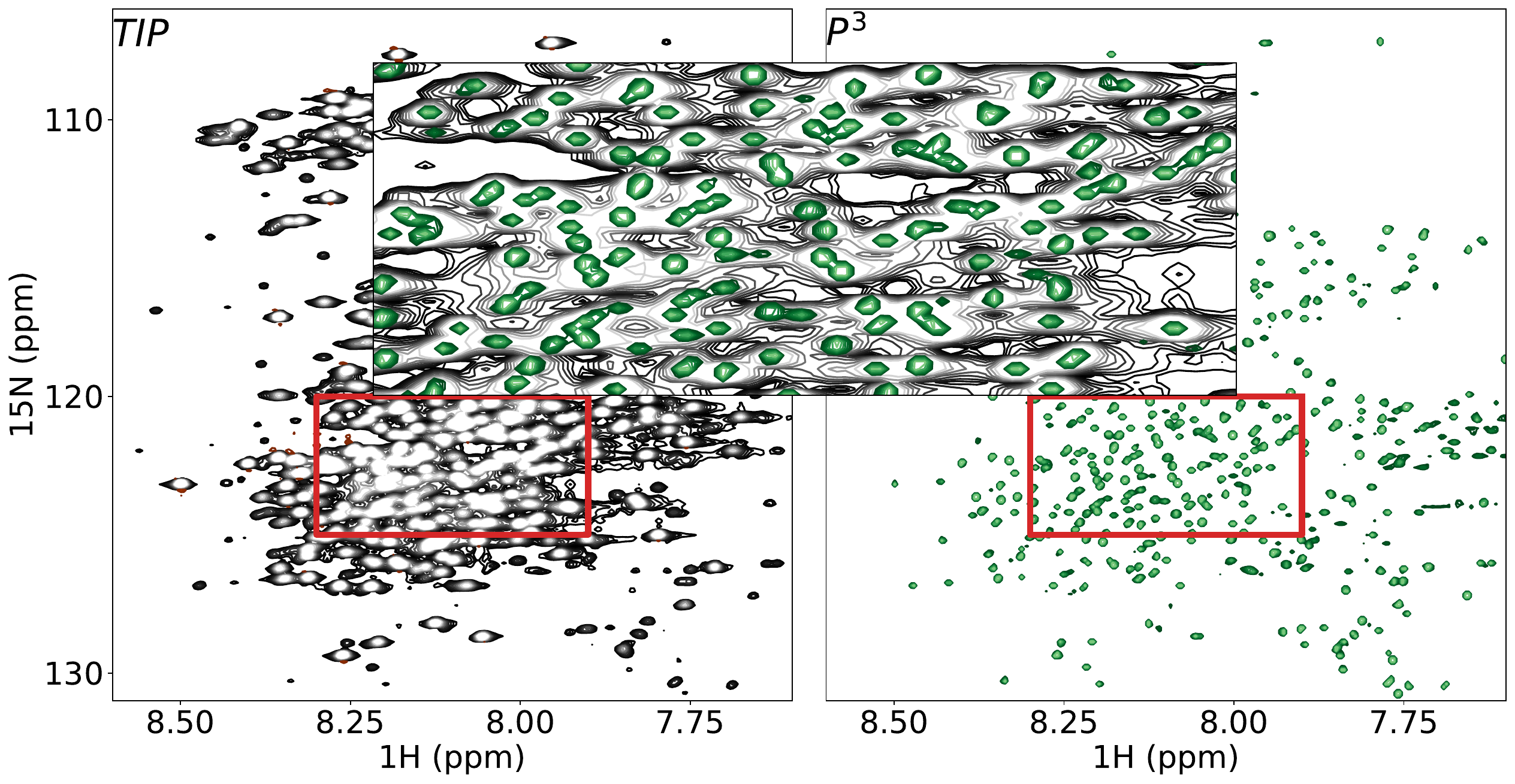}
    \caption{\textbf{2D $^{1}$H-$^{15}$N --- TROSY spectra of Tau protein.} Traditional intensity presentation, $TIP$, in black and peak probability presentations, $P^3$ by using MR-Ai, in green color.}
    \label{fig:HSQC_Tau}
\end{figure}

\begin{figure}[htbp]
    \centering
    \includegraphics[width=0.75\textwidth]{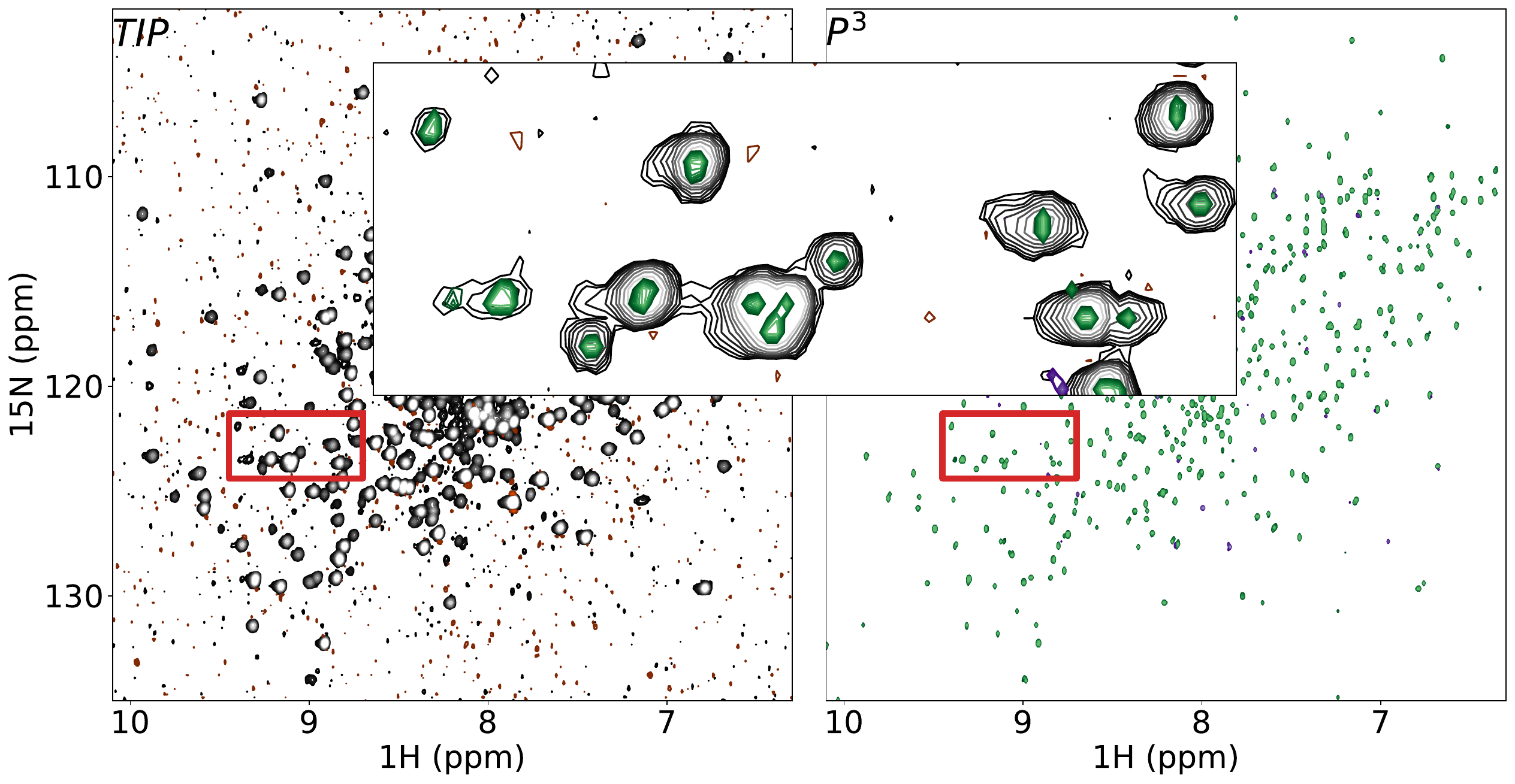}
    \includegraphics[width=0.75\textwidth]{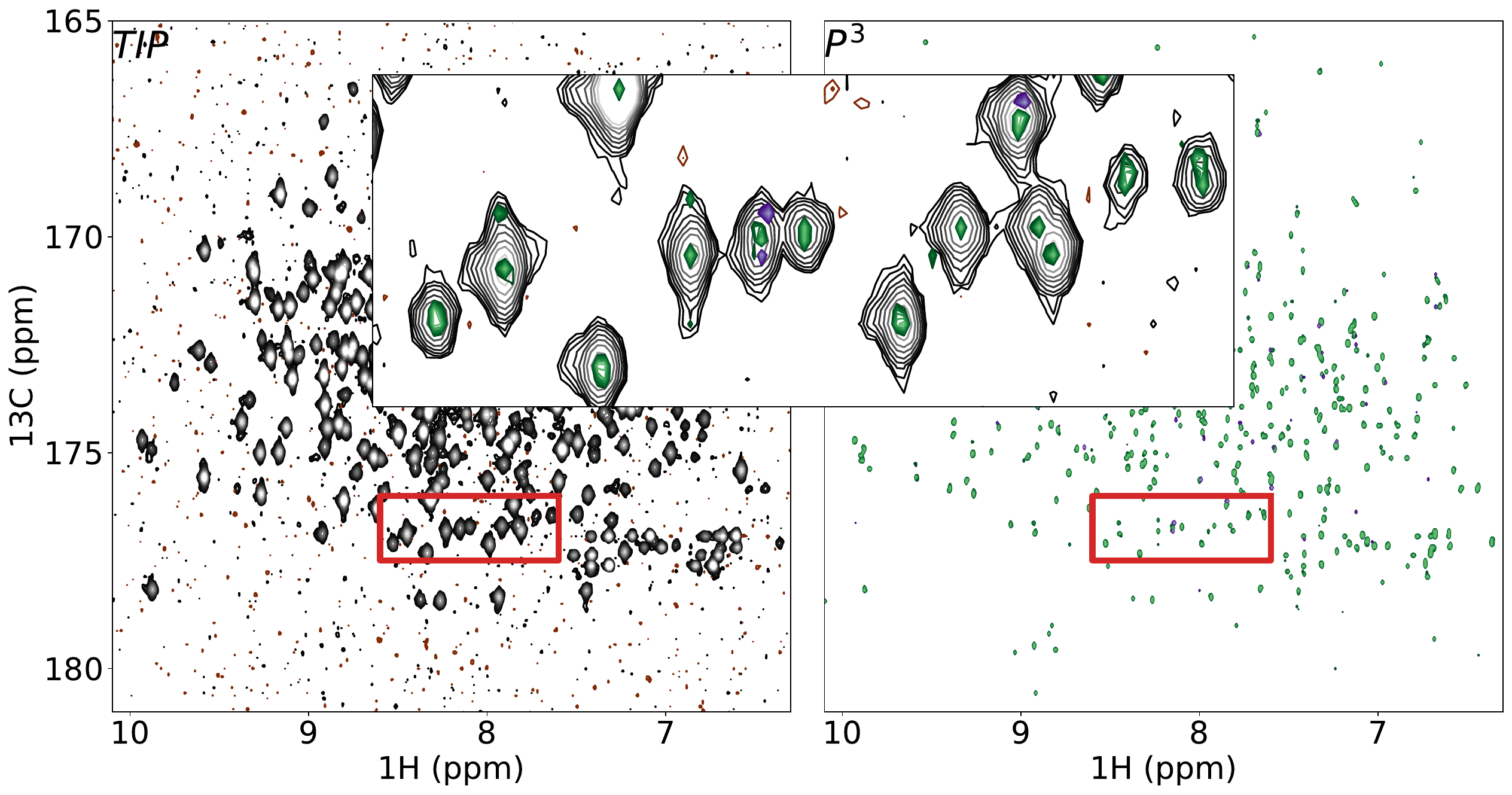}
    \includegraphics[width=0.75\textwidth]{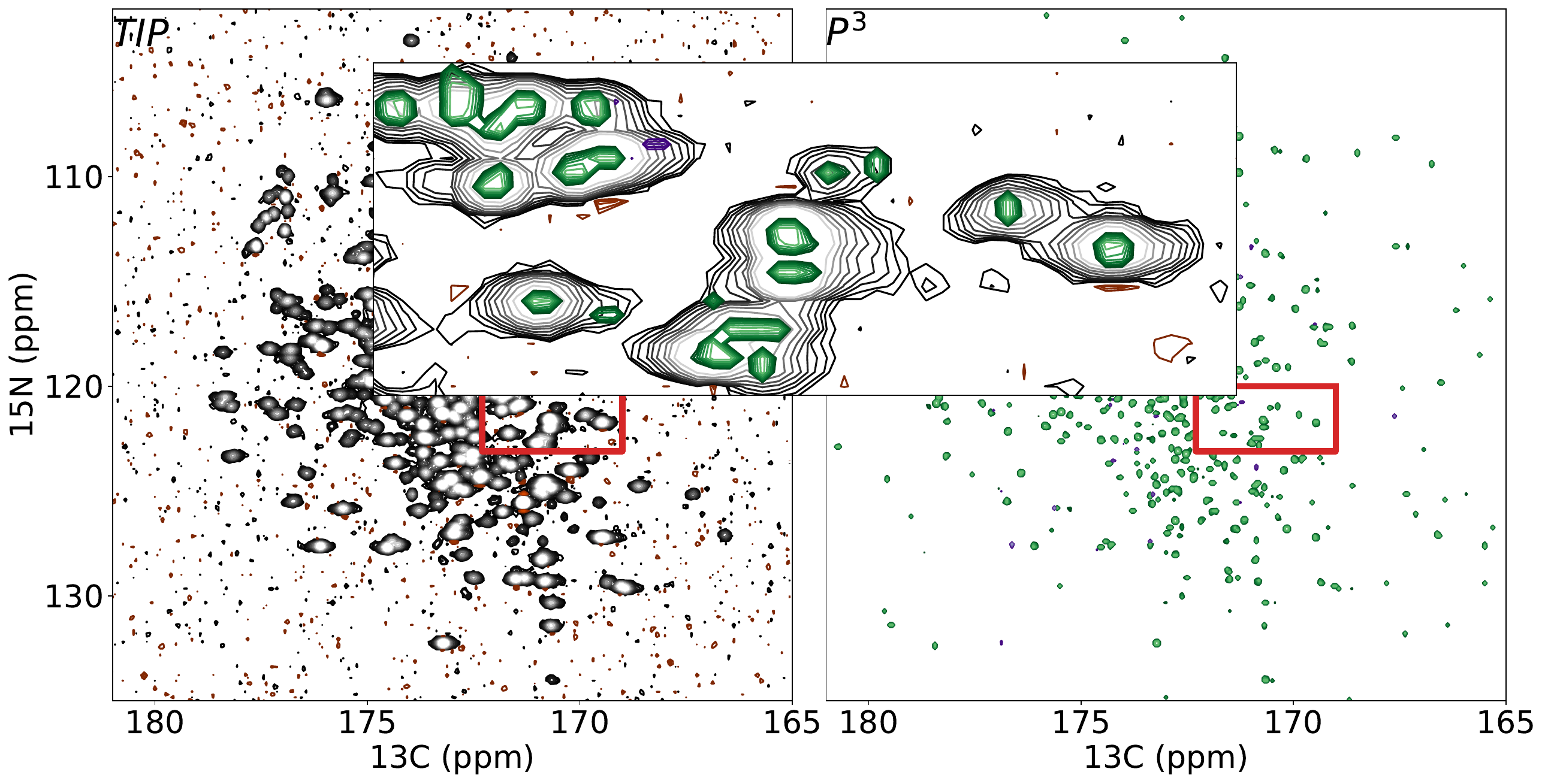}
    \caption{\textbf{3D HNCO NUS spectrum of MALT1 reconstructed with CS-IST.} Three orthogonal 2D spectrum projections are shown in the intensity presentation, $TIP$, in black ( orange for negative) and peak probability presentations, $P^3$ by using MR-Ai, in green (purple for negative) color of 2D projections.}
    \label{fig:HNCO_Malt}
\end{figure}

\begin{figure}[htbp]
    \centering
    \includegraphics[width=0.75\textwidth]{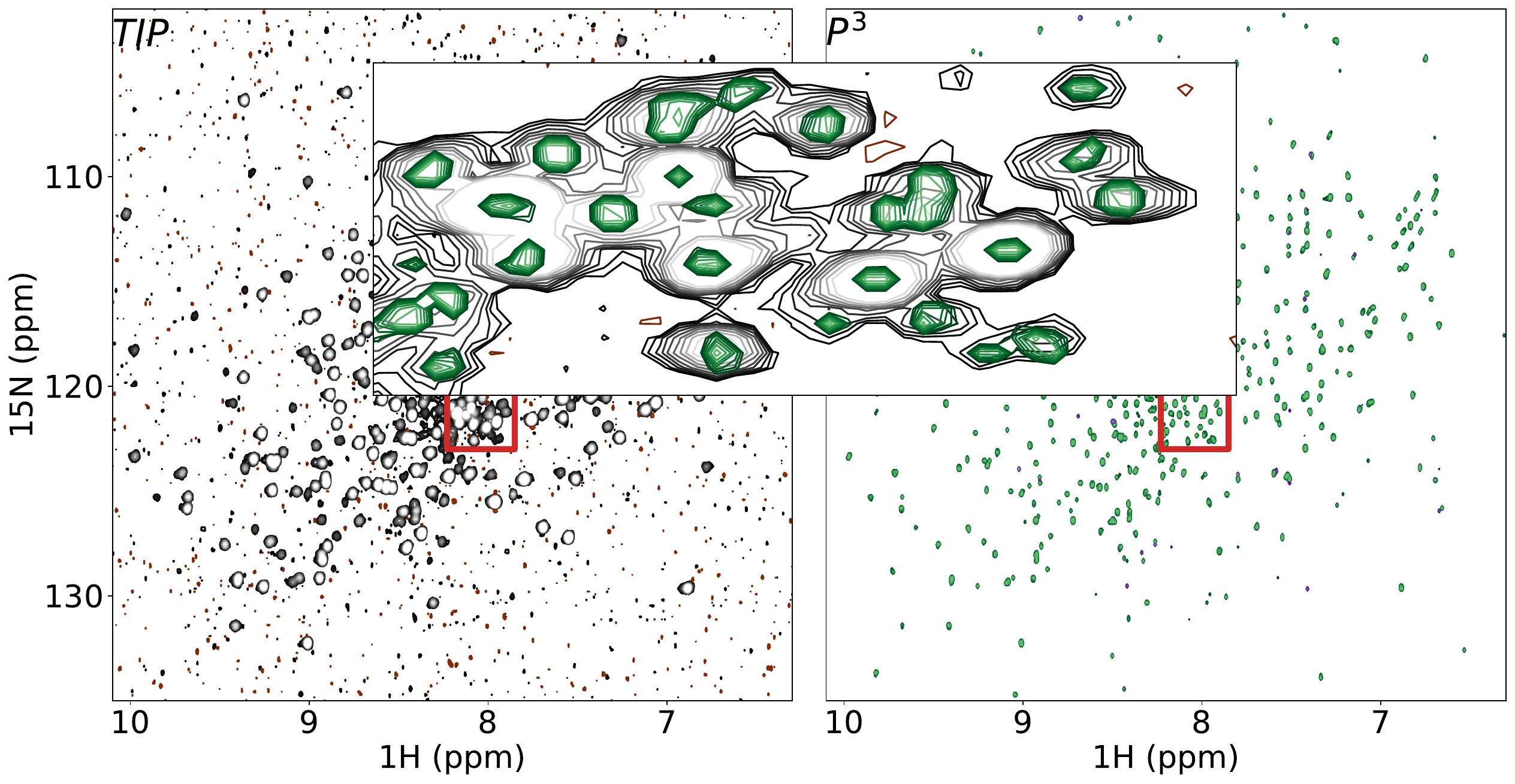}
    \includegraphics[width=0.75\textwidth]{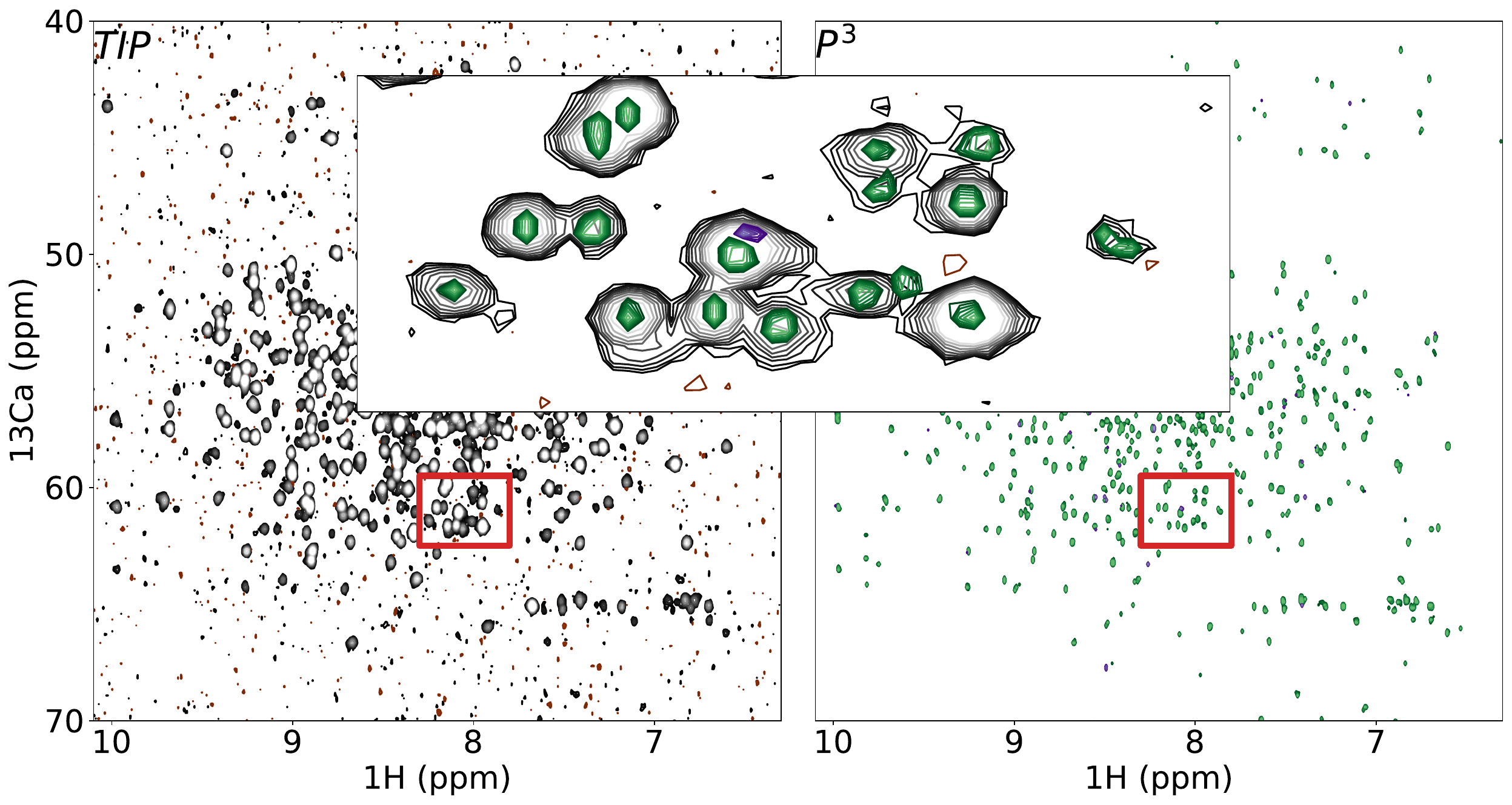}
    \includegraphics[width=0.75\textwidth]{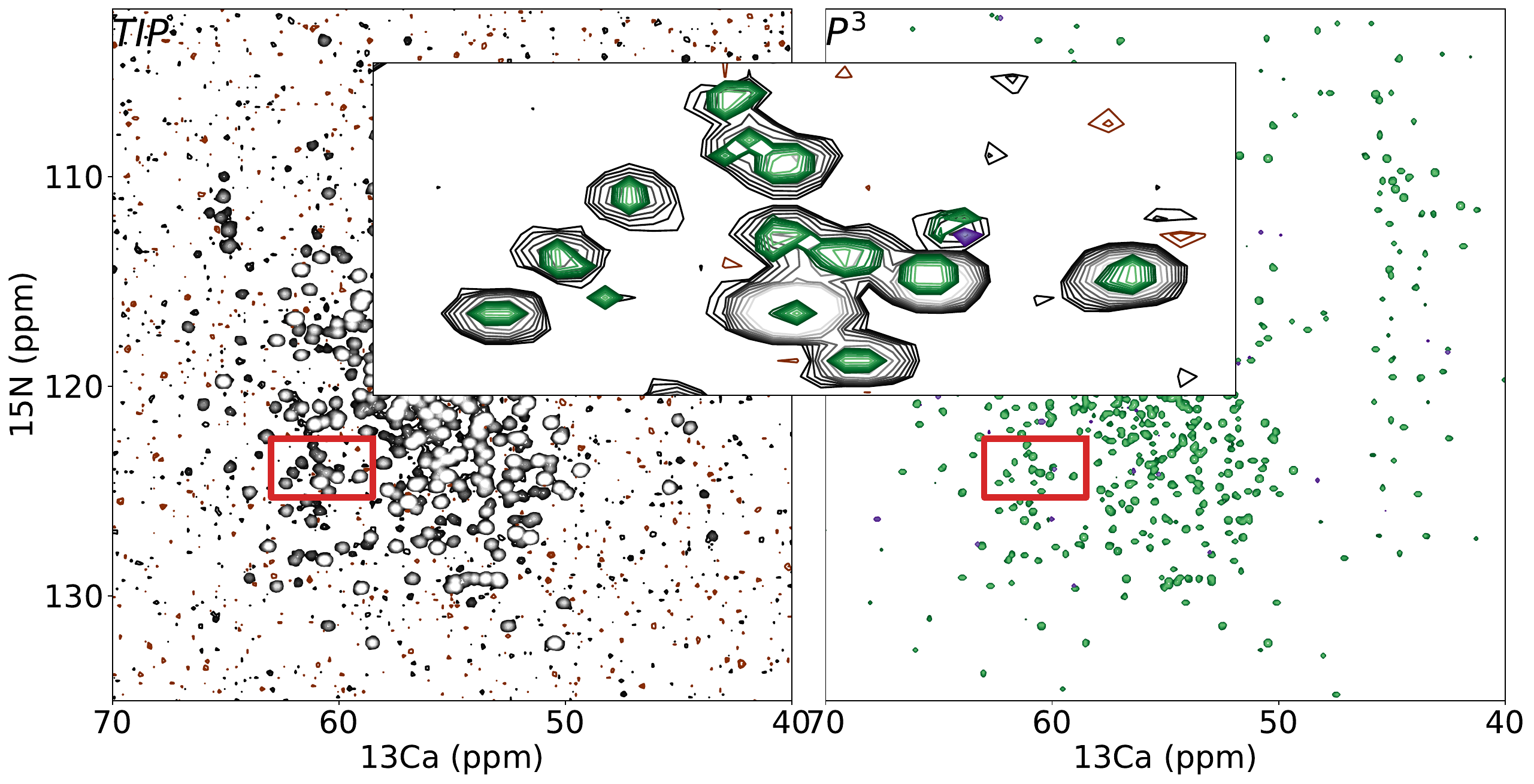}
    \caption{\textbf{3D HNCA NUS reconstructed spectrum with CS-IST of MALT1 protein.} Intensity presentation, $TIP$, in black ( orange for negative) and peak probability presentations, $P^3$ by using MR-Ai, in green (purple for negative) color of 2D projections.}
    \label{fig:HNCA_Malt}
\end{figure}

\begin{figure}[htbp]
    \centering
    \includegraphics[width=0.75\textwidth]{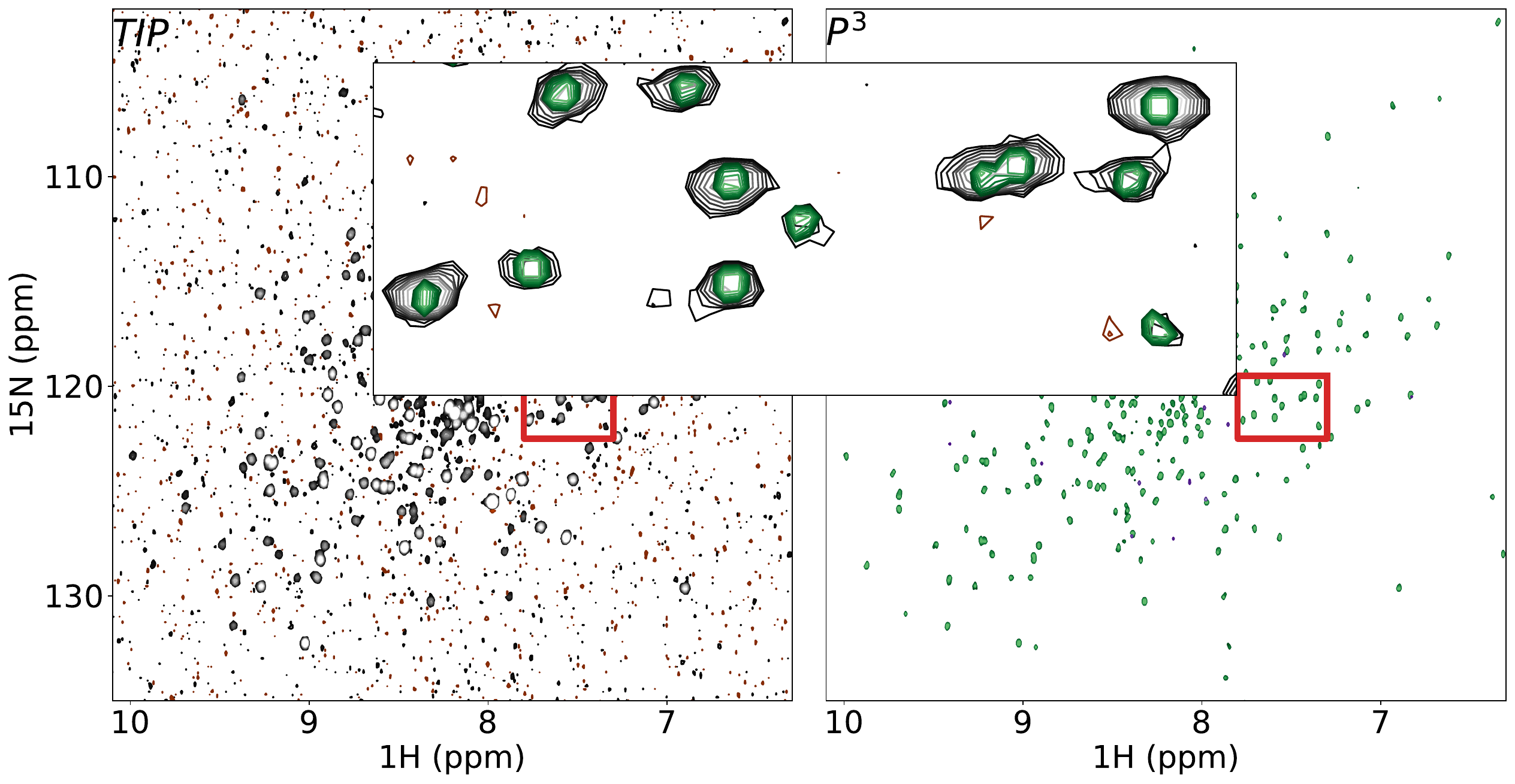}
    \includegraphics[width=0.75\textwidth]{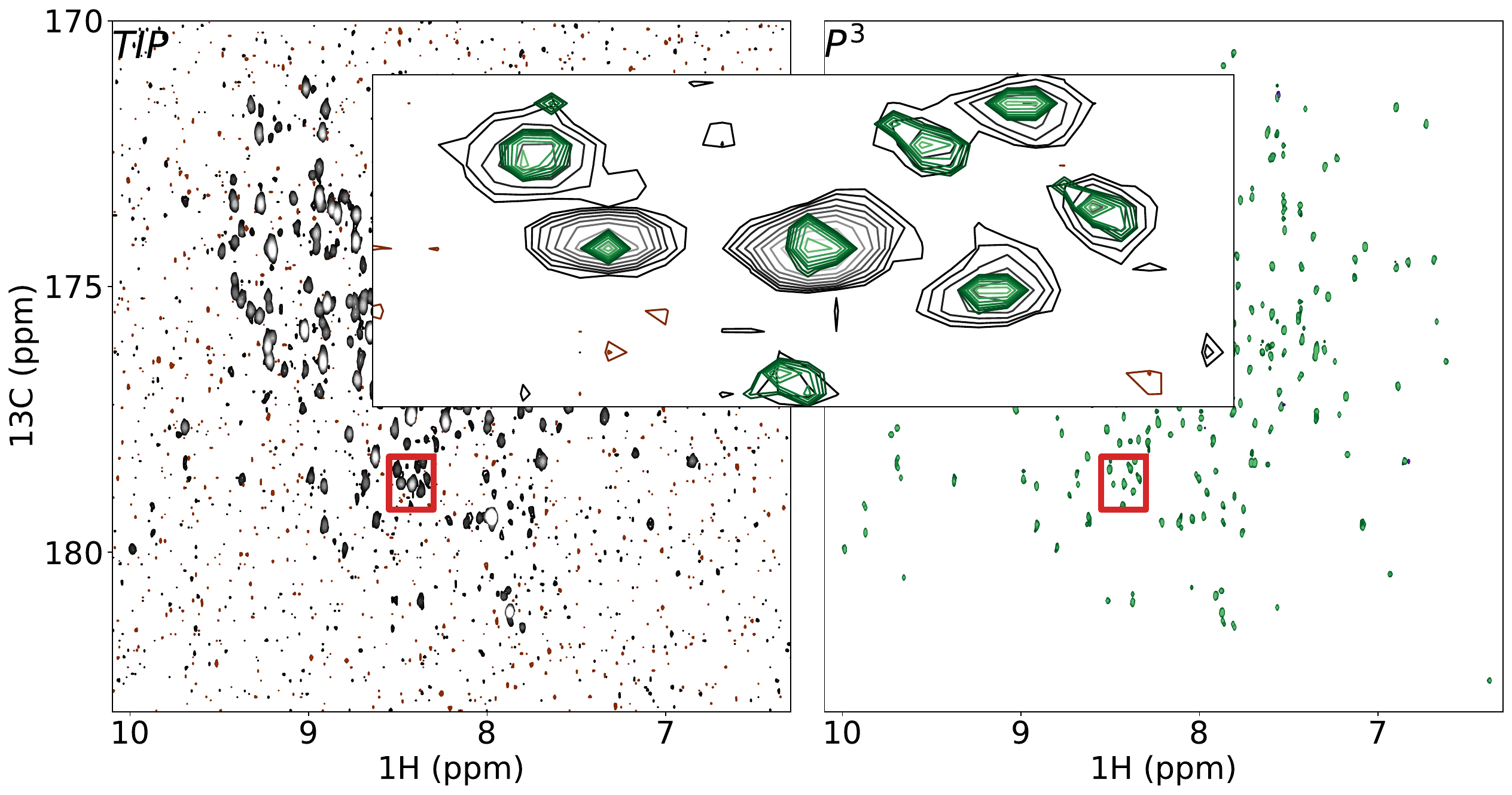}
    \includegraphics[width=0.75\textwidth]{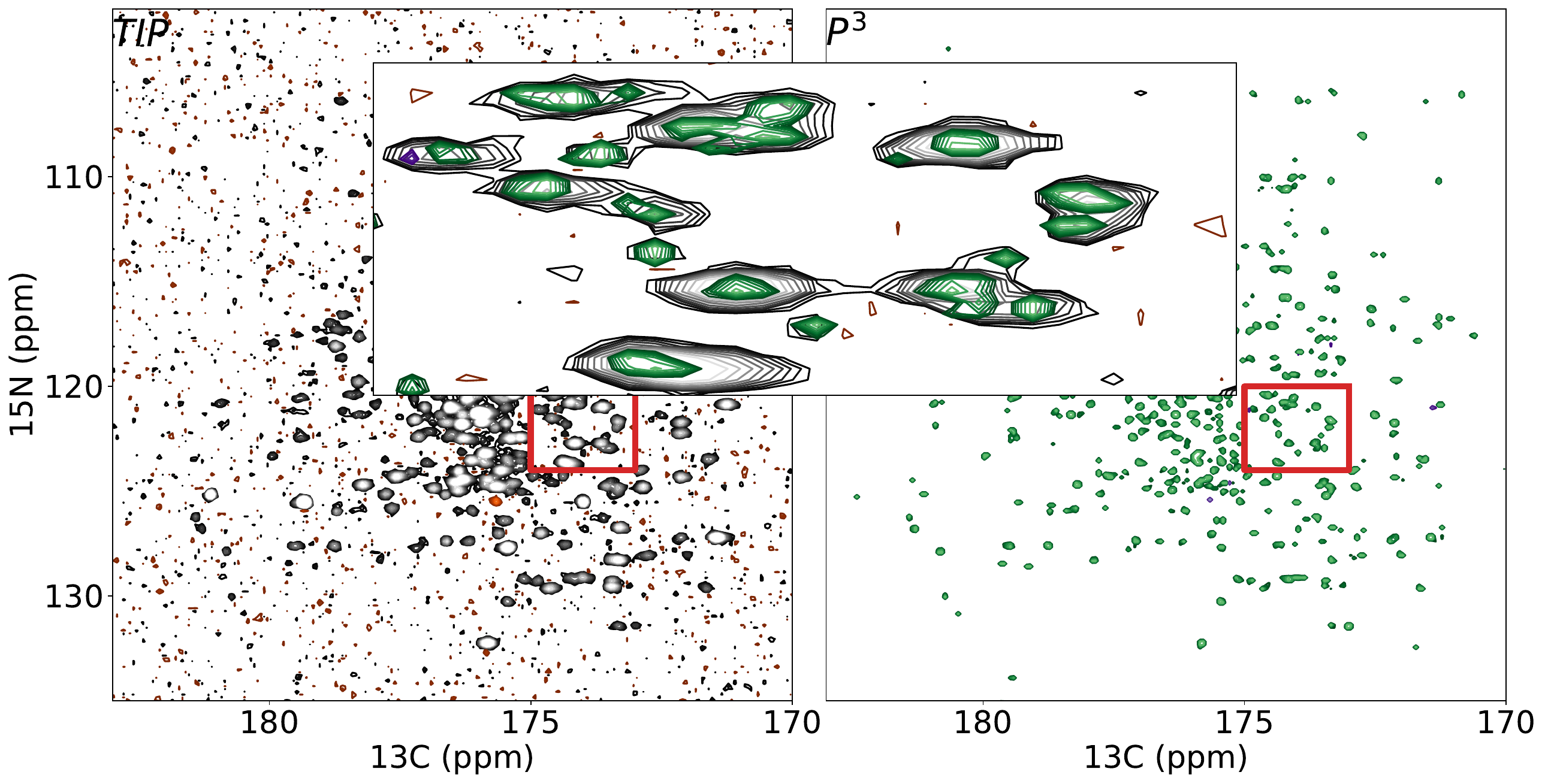}
    \caption{\textbf{3D HN(CA)CO NUS reconstructed spectrum with CS-IST of MALT1 protein.} Intensity presentation, $TIP$, in black ( orange for negative) and peak probability presentations, $P^3$ by using MR-Ai, in green (purple for negative) color of 2D projections.}
    \label{fig:HN(CA)CO_Malt}
\end{figure}

\begin{figure}[htbp]
    \centering
    \includegraphics[width=0.75\textwidth]{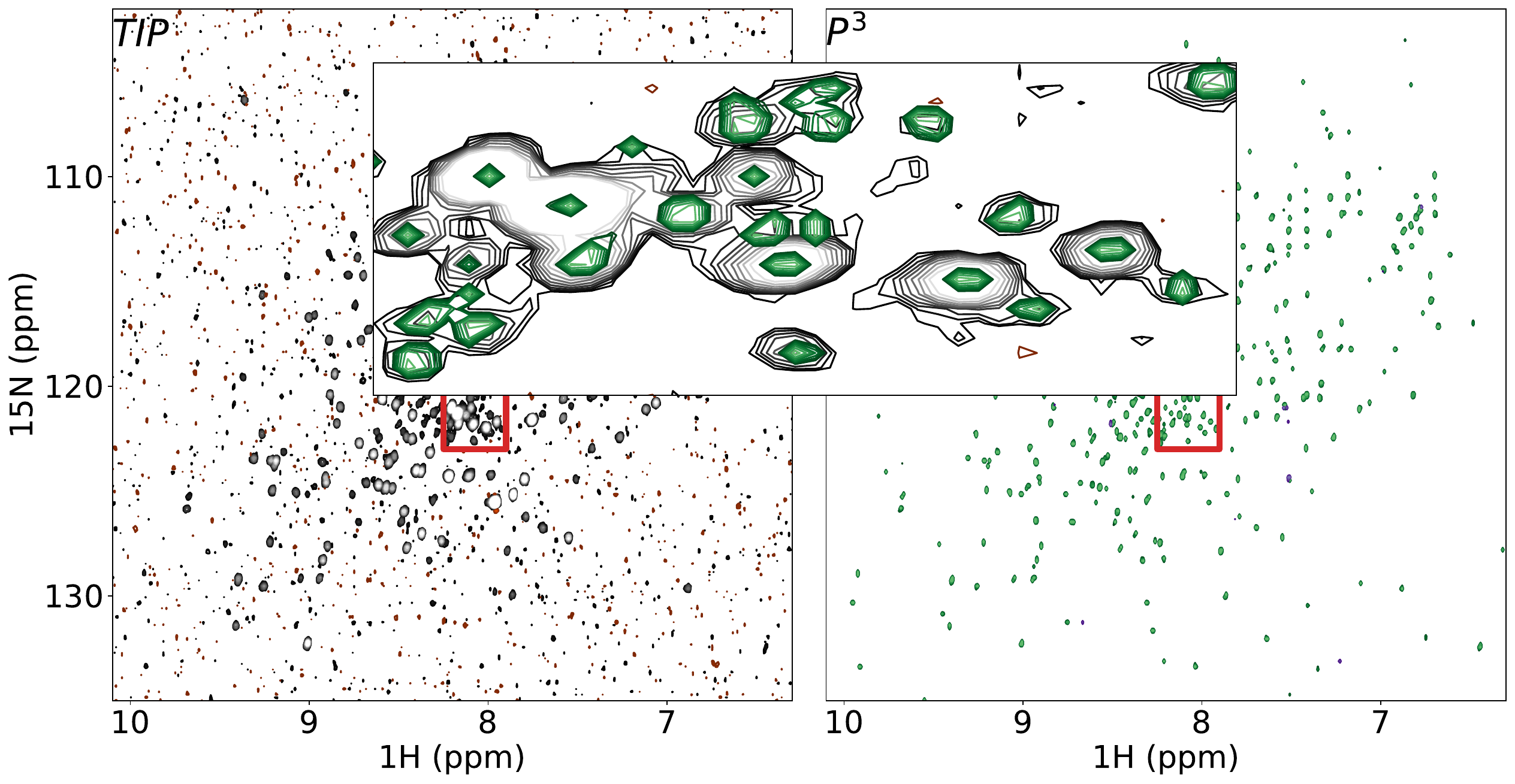}
    \includegraphics[width=0.75\textwidth]{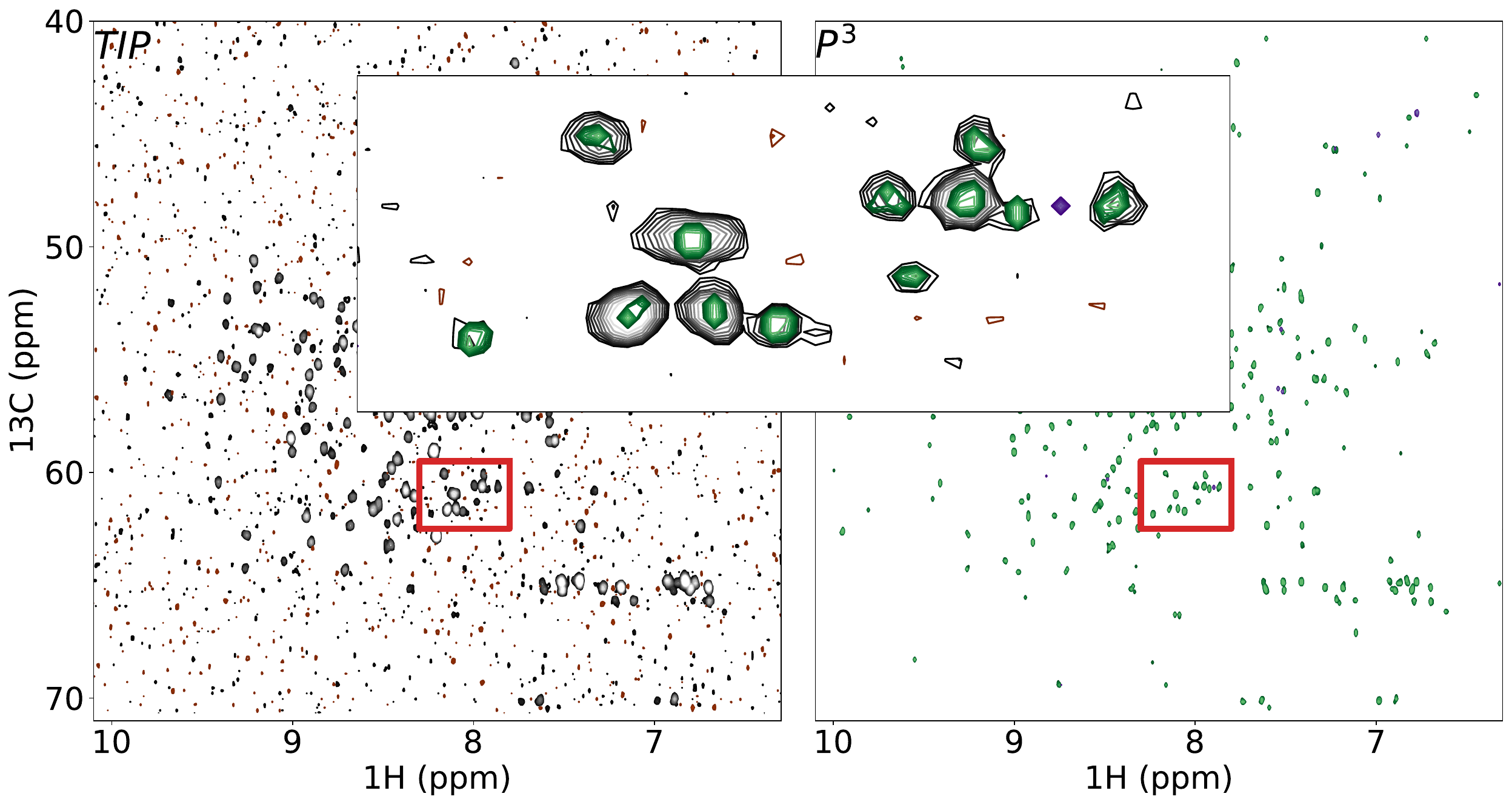}
    \includegraphics[width=0.75\textwidth]{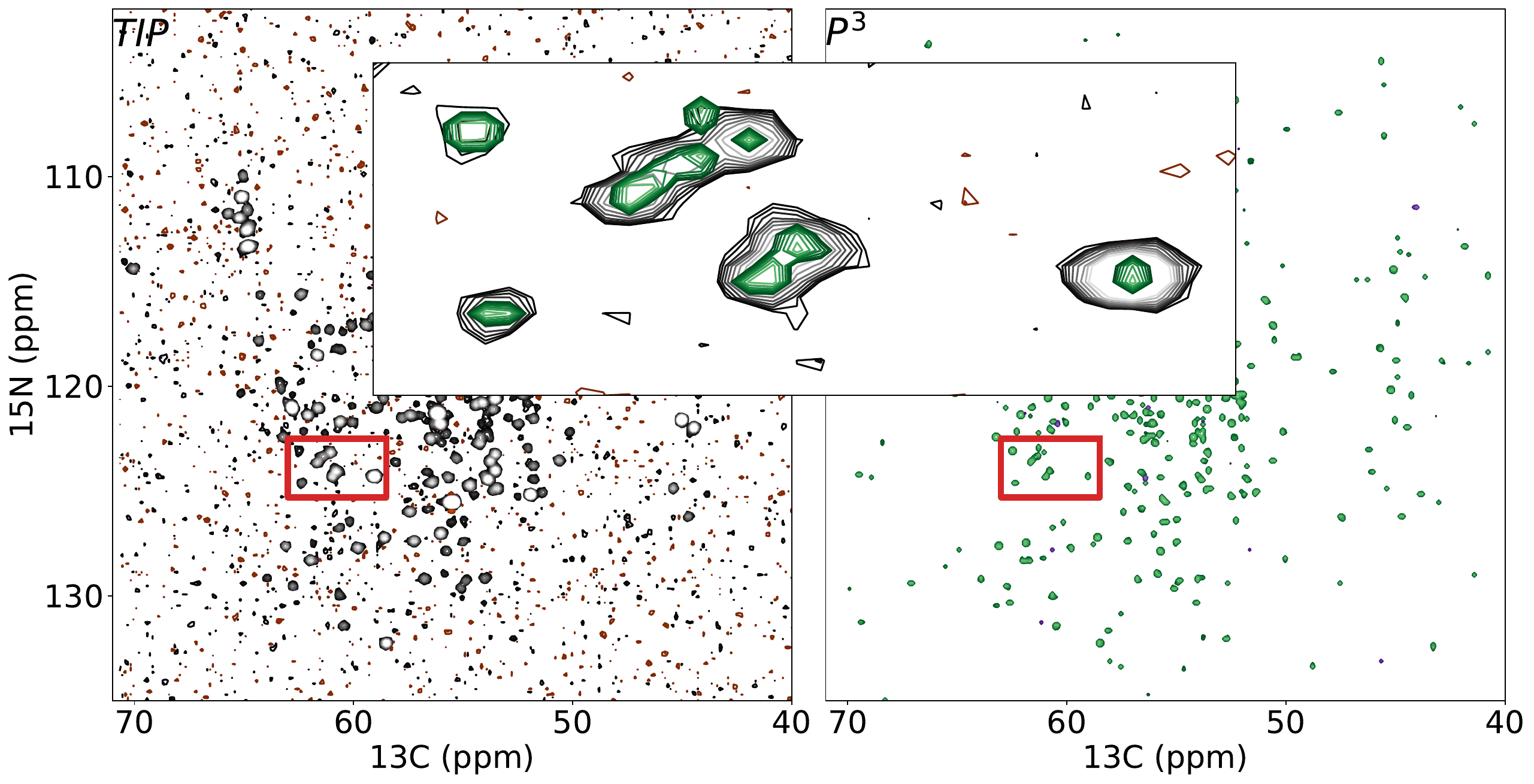}
    \caption{\textbf{3D HN(CO)CA NUS reconstructed spectrum with CS-IST of MALT1 protein.} Intensity presentation, $TIP$, in black ( orange for negative) and peak probability presentations, $P^3$ by using MR-Ai, in green (purple for negative) color of 2D projections.}
    \label{fig:HN(CO)CA_Malt}
\end{figure}

\begin{figure}[htbp]
    \centering
    \includegraphics[width=0.75\textwidth]{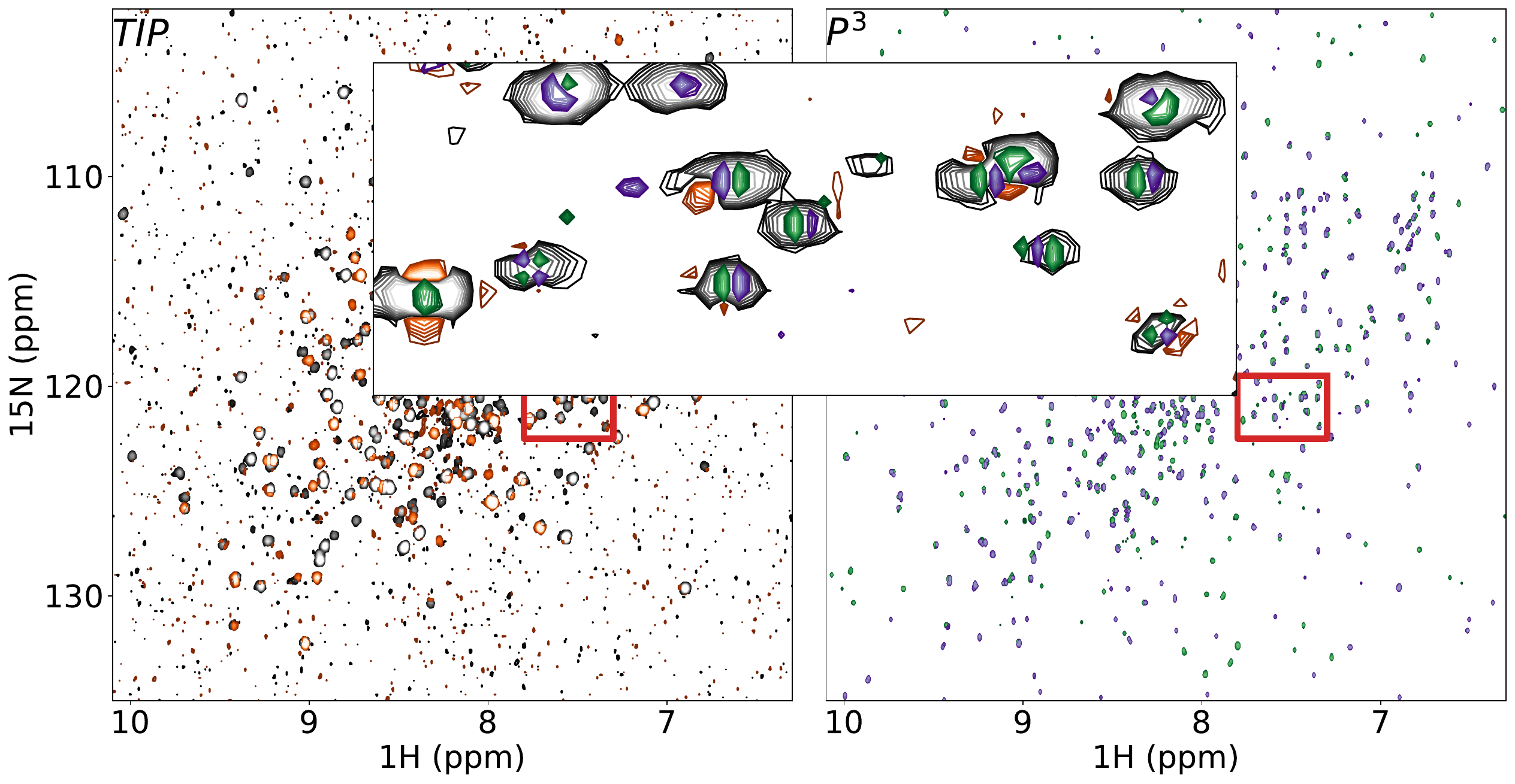}
    \includegraphics[width=0.75\textwidth]{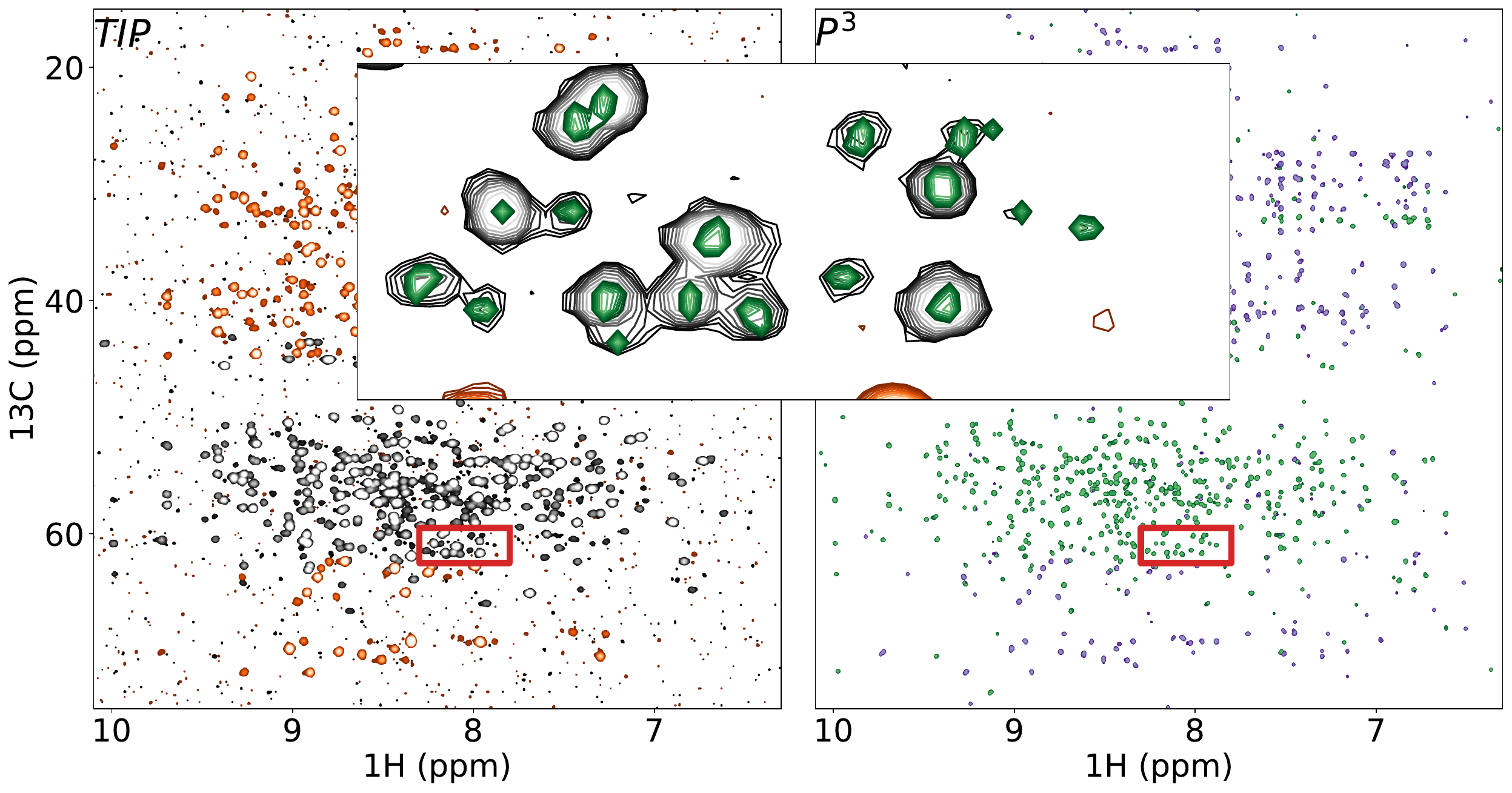}
    \includegraphics[width=0.75\textwidth]{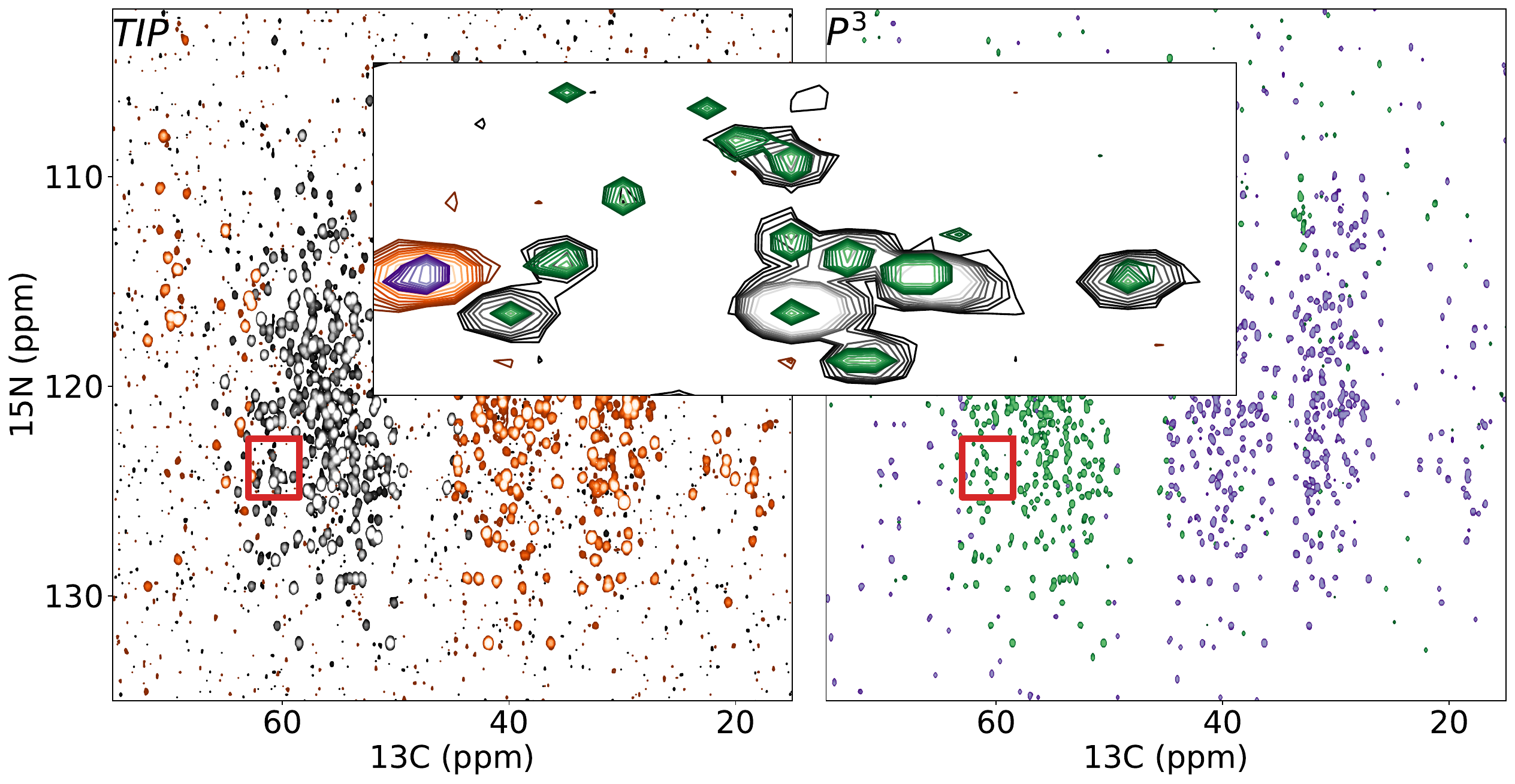}
    \caption{\textbf{3D HNCACB NUS reconstructed spectrum with CS-IST of MALT1 protein.} Intensity presentation, $TIP$, in black ( orange for negative) and peak probability presentations, $P^3$ by using MR-Ai, in green (purple for negative) color of 2D projections.}
    \label{fig:HNCACB_Malt}
\end{figure}

\begin{figure}[htbp]
    \centering
    \includegraphics[width=0.75\textwidth]{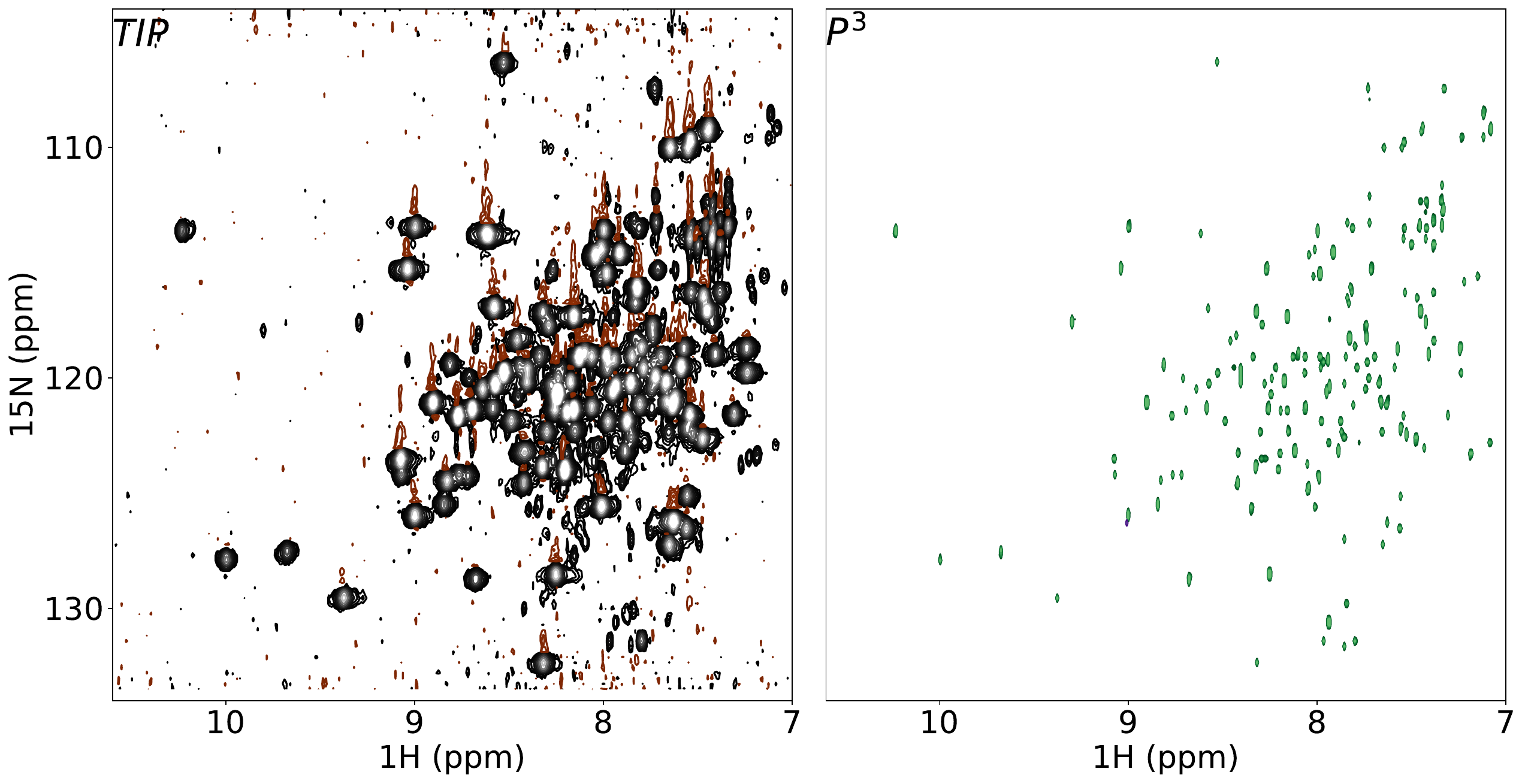}
    \includegraphics[width=0.75\textwidth]{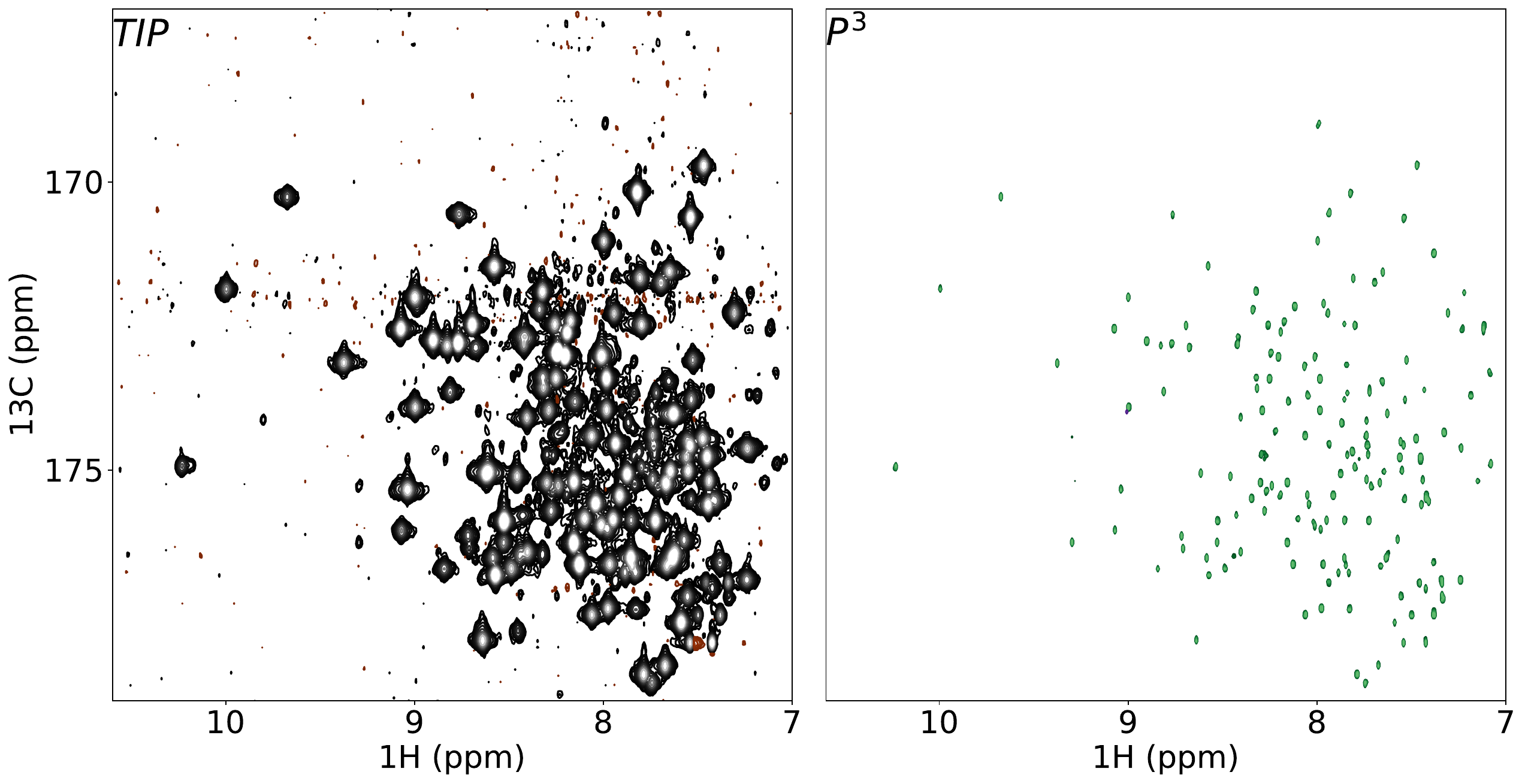}
    \includegraphics[width=0.75\textwidth]{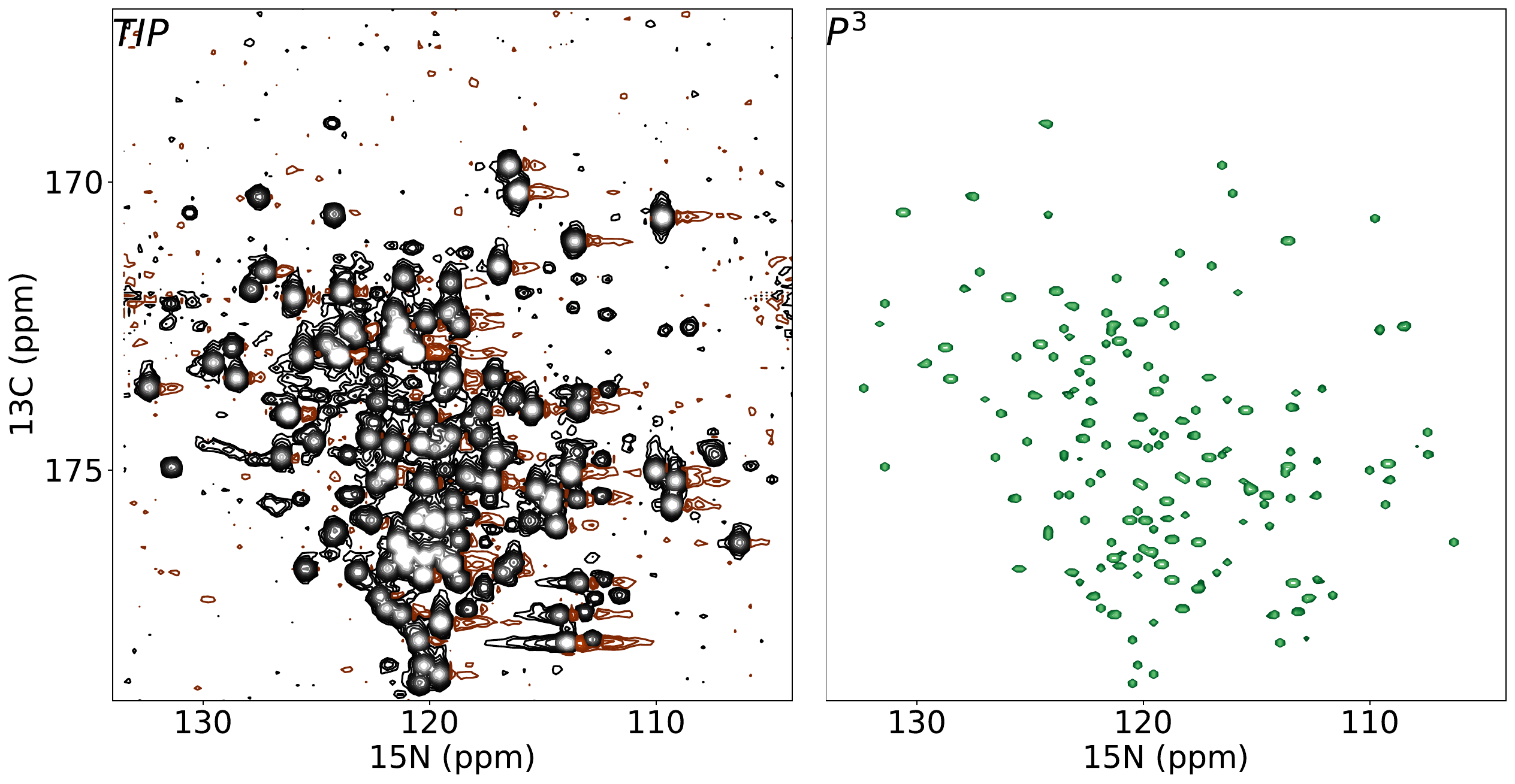}
    \caption{\textbf{3D HNCO NUS spectrum of Calmodulin reconstructed with CS-IST.} Intensity presentation, $TIP$, in black ( orange for negative) and peak probability presentations, $P^3$ by using MR-Ai, in green (purple for negative) color of 2D projections. Note a small phase error in $^{15}N$ dimension, which is however within the training parameter limits (Table 1) and does not affect the resulting $P^3$.}
    \label{fig:HNCO_Cam}
\end{figure}

\begin{figure}[htbp]
    \centering
    \includegraphics[width=0.75\textwidth]{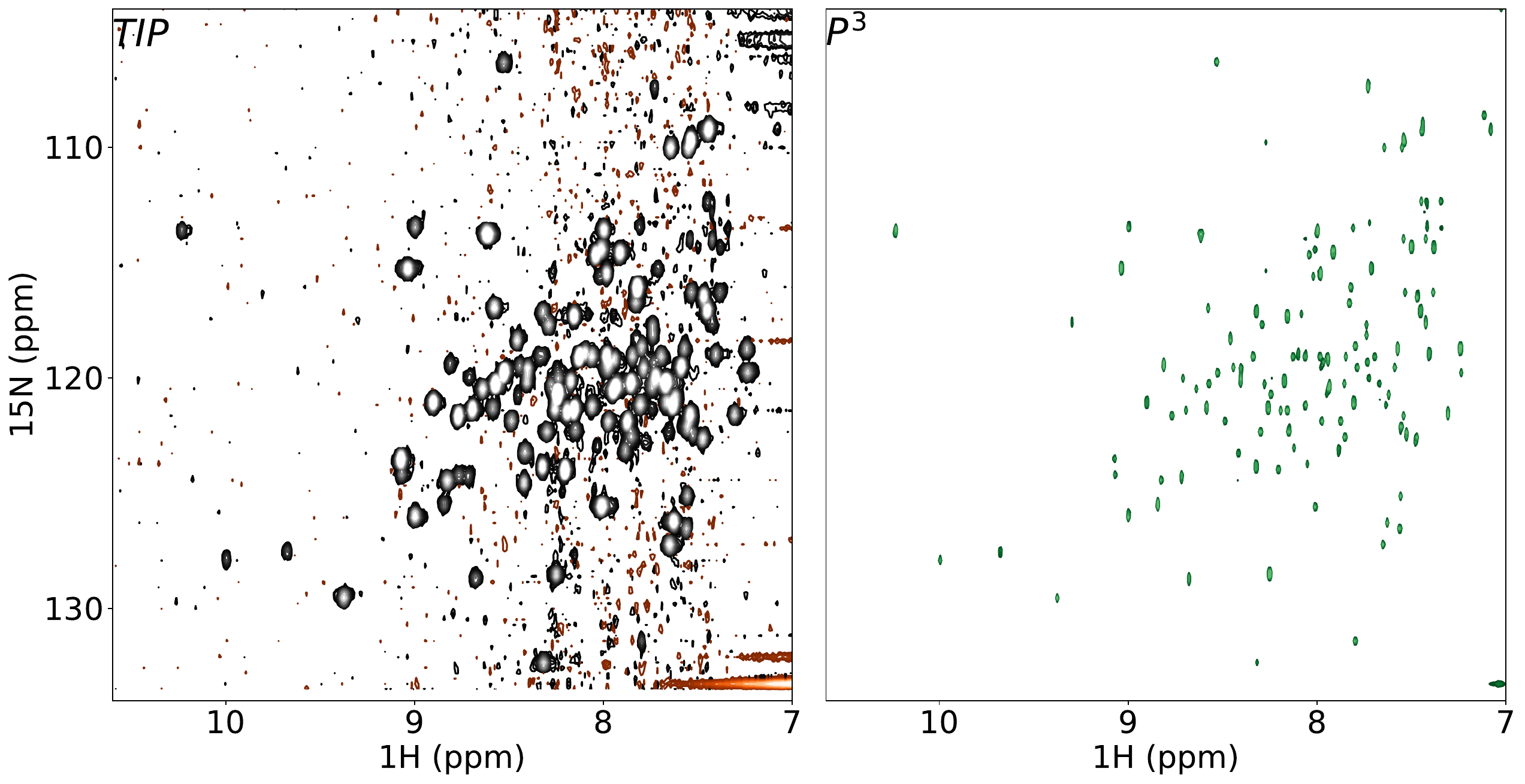}
    \includegraphics[width=0.75\textwidth]{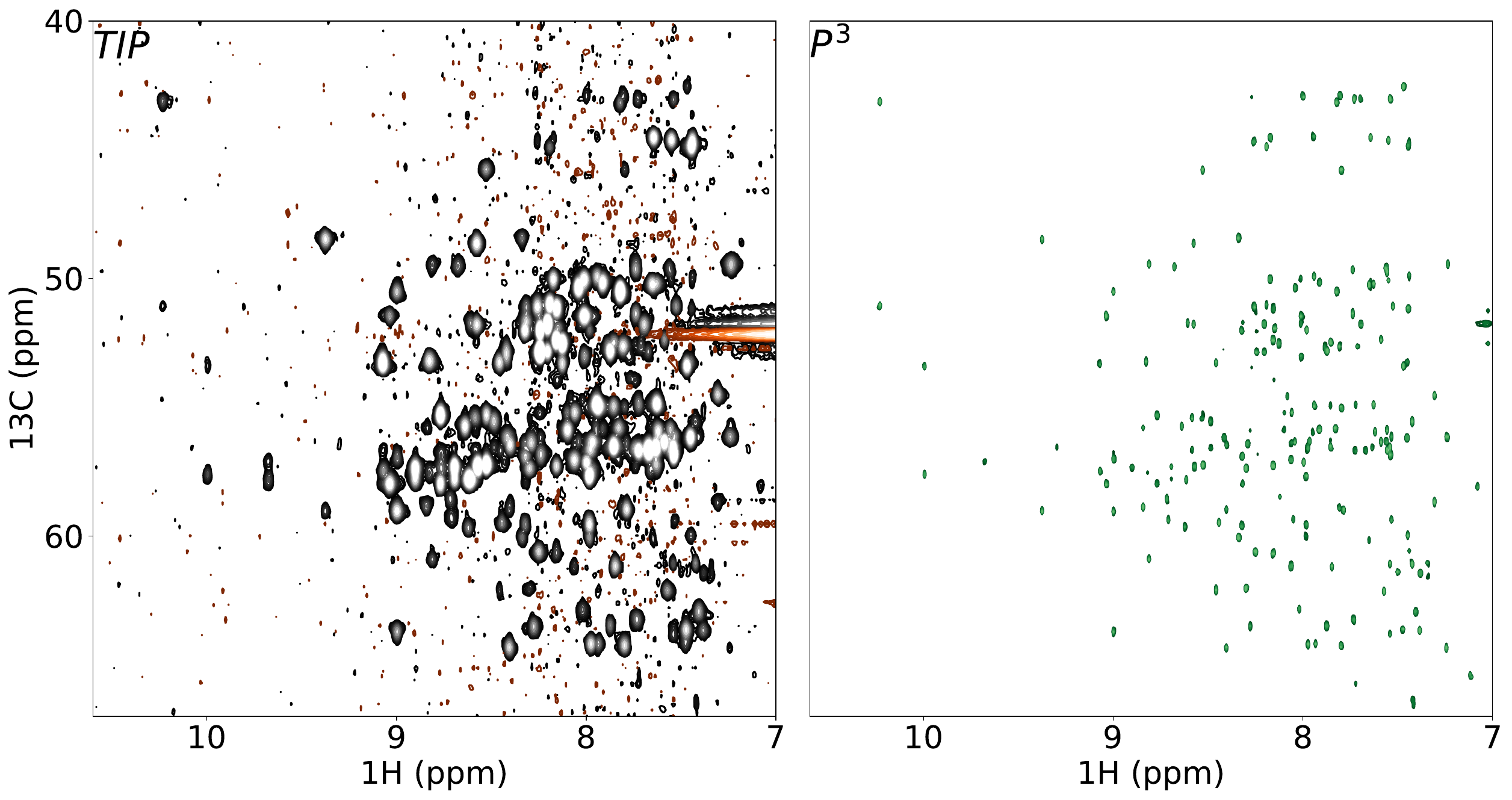}
    \includegraphics[width=0.75\textwidth]{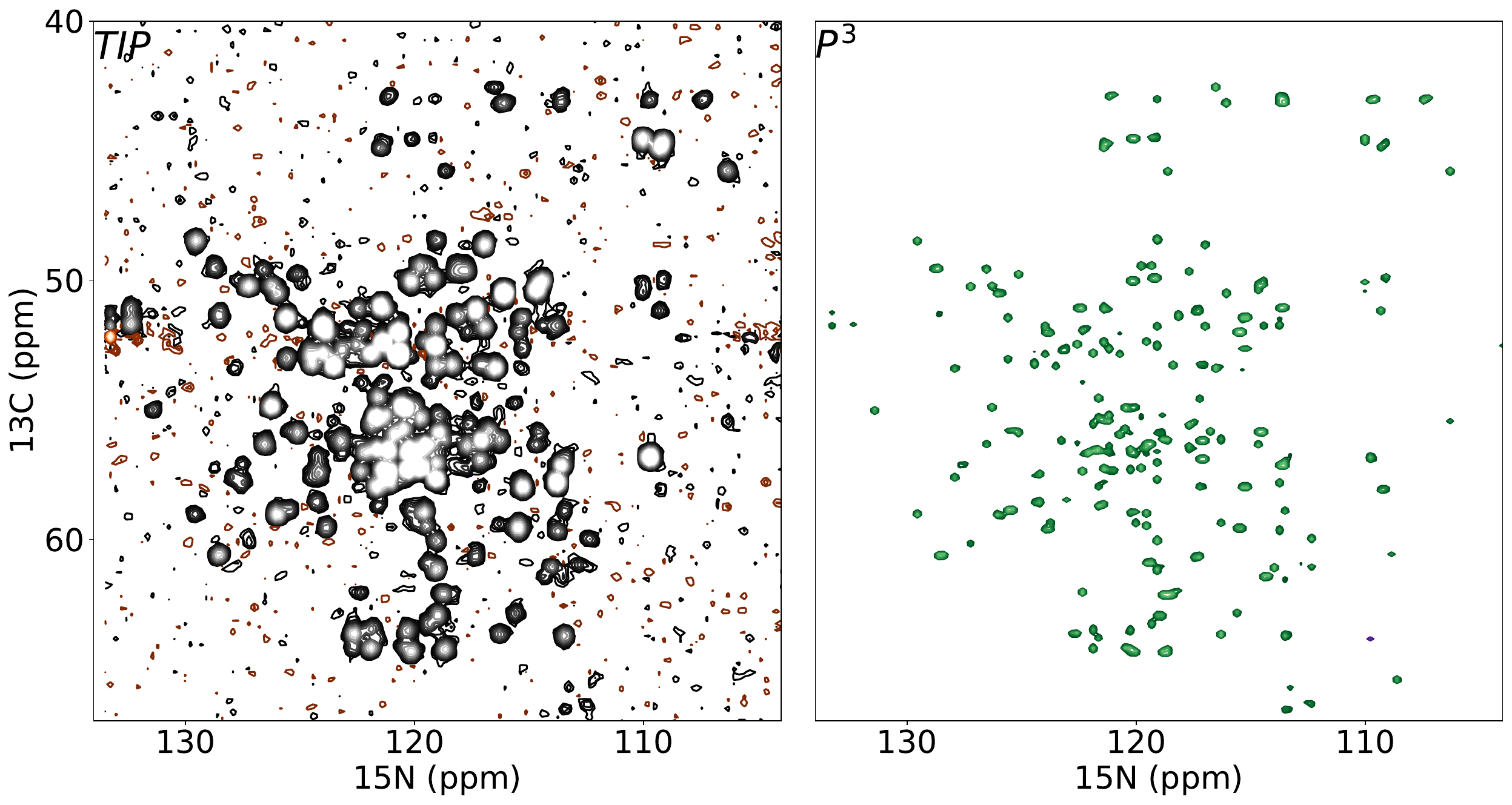}
    \caption{\textbf{3D HNCA NUS reconstructed spectrum with CS-IST of Calmodulin protein.} Intensity presentation, $TIP$, in black ( orange for negative) and peak probability presentations, $P^3$ by using MR-Ai, in green (purple for negative) color of 2D projections.}
    \label{fig:HNCA_Cam}
\end{figure}

\begin{figure}[htbp]
    \centering
    \includegraphics[width=0.75\textwidth]{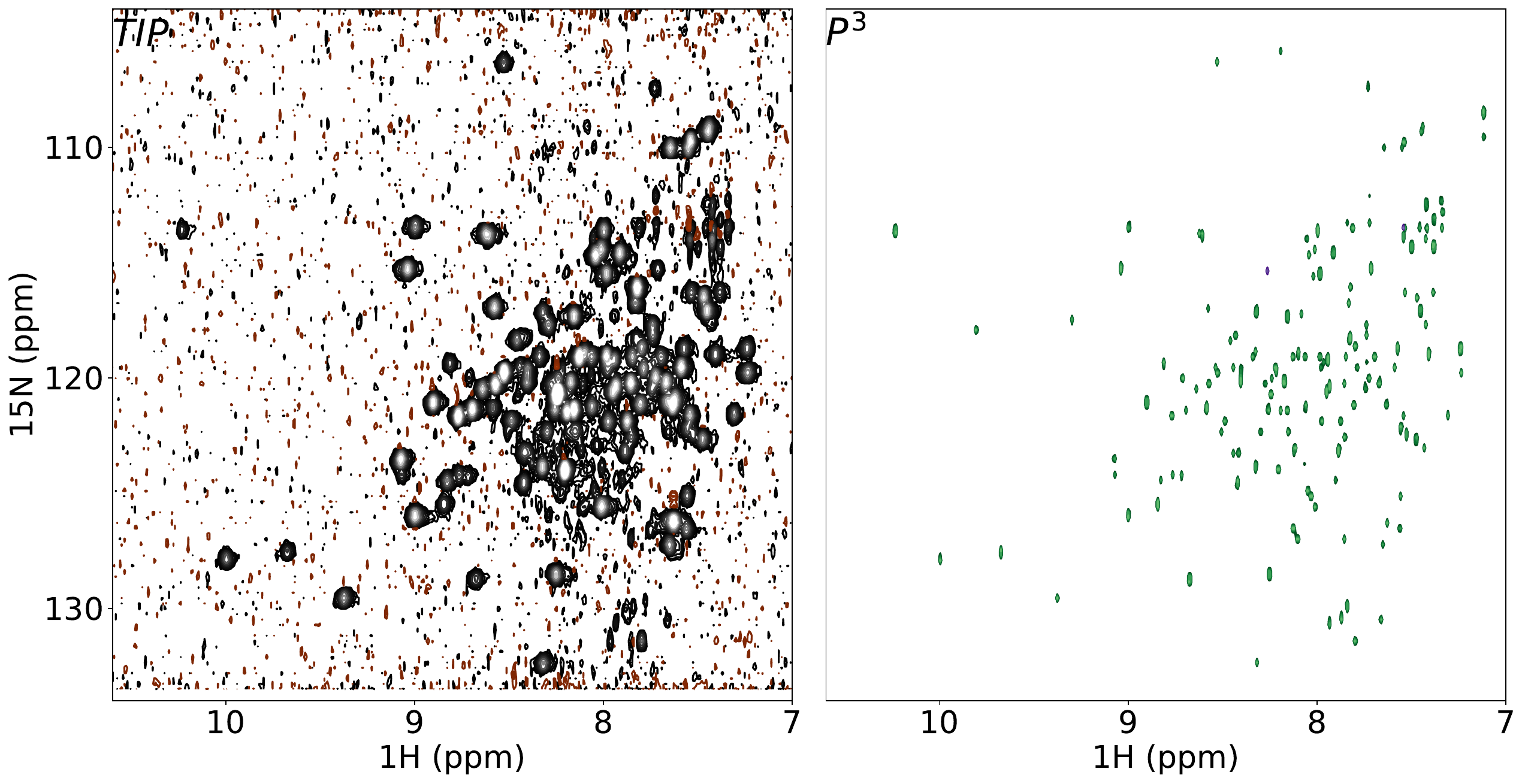}
    \includegraphics[width=0.75\textwidth]{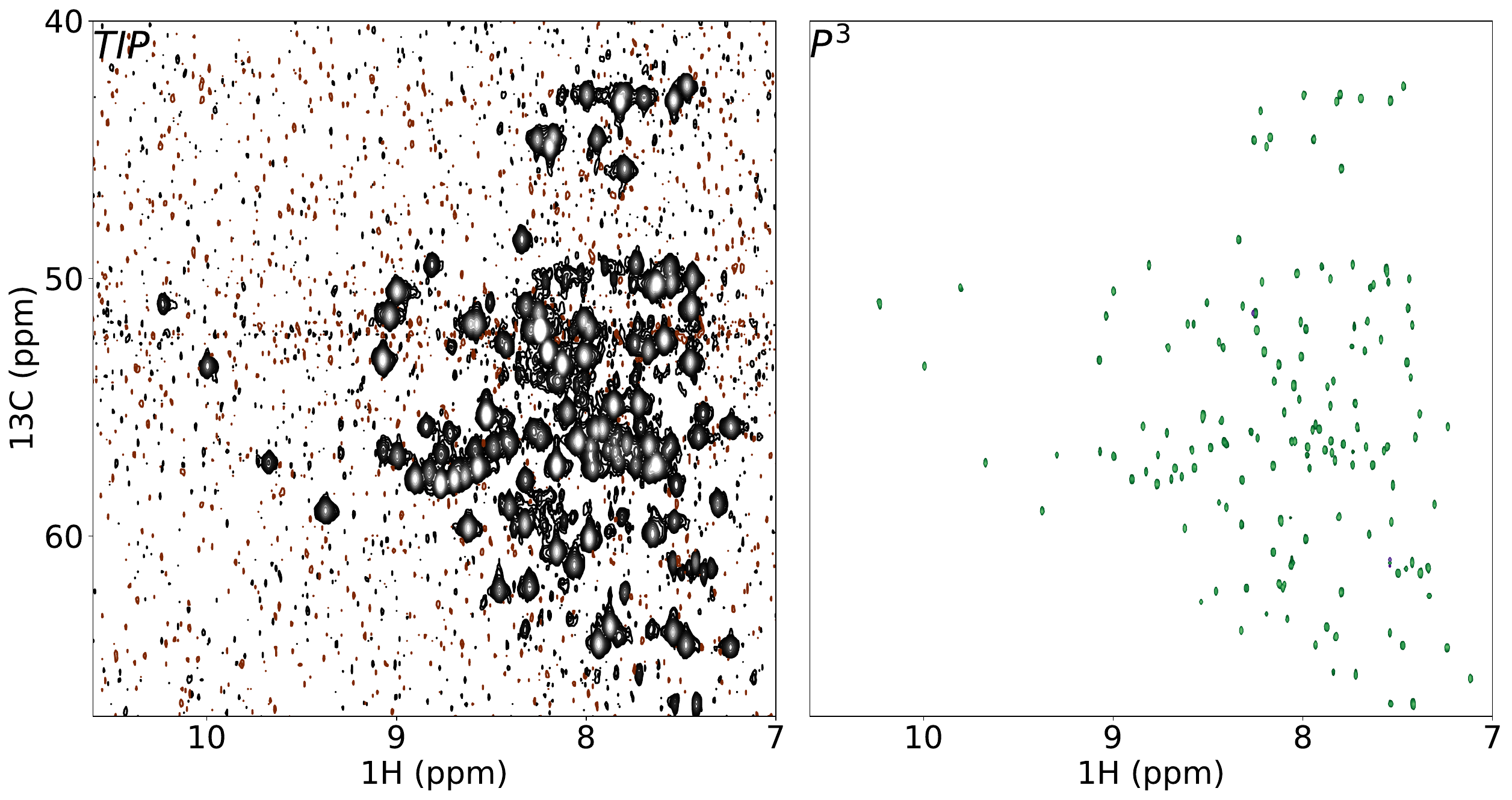}
    \includegraphics[width=0.75\textwidth]{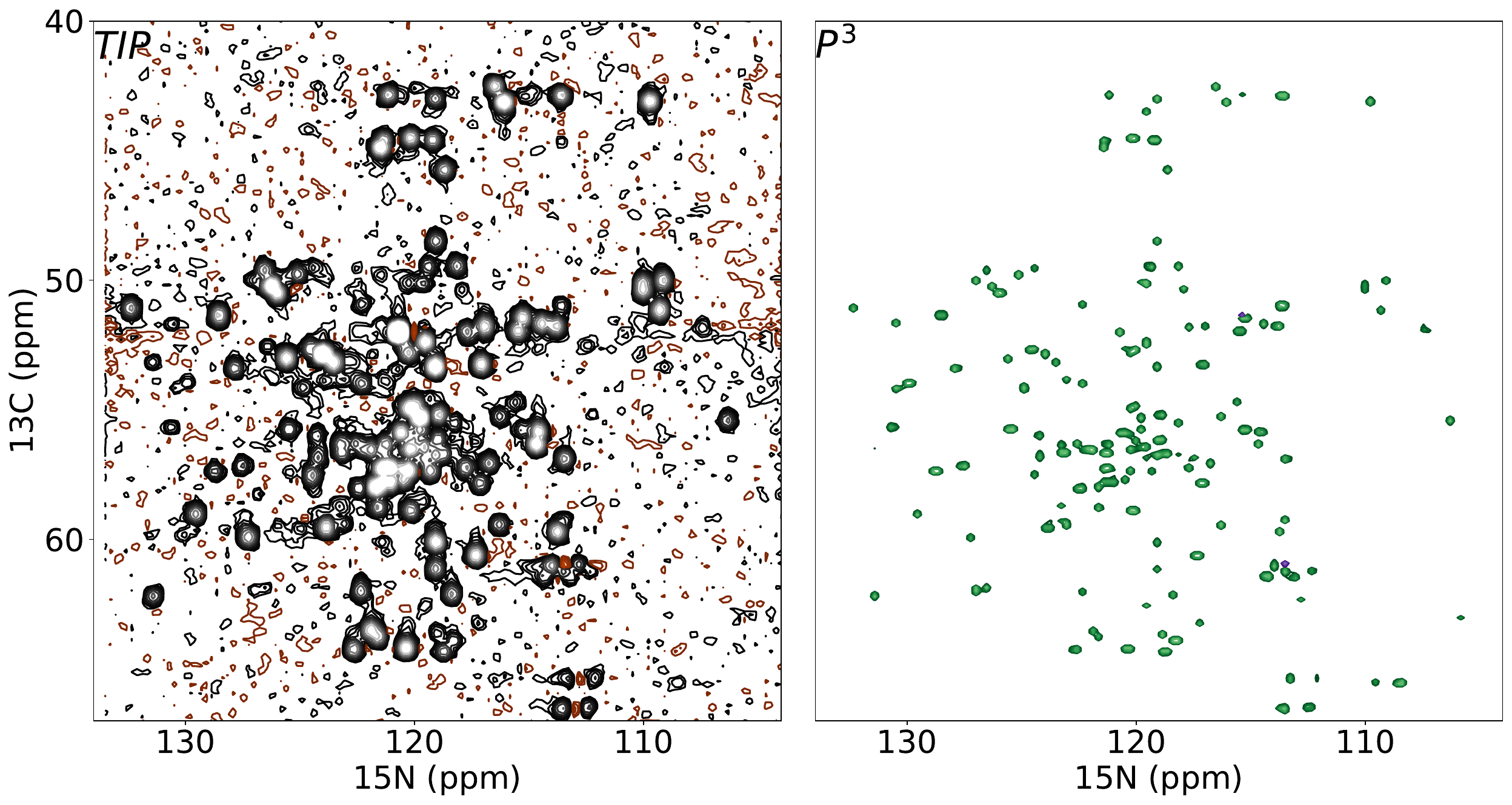}
    \caption{\textbf{3D HN(CO)CA NUS reconstructed spectrum with CS-IST of Calmodulin protein.} Intensity presentation, $TIP$, in black ( orange for negative) and peak probability presentations, $P^3$ by using MR-Ai, in green (purple for negative) color of 2D projections.}
    \label{fig:HN(CO)CA_Cam}
\end{figure}

\begin{figure}[htbp]
    \centering
    \includegraphics[width=0.75\textwidth]{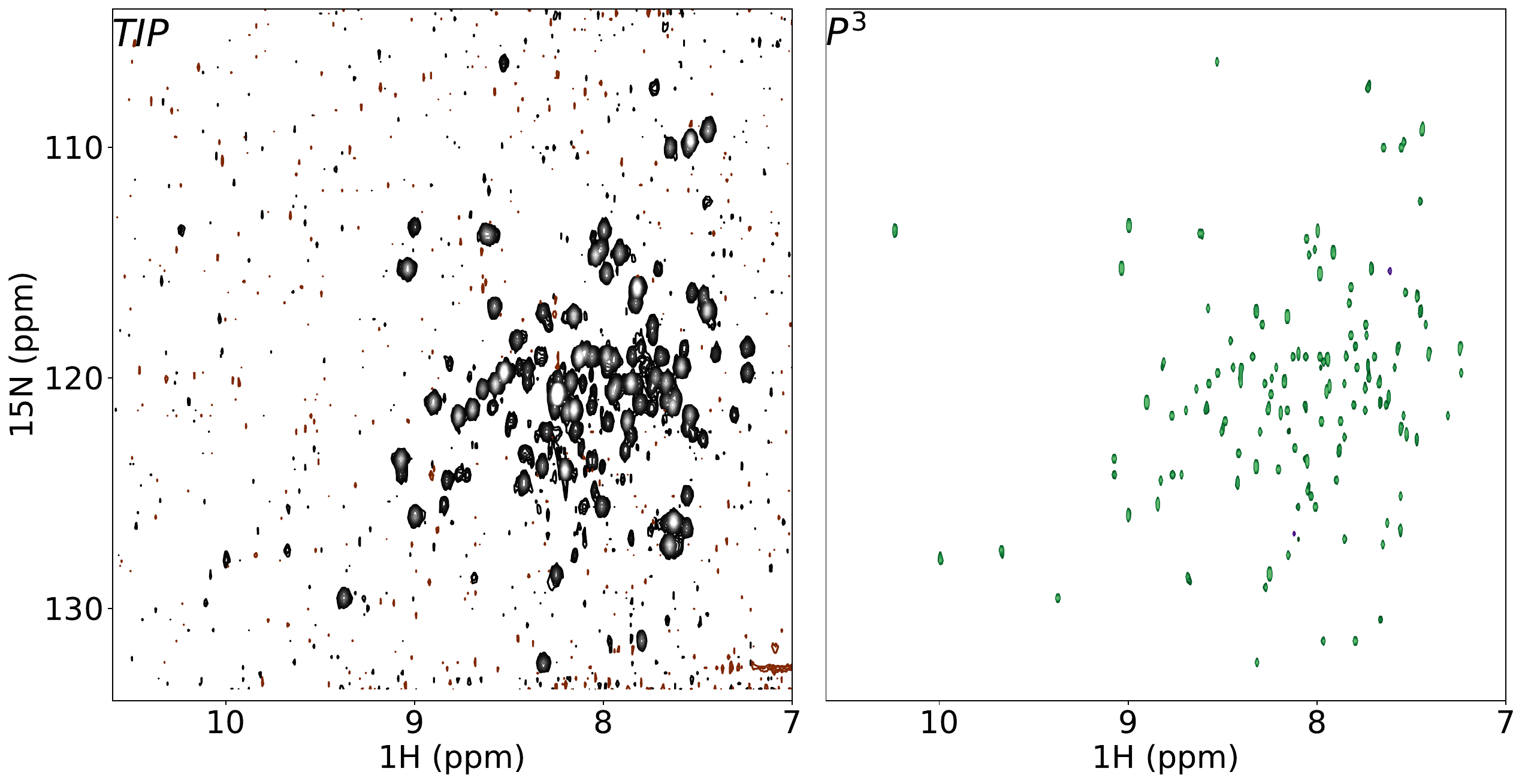}
    \includegraphics[width=0.75\textwidth]{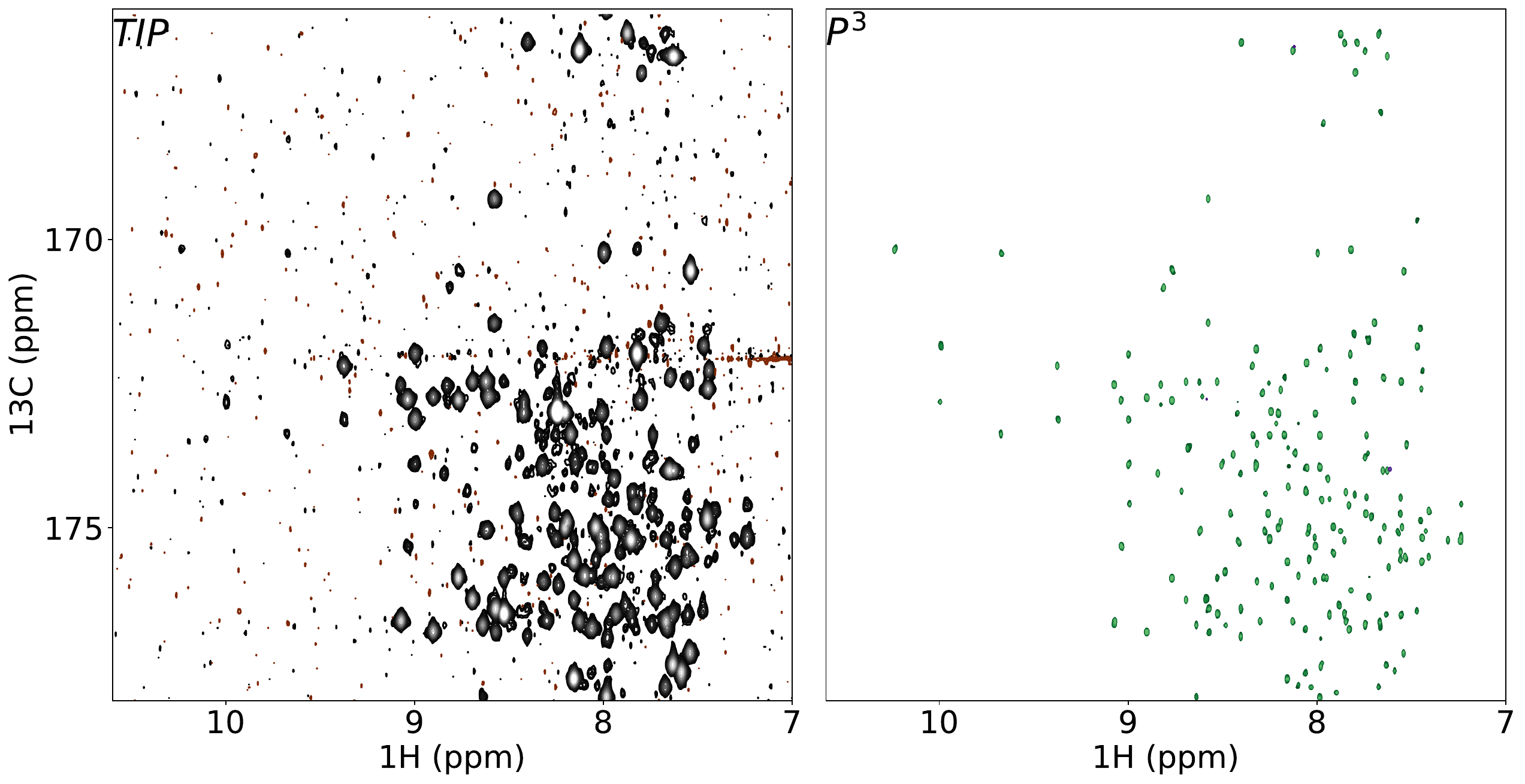}
    \includegraphics[width=0.75\textwidth]{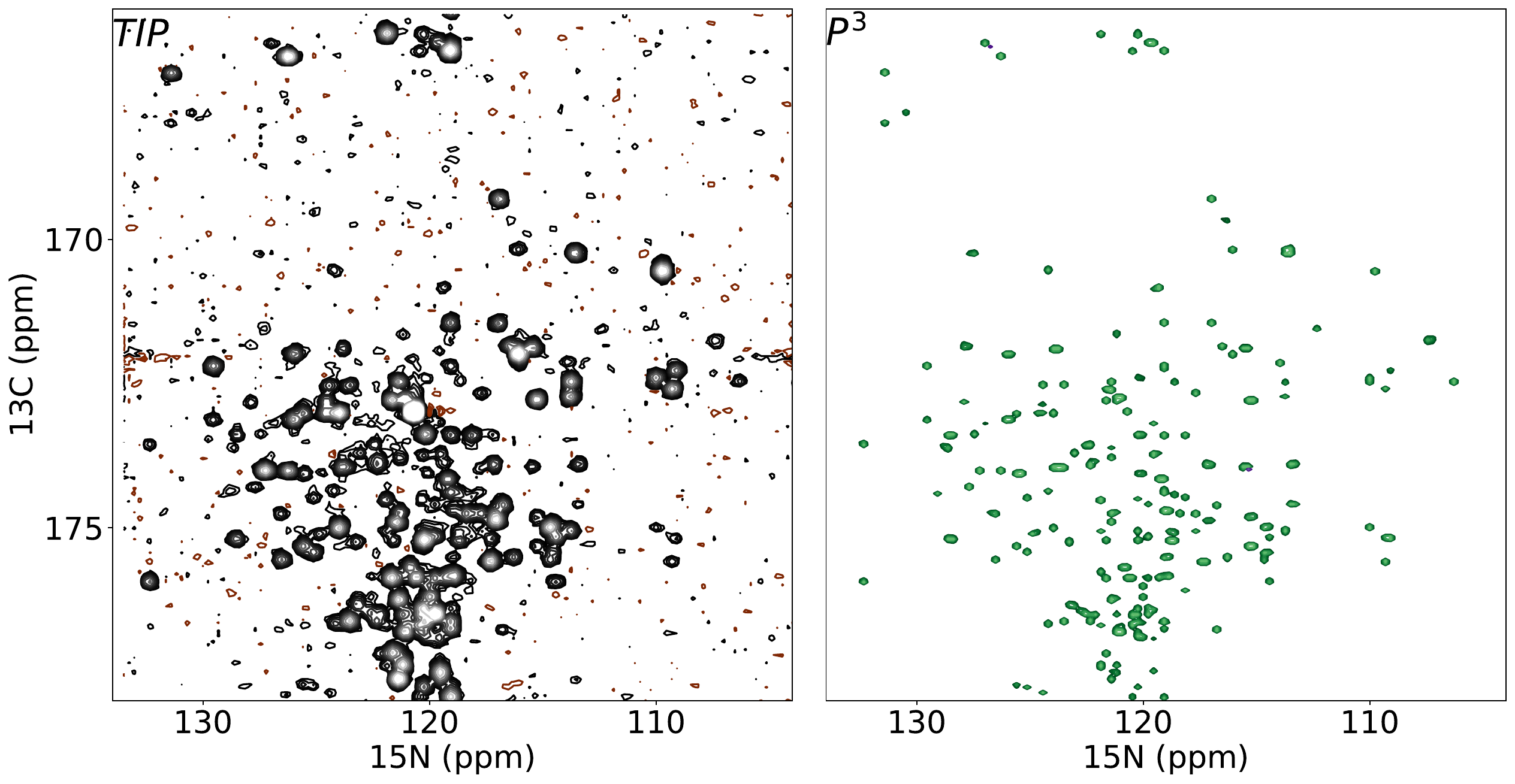}
    \caption{\textbf{3D HN(CA)CO NUS reconstructed spectrum with CS-IST of Calmodulin protein.} Intensity presentation, $TIP$, in black ( orange for negative) and peak probability presentations, $P^3$ by using MR-Ai, in green (purple for negative) color of 2D projections.}
    \label{fig:HN(CA)CO_Cam}
\end{figure}

\begin{figure}[htbp]
    \centering
    \includegraphics[width=0.75\textwidth]{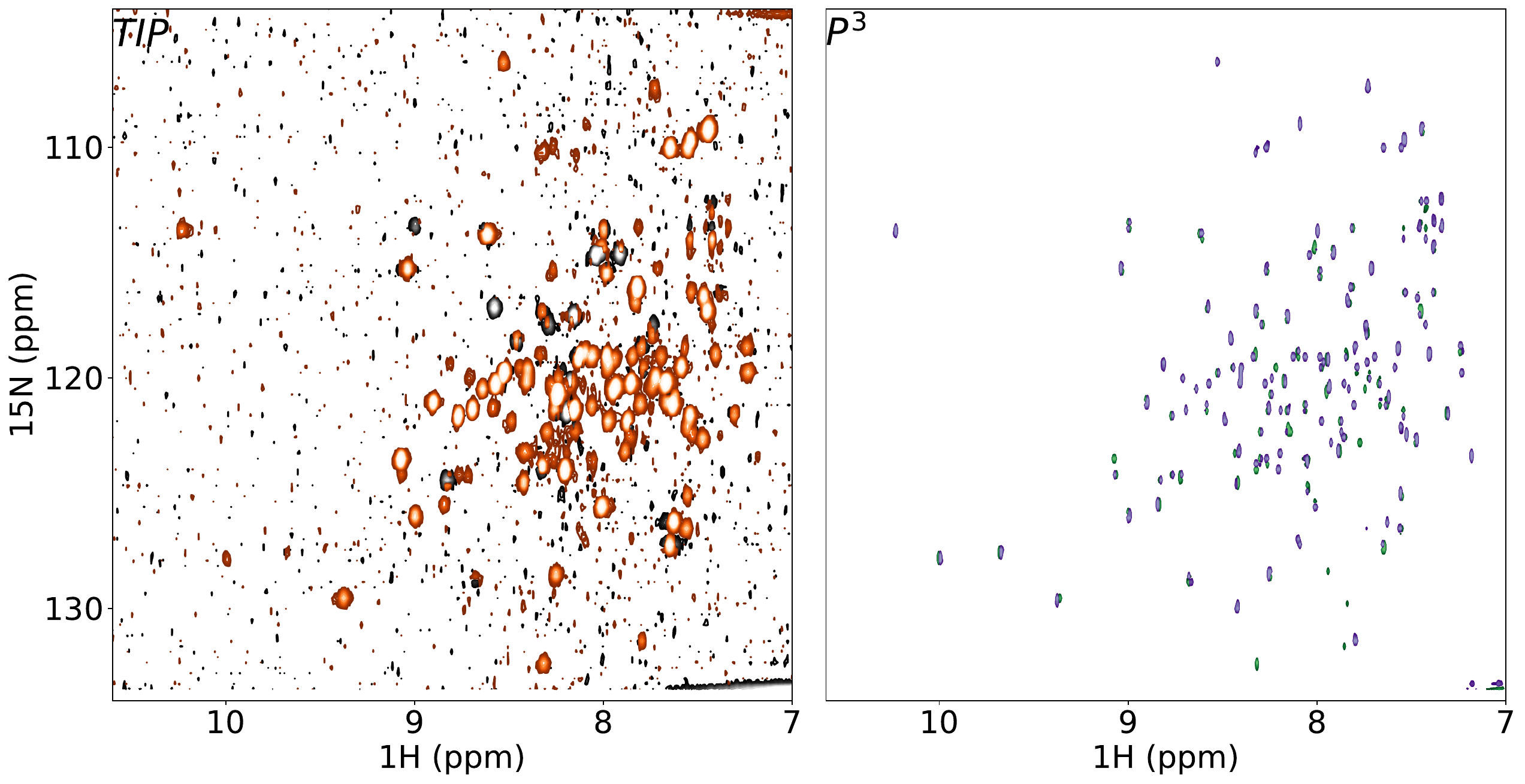}
    \includegraphics[width=0.75\textwidth]{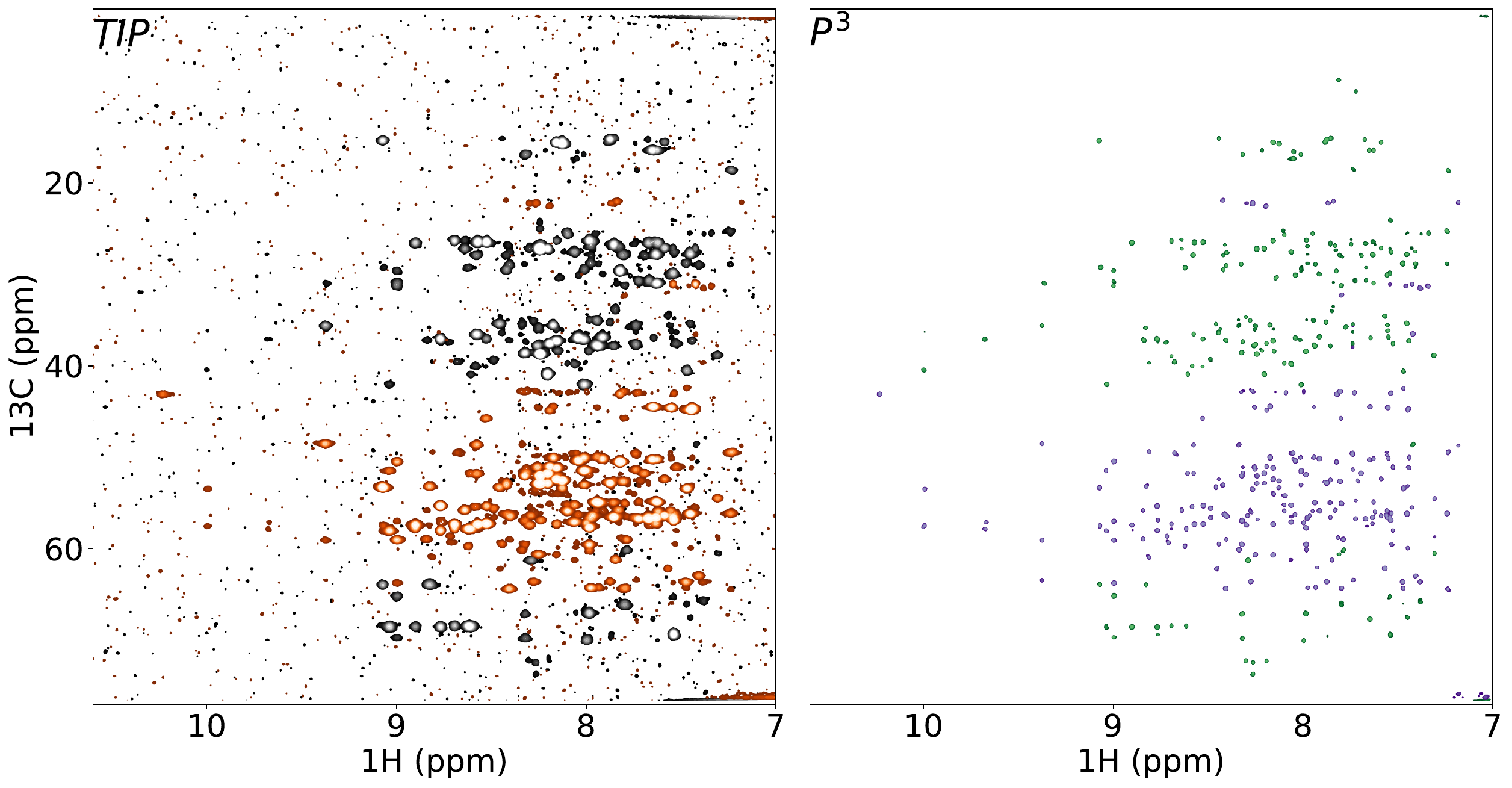}
    \includegraphics[width=0.75\textwidth]{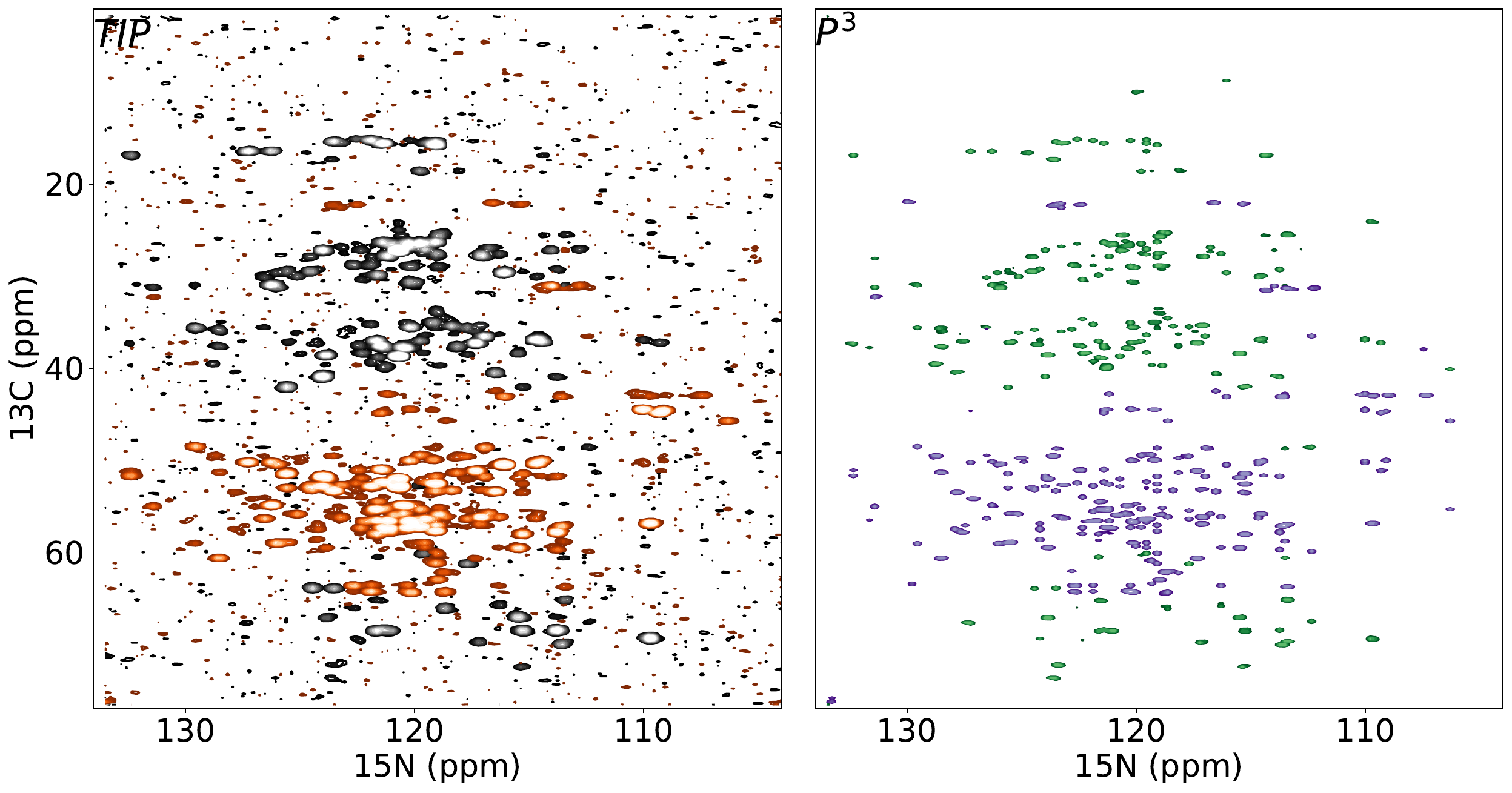}
    \caption{\textbf{3D HNCACB NUS reconstructed spectrum with CS-IST of Calmodulin protein.} Intensity presentation, $TIP$, in black ( orange for negative) and peak probability presentations, $P^3$ by using MR-Ai, in green (purple for negative) color of 2D projections.}
    \label{fig:HNCACB_Cam}
\end{figure}

\begin{figure}[htbp]
    \centering
    \includegraphics[width=0.75\textwidth]{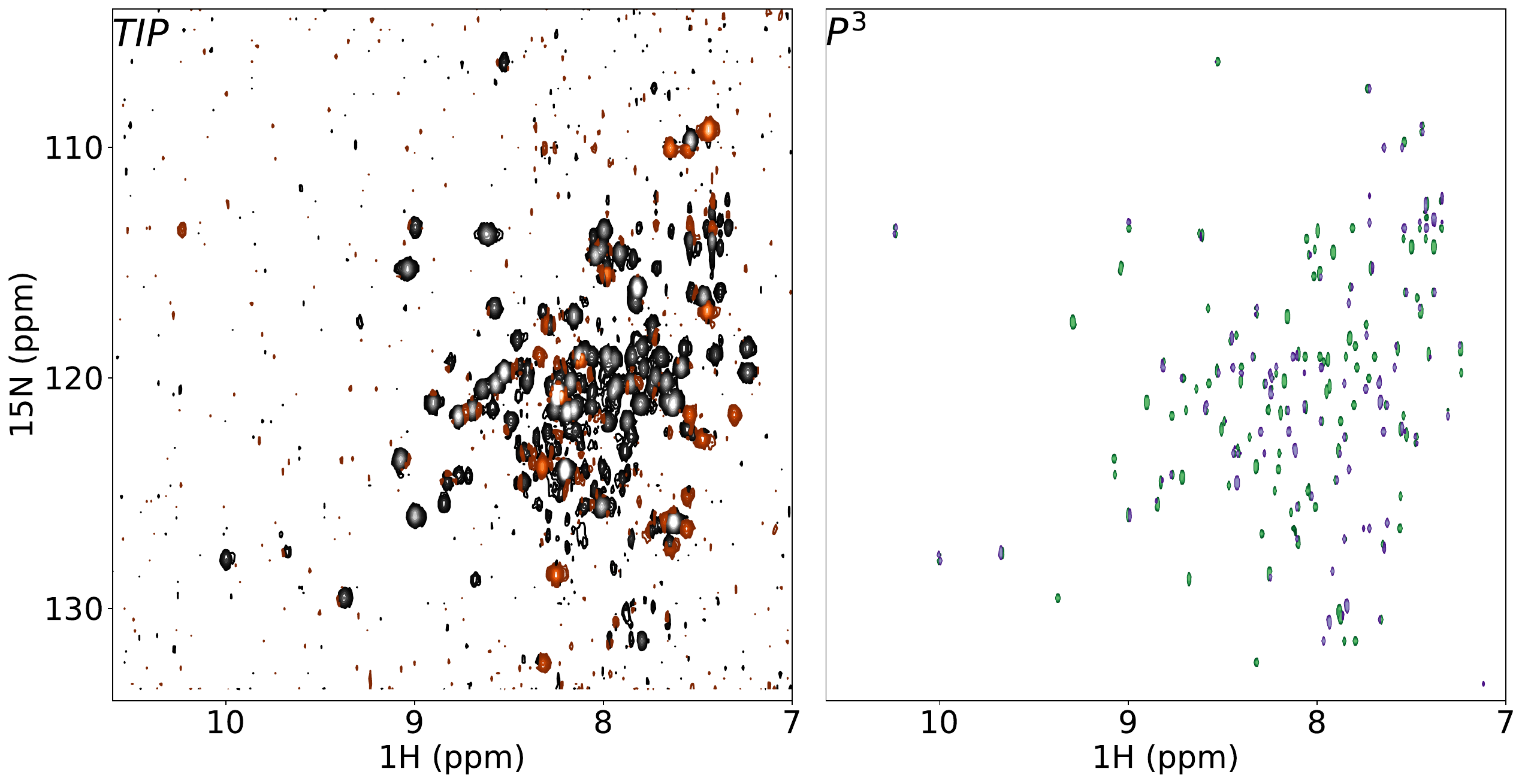}
    \includegraphics[width=0.75\textwidth]{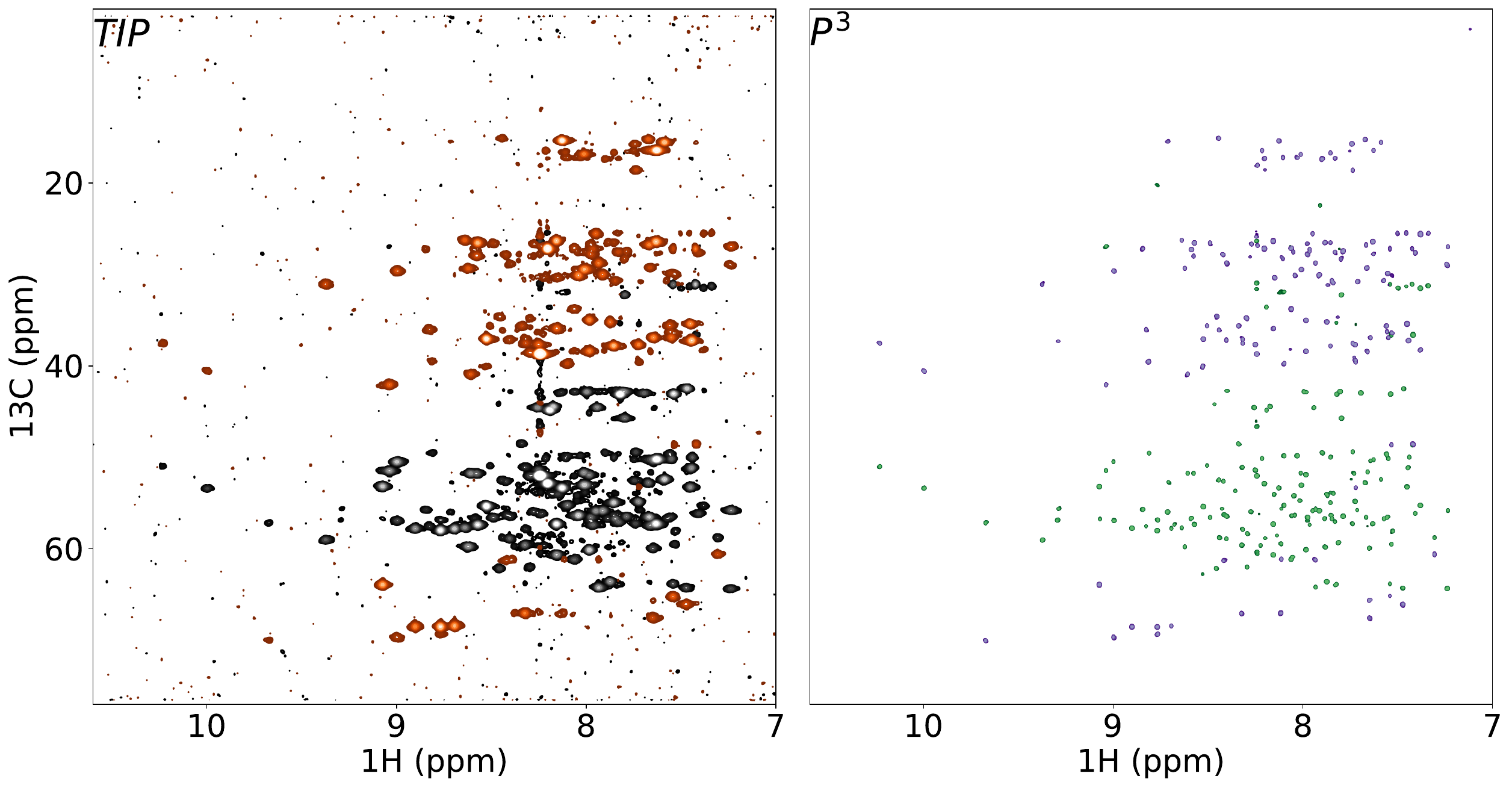}
    \includegraphics[width=0.75\textwidth]{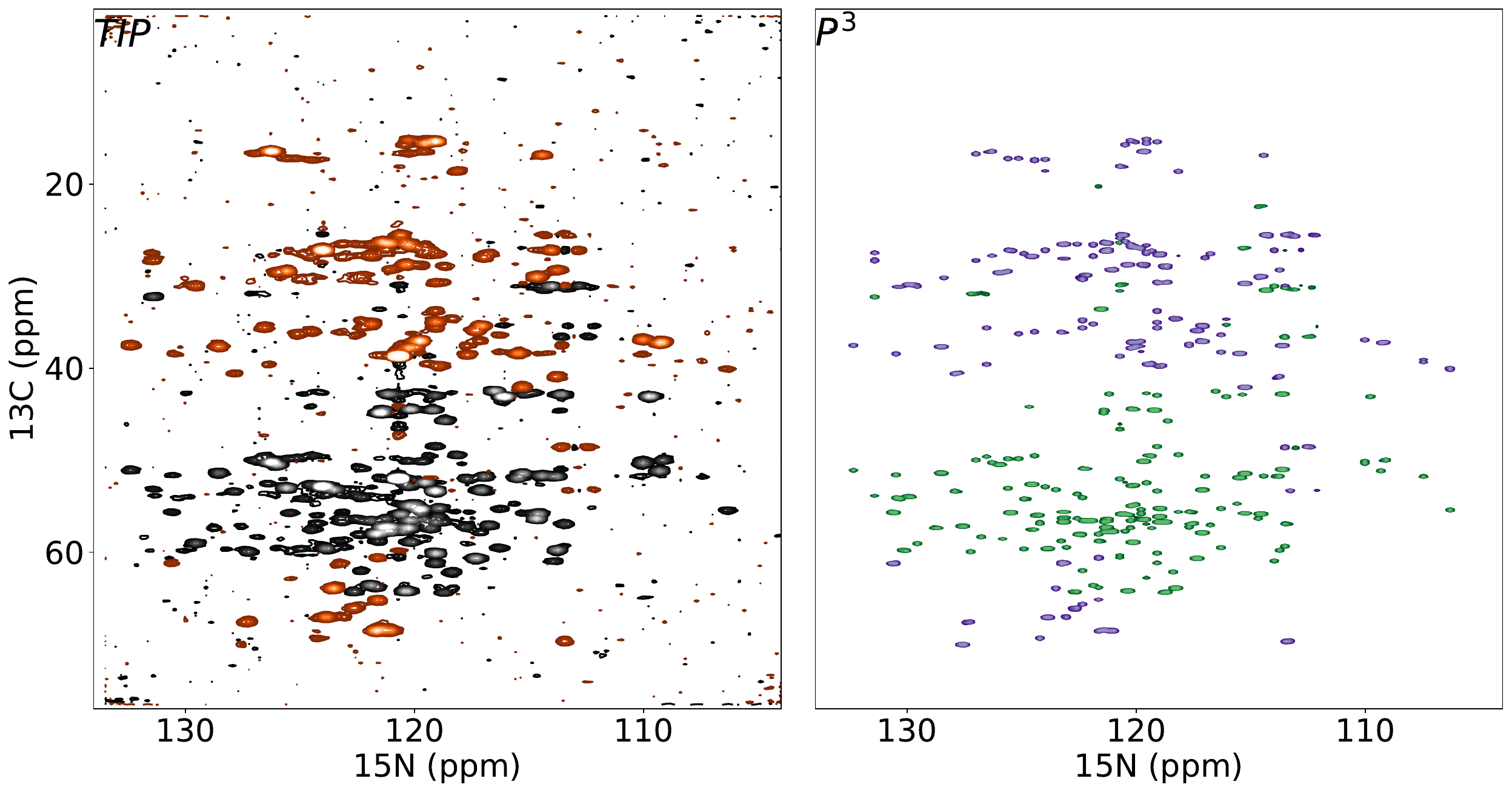}
    \caption{\textbf{3D HN(CO)CACB NUS reconstructed spectrum with CS-IST of Calmodulin protein.} Intensity presentation, $TIP$, in black ( orange for negative) and peak probability presentations, $P^3$ by using MR-Ai, in green (purple for negative) color of 2D projections.}
    \label{fig:HN(CO)CACB_Cam}
\end{figure}

\begin{figure}[htbp]
            \centering
            \includegraphics[width=\textwidth]{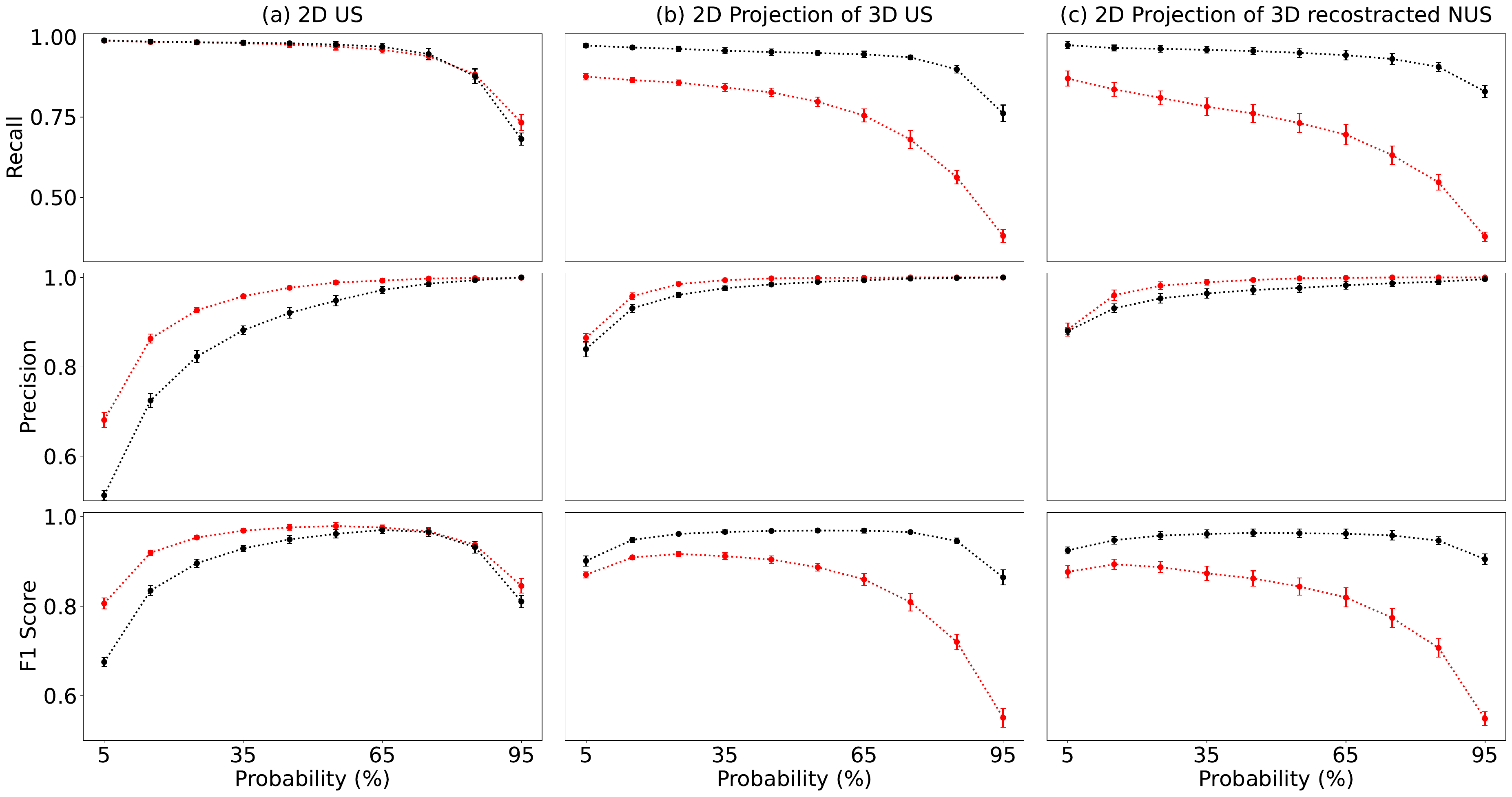}
            \caption{\textbf{Comparison of Peak Detection $P^3$ in Synthetic 2D Spectra Between Two MR-Ai Models Trained with Different Noise Distributions on Training Dataset.} Black and red lines with error bars represent $P^3$ results from MR-Ai trained on double Gaussian and normal Gaussian noise distributions, respectively, across 10 synthetic NMR spectra, each containing 256 peaks, for (a) 2D US, (b) 2D projection of 3D US, and (c) 2D projection of 3D 15\% NUS reconstructed using CS-IST. F1 score is defined as the harmonic mean of the precision and recall scores with $F1=2\frac{Precision\times Recall}{Precision+ Recall}$. Recall is defined as the ratio of the correctly $detected$ pixels to all $detectable$ pixels, while precision is the ratio of correctly $detected$ pixels to all $detected$ pixels. A pixel is considered as $detected$ when its $P^3$ value is above the probability threshold indicated on the horizontal axis of the chart. A pixel is correctly $detected$ if it is found in the vicinity of a $detectable$ pixel. The $detectable$ pixels are defined as those near the maxima of the ground truth peaks with intensities higher than $2\sigma$-noise for the US spectra. To account for the shorter experiment time in the 15\% NUS spectra, a threshold of $5\sigma$-noise from the corresponding US spectra was used.}
            \label{fig:2D_GC}
        \end{figure}

        \begin{figure}[htbp]
            \centering
            \includegraphics[width=\textwidth]{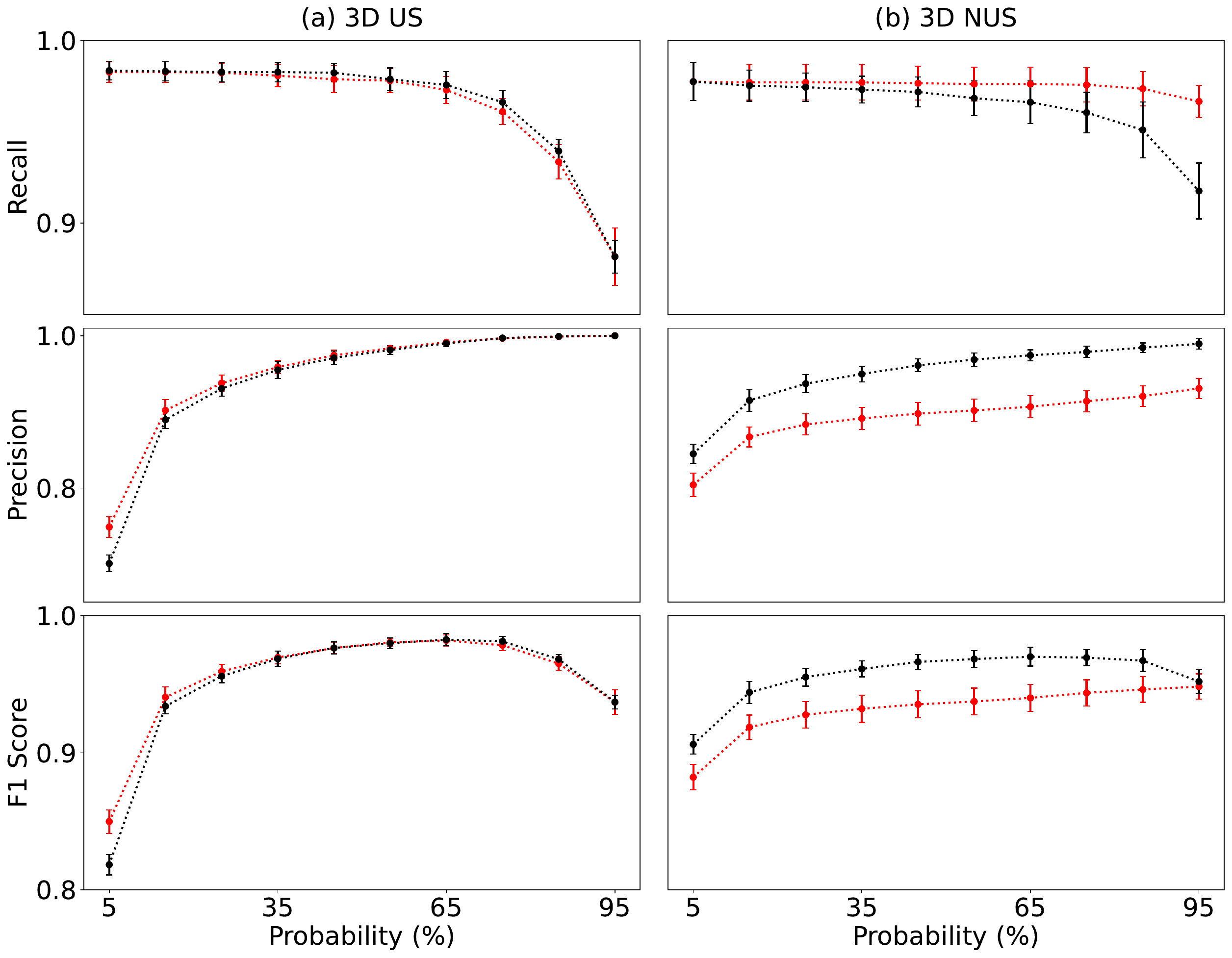}
            \caption{\textbf{Comparison of Peak Detection $P^3$ in Synthetic 3D Spectra Between Two MR-Ai Models Trained with Different Noise Distributions on Training Dataset.} Black and red lines with error bars represent $P^3$ results from MR-Ai trained on Cushy-Gaussian and Gaussian noise distributions, respectively, across 10 synthetic NMR spectra, each containing 256 peaks, for (a) 3D US and (b) 3D 15\% NUS reconstructed using CS-IST. F1 score is defined as the harmonic mean of the precision and recall scores with $F1=2\frac{Precision\times Recall}{Precision+ Recall}$. Recall is defined as the ratio of the correctly $detected$ pixels to all $detectable$ pixels, while precision is the ratio of correctly $detected$ pixels to all $detected$ pixels. A pixel is considered as $detected$ when its $P^3$ value is above the probability threshold indicated on the horizontal axis of the chart. A pixel is correctly $detected$ if it is found in the vicinity of a $detectable$ pixel. The $detectable$ pixels are defined as those near the maxima of the ground truth peaks with intensities higher than $2\sigma$-noise for the US spectra. To account for the shorter experiment time in the 15\% NUS spectra, a threshold of $5\sigma$-noise from the corresponding US spectra was used.}
            \label{fig:3D_GC}
        \end{figure}

\end{document}